\documentclass[twocolumn]{aastex62}

\usepackage{graphicx}

\usepackage{aas_macros}
\usepackage{amsmath}
\usepackage{cases}
\usepackage{apjfonts}
\usepackage{courier}
\usepackage[caption=false]{subfig}
\usepackage{multirow}

\usepackage[makeroom]{cancel}

\usepackage{booktabs}
\usepackage{threeparttable}
\usepackage{tabularx}

\usepackage{color}

\DeclareMathAlphabet{\mathsc}{OT1}{cmr}{m}{sc}
\def\testbx{bx}%
\DeclareRobustCommand{\ion}[2]{%
\relax\ifmmode
\ifx\testbx\f@series
{\mathbf{#1\,\mathsc{#2}}}\else
{\mathrm{#1\,\mathsc{#2}}}\fi
\else\textup{#1\,{\mdseries\textsc{#2}}}%
\fi}

\begin{document}


\title{Probing Cosmic Reionization and Molecular Gas Growth with TIME}

\author{G.~Sun}
\correspondingauthor{Guochao Sun}
\email{gsun@astro.caltech.edu}
\affiliation{California Institute of Technology, 1200 E. California Blvd., Pasadena, CA 91125, USA}

\author{T.-C.~Chang}
\affiliation{Jet Propulsion Laboratory, California Institute of Technology, 4800 Oak Grove Dr., Pasadena, CA 91109, USA}
\affiliation{California Institute of Technology, 1200 E. California Blvd., Pasadena, CA 91125, USA}
\affiliation{Institute of Astronomy and Astrophysics, Academia Sinica, Taipei 10617, Taiwan}

\author{B.~D.~Uzgil}
\affiliation{California Institute of Technology, 1200 E. California Blvd., Pasadena, CA 91125, USA}

\author{J.~J.~Bock}
\affiliation{California Institute of Technology, 1200 E. California Blvd., Pasadena, CA 91125, USA}
\affiliation{Jet Propulsion Laboratory, California Institute of Technology, 4800 Oak Grove Dr., Pasadena, CA 91109, USA}

\author{C.~M.~Bradford}
\affiliation{Jet Propulsion Laboratory, California Institute of Technology, 4800 Oak Grove Dr., Pasadena, CA 91109, USA}
\affiliation{California Institute of Technology, 1200 E. California Blvd., Pasadena, CA 91125, USA}

\author{V.~Butler}
\affiliation{Rochester Institute of Technology, 1 Lomb Memorial Drive, Rochester, NY 14623, USA}

\author{T.~Caze-Cortes}
\affiliation{Rochester Institute of Technology, 1 Lomb Memorial Drive, Rochester, NY 14623, USA}

\author{Y.-T.~Cheng}
\affiliation{California Institute of Technology, 1200 E. California Blvd., Pasadena, CA 91125, USA}

\author{A.~Cooray}
\affiliation{Department of Physics \& Astronomy, University of California, Irvine, CA 92697, USA}

\author{A.~T.~Crites}
\affiliation{Dunlap Institute for Astronomy \& Astrophysics, University of Toronto, 50 St George St, Toronto, ON, M5S 3H4, Canada}
\affiliation{Department of Astronomy \& Astrophysics, University of Toronto,
 50 St George St, Toronto, ON, M5S 3H4, Canada}
 \affiliation{California Institute of Technology, 1200 E. California Blvd., Pasadena, CA 91125, USA}

\author{S.~Hailey-Dunsheath}
\affiliation{California Institute of Technology, 1200 E. California Blvd., Pasadena, CA 91125, USA}

\author{N.~Emerson}
\affiliation{Steward Observatory, University of Arizona, 933 North Cherry Avenue, Tucson, Arizona 85721, USA}

\author{C.~Frez}
\affiliation{Jet Propulsion Laboratory, California Institute of Technology, 4800 Oak Grove Dr., Pasadena, CA 91109, USA}

\author{B.~L~.Hoscheit}
\affiliation{California Institute of Technology, 1200 E. California Blvd., Pasadena, CA 91125, USA}

\author{J.~Hunacek}
\affiliation{California Institute of Technology, 1200 E. California Blvd., Pasadena, CA 91125, USA}

\author{R.~P.~Keenan}
\affiliation{Steward Observatory, University of Arizona, 933 North Cherry Avenue, Tucson, Arizona 85721, USA}

\author{C.~T.~Li}
\affiliation{Institute of Astronomy and Astrophysics, Academia Sinica, Taipei 10617, Taiwan}

\author{P.~Madonia}
\affiliation{California Institute of Technology, 1200 E. California Blvd., Pasadena, CA 91125, USA}

\author{D.~P.~Marrone}
\affiliation{Steward Observatory, University of Arizona, 933 North Cherry Avenue, Tucson, Arizona 85721, USA}

\author{L.~Moncelsi}
\affiliation{California Institute of Technology, 1200 E. California Blvd., Pasadena, CA 91125, USA}

\author{C.~Shiu}
\affiliation{Department of Physics, Princeton University, Princeton, NJ 08544, USA}

\author{I.~Trumper}
\affiliation{College of Optical Sciences, University of Arizona, 1630 E. University Blvd., Tucson, AZ 85721, USA}

\author{A.~Turner}
\affiliation{Jet Propulsion Laboratory, California Institute of Technology, 4800 Oak Grove Dr., Pasadena, CA 91109, USA}

\author{A.~Weber}
\affiliation{Jet Propulsion Laboratory, California Institute of Technology, 4800 Oak Grove Dr., Pasadena, CA 91109, USA}

\author{T.~S.~Wei}
\affiliation{Institute of Astronomy and Astrophysics, Academia Sinica, Taipei 10617, Taiwan}

\author{M.~Zemcov}
\affiliation{Rochester Institute of Technology, 1 Lomb Memorial Drive, Rochester, NY 14623, USA}
\affiliation{Jet Propulsion Laboratory, California Institute of Technology, 4800 Oak Grove Dr., Pasadena, CA 91109, USA}

\begin{abstract}
Line intensity mapping (LIM) provides a unique and powerful means to probe cosmic structures by measuring the aggregate line emission from all galaxies across redshift. The method is complementary to conventional galaxy redshift surveys that are object-based and demand exquisite point-source sensitivity. The Tomographic Ionized-carbon Mapping Experiment (TIME) will measure the star formation rate (SFR) during cosmic reionization by observing the redshifted [\ion{C}{ii}] 158\,$\mu$m line ($6 \la z \la 9$) in the LIM regime. TIME will simultaneously study the abundance of molecular gas during the era of peak star formation by observing the rotational CO lines emitted by galaxies at $0.5 \la z \la 2$. We present the modeling framework that predicts the constraining power of TIME on a number of observables, including the line luminosity function, and the auto- and cross-correlation power spectra, including synergies with external galaxy tracers. Based on an optimized survey strategy and fiducial model parameters informed by existing observations, we forecast constraints on physical quantities relevant to reionization and galaxy evolution, such as the escape fraction of ionizing photons during reionization, the faint-end slope of the galaxy luminosity function at high redshift, and the cosmic molecular gas density at cosmic noon. We discuss how these constraints can advance our understanding of cosmological galaxy evolution at the two distinct cosmic epochs for TIME, starting in 2021, and how they could be improved in future phases of the experiment. 
\end{abstract}

\keywords{cosmology: observations -- theory -- dark ages, reionization, first stars -- large-scale structure of universe -- galaxies: ISM}


\section{Introduction}

Marked by the emergence of a substantial hydrogen-ionizing background sourced by the first generations of galaxies, the epoch of reionization (EoR) at $6 \la z \la 10$ represents a mysterious chapter in the history of the universe \citep{BL_2001, LF_2013, Stark_2016}. How the formation and evolution of the first, star-forming galaxies explains the history of reionization is a key question to be addressed. The answer lies in the cosmic star formation history (SFH) required to complete reionization by $z\sim6$, from which the net production and escaping of ionizing photons can be inferred. The study of the SFH also involves understanding how efficiently generations of stars formed out of the cold molecular gas supply regulated by feedback processes \citep{BY_2011, CW_2013}. A census of the molecular gas content across cosmic time offers a different perspective on the redshift evolution of cosmic star formation and is amenable to study at later times, including the pronounced peak (sometimes dubbed as the ``cosmic noon'') at $1 \lesssim z \lesssim 3$. 

Over the past decades, our understanding of the EoR has deepened from advances in the observational frontier of galaxies in the early universe. Dedicated surveys of high-redshift galaxies using the \textit{Hubble Space Telescope} (HST) have measured a large sample of galaxies out to redshift as high as $z \sim 8$ \cite[][]{Bouwens_2015LF, Finkelstein_2015}, which with the help of gravitational lensing has allowed the rest-frame ultraviolet (UV) galaxy luminosity function (LF) to be accurately constrained to a limiting magnitude of $M^{\rm AB}_{\rm UV} \ga -15$ \cite[][]{Atek_2015, Bouwens_2017, Yue_2018}. It is expected that, by the advent of the \textit{James Webb Space Telescope} (JWST), not only the currently limited sample size of $9 \la z \la 12$ galaxies and candidates \cite[][]{Ellis_2013, Oesch_2014, Oesch_2016, Oesch_2018}, but also constraints on the faint-end slope evolution of the UVLF, will be considerably enhanced \cite[][]{Mason_2015, Yung_2019}. Combined with the Thomson scattering optical depth $\tau_{\rm es} = 0.055\pm0.009$ inferred from the CMB temperature and polarization power spectra by \cite{Planck_XLVI}, the SFH based on a plausible faint-end extrapolation of the luminosity function suggests that the global reionization history could be explained by the ``known'' high-$z$ galaxy population. If the average escape fraction of their ionizing photons into the intergalactic medium (IGM) is in the range of 10--20\% \cite[e.g.,][]{Robertson_2015, Bouwens_2015EoR, Mason_2015, SF_2016, Madau_2017, Naidu_2020}, there will be no need to invoke additional ionizing sources such as Population~III stars and quasars. Nevertheless, the uncertainty associated with such an extrapolation indicates a fundamental limitation of surveys of individual objects---sources too faint compared with the instrument sensitivity, such as dwarf galaxies, are entirely missed by galaxy surveys, even though a significant fraction, if not the majority, of the ionizing photons are contributed by them (\citealt{Wise_2014}, \citealt{Trebitsch_2018}; but see also \citealt{Naidu_2020}). 

On the other hand, despite being subject to different sources of systematics (e.g., dust attenuation, source confusion, etc.), surveys at optical to far-infrared (FIR) wavelengths have revealed a general picture of the cosmic evolution of the star formation rate density \cite[SFRD, e.g.,][]{Cucciati_2012, Gruppioni_2013, Bourne_2017} and the stellar mass density \cite[SMD, e.g.,][]{HB_2006, PG_2008, Muzzin_2013}. Since the onset of galaxy formation at $z \gtrsim 10$, the star formation in galaxies first increased steadily with redshift as a result of continuous accretion of gas and mergers. The SFRD then reached a peak at redshift $z\sim2$ and declined by roughly a factor of 10 towards $z=0$. Changes in the supply of cold molecular gas as the fuel of star formation may be responsible for the decline in the cosmic star formation at $z \lesssim 2$. The coevolution of the cosmic molecular gas density and the SFRD is therefore of significant interest \cite[][]{Popping_2014, Decarli_2016}. Unfortunately, the faintness of cold ISM tracers, such as rotational lines of carbon monoxide (CO), has restricted observations to only the more luminous galaxies \cite[][]{Tacconi_2013, Decarli_2016, Riechers_2019, Decarli_2020}. A census of the bulk molecular gas, however, requires a complete CO survey down to the very faint end of the line luminosity function \cite[see e.g.,][]{Uzgil_2019}.

As an alternative method complementary to sensitivity-limited surveys of point sources, line intensity mapping (LIM) measures statistically the aggregate line emission from the entire galaxy population \cite[][]{VL_2010}, including those at the very faint end of the luminosity distribution that are difficult to detect individually. First pioneered in the deep survey of \ion{H}{i} 21cm line at $z\sim1$ to probe the baryon acoustic oscillation (BAO) peak as a cosmological standard ruler \citep{Chang_2008, Chang_2010}, LIM provides an economical way to survey large-scale structure (LSS) without detecting individual line emitters. Over the past decade, LIM has received increasing attention in a variety of topics in astrophysics and cosmology (see the recent review by \citealt{Kovetz_2017}, and references therein). 

\begin{figure}[h!]
\centering
\includegraphics[width=0.48\textwidth]{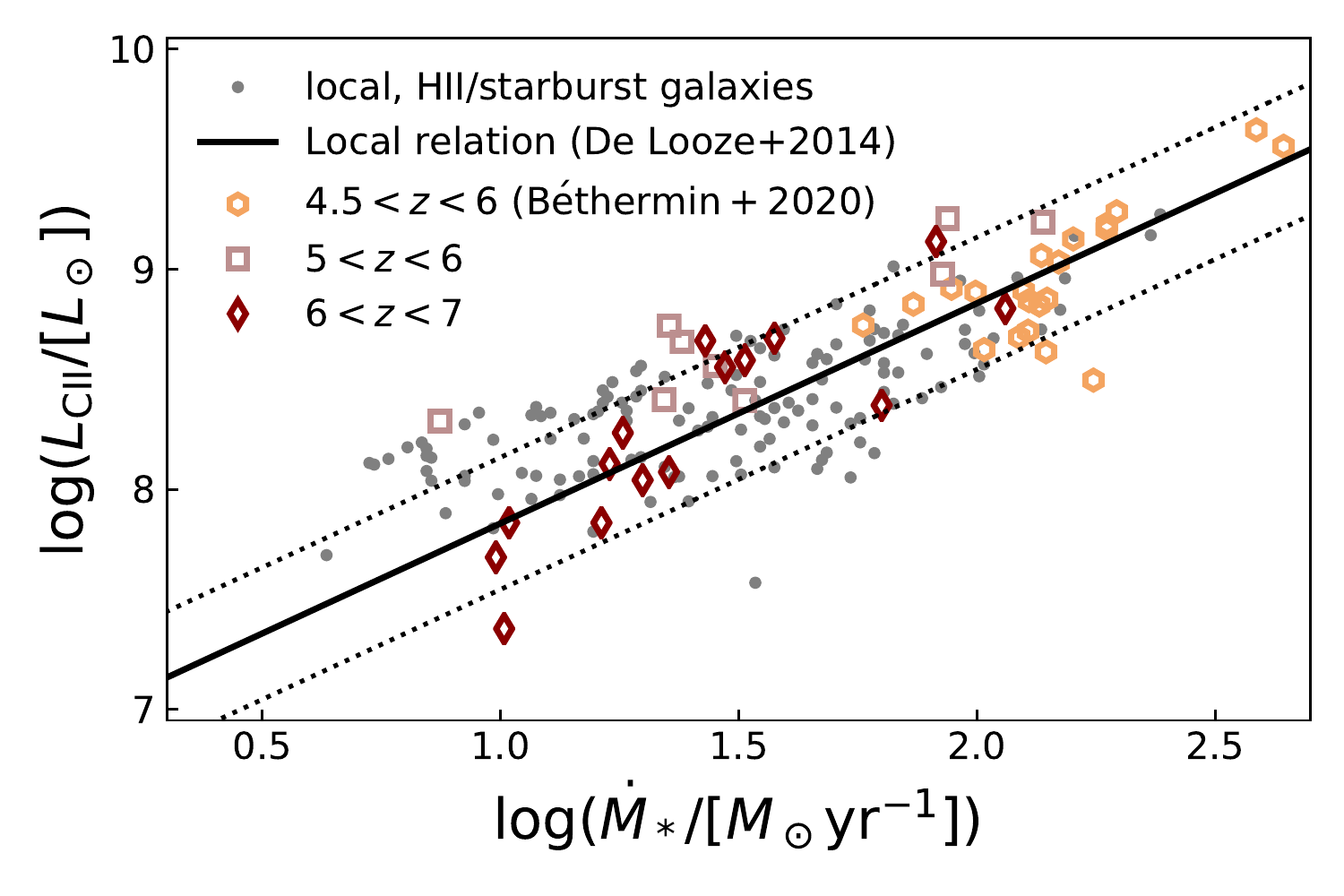}
\caption{The observed correlation between [\ion{C}{ii}] luminosity and the total SFR (UV + IR) of galaxies in local universe and $z\ga5$. Measurements from the ALPINE survey are shown by the hexagons for sources with dust continuum detection \citep{Bethermin_2020}. Additional $z\ga5$ data shown by the squares and diamonds are compiled by \citet{Matthee_2019}. The solid line represents the best-fit relation to local, \ion{H}{ii}/starburst galaxies from \citet{De_Looze_2014}, which has a scatter of about 0.3\,dex as indicated by the dotted lines. Both the fitting relation and data points are homogenized to be consistent with the same Salpeter IMF assumed throughout this paper.}
\label{fig:lcii_sfr}
\end{figure}

In addition to the 21cm line, a number of other emission lines have also been proposed as tracers for different phases of the ISM and the IGM, including Ly$\alpha$, H$\alpha$, [\ion{C}{ii}], CO and so forth. Among these lines, [\ion{C}{ii}] is particularly interesting for constraining the global SFH. Thanks to the abundance of carbon, its low ionization potential (11.3\,eV), and the modest equivalent temperature of fine-structure splitting (91\,K), the 157.7\,$\mu$m ${^2P_{3/2}} \rightarrow {^2P_{1/2}}$ transition of [\ion{C}{ii}] is the major coolant of neutral ISM and can comprise up to 1\% of the total FIR luminosity of galaxies. As illustrated in Figure~\ref{fig:lcii_sfr}, a tight, nearly redshift-independent correlation between [\ion{C}{ii}] line luminosity and the SFR has been identified in both nearby galaxies \cite[e.g.,][]{De_Looze_2011, De_Looze_2014, HerreraCamus_2015} and distant galaxies at redshift up to $z\sim5$ as revealed by deep ALMA observations \cite[e.g.,][]{Capak_2015, Matthee_2019, Schaerer_2020}, which makes [\ion{C}{ii}] a promising SFR tracer. Even though some observations \cite[][]{Willott_2015, Bradac_2017} and semi-analytical models \cite[][]{Lagache_2018} suggest a larger scatter in $L_{\ion{C}{ii}}$--SFR relation at high redshifts, the general reliability of using [\ion{C}{ii}] to trace star formation has motivated a number of LIM experiments targeting at the redshifted [\ion{C}{ii}] signal from the EoR, including TIME \citep{Crites_2014SPIE}, CONCERTO \cite[CarbON \ion{C}{ii} line in post-rEionization and ReionizaTiOn epoch;][]{CONCERTO_2020}, the Cerro Chajnantor Atacama Telescope-prime \cite[CCAT-prime;][]{Stacey_2018SPIE}, and the Deep Spectroscopic High-redshift Mapper \cite[DESHIMA;][]{Endo_2019}. Meanwhile, on large scales [\ion{C}{ii}] intensity maps complement surveys of the 21cm line tracing the neutral IGM. The [\ion{C}{ii}]--21cm cross-correlation provides a promising means to overcome foregrounds of 21cm data and to measure the size evolution of ionized bubbles during reionization \cite[e.g.,][]{Gong_2012, Dumitru_2019}. 

TIME is a wide-bandwidth, imaging spectrometer array \citep{Crites_2014SPIE, Hunacek_2016JLTP, Hunacek_2018JLTP} designed for simultaneously (1) conducting the first tomographic measurement of [\ion{C}{ii}] intensity fluctuations during the EoR, and (2) investigating the molecular gas growth at cosmic noon by measuring the intensity fluctuations of rotational CO lines, which also present a source of foreground contamination \cite[][]{LT_2016, Cheng_2016, Sun_2018, CCB_2020}. TIME will operate at the ALMA 12-m Prototype Antenna (APA) at the Arizona Radio Observatory (ARO) in Kitt Peak, Arizona, for 1000 hours of winter observing time, starting in 2021. Meanwhile, the instrument may observe from the Leighton Chajnantor Telescope (LCT) in Chile in the future, enabling a significantly longer observing time and lower loading. We refer to this phase as TIME-EXT hereafter, which, as will be discussed in Section~\ref{sec:results}, represents a case where the constraining power from [\ion{C}{ii}] auto-power spectrum is pushed to the limit. In this paper, we will describe in detail the modeling framework that allows us to demonstrate the science cases and forecast parameter constraints for the two important cosmic epochs. 

The remainder of this paper is structured as follows. In Section~\ref{sec:observables}, we first provide an overview of the types of measurements TIME (and TIME-EXT) performs, together with the corresponding observables. In Section~\ref{sec:model}, we describe the modeling framework for the various signals TIME will observe, which provide physical constraints on the galaxy evolution during reionization and the molecular gas growth history near cosmic noon. We then describe the survey strategy of TIME in Section~\ref{sec:survey}. In Section~\ref{sec:results} we present the predicted sensitivities to different observables as well as TIME's constraining power on various physical quantities. We elaborate the issue of foreground contamination and our mitigation strategies in Section~\ref{sec:foreground}. We discuss the implications and limitations of TIME(-EXT) measurements, and briefly describe the scientific opportunities for a next-generation experiment, TIME-NG, in synergy with other EoR probes in Section~\ref{sec:discuss}, before summarizing our main conclusions in Section~\ref{sec:conclusions}. Throughout the paper, we assume cosmological parameters consistent with recent measurements by Planck Collaboration XIII \citep{Planck_XIII}.

\begin{table*}[t]
\centering
\caption{Emission lines observable to TIME}
\label{tb:lines_time}
\begin{tabular}{cccccc}
\toprule
\toprule
Line & Wavelength, $\lambda_{\rm rf}$ & Observable $z$ range & Intensity, $\bar{I}_{\rm 250\,GHz}$ & ($K_{\rm \perp, min}$, $K_{\rm \perp, max}$) & ($K_{\rm \parallel, min}$, $K_{\rm \parallel, max}$) \\
& ($\mu$m) & & (Jy/sr) & ($h$/Mpc) & ($h$/Mpc) \\
\hline
$[\ion{C}{ii}]$ & 158 & (5.29, 8.51) & 384 & (0.061, 5.471) & (0.023, 0.511) \\
$[\ion{C}{i}]$ & 609 & (0.63, 1.46) & 198 & (0.186, 16.78) & (0.005, 0.100) \\
CO(3--2) & 867 & (0.15, 0.73) & 234 & (0.535, 48.11) & (0.005, 0.100) \\
CO(4--3) & 650 & (0.53, 1.31) & 510 & (0.212, 19.12) & (0.004, 0.099) \\
CO(5--4) & 520 & (0.91, 1.88) & 544 & (0.144, 13.00) & (0.005, 0.103) \\
CO(6--5) & 434 & (1.29, 2.46) & 482 & (0.115, 10.38) & (0.005, 0.109) \\
CO(7--6) & 372 & (1.67, 3.04) & 320 & (0.099, 8.928) & (0.005, 0.116) \\
\bottomrule
\end{tabular}
\end{table*}


\section{Observables for TIME} \label{sec:observables}

\subsection{Observables Internal to TIME Datasets}

The primary goal of TIME is to constrain the SFH during the EoR by measuring the spatial fluctuations of [\ion{C}{ii}] line intensity. In particular, we will extract physical information of interest from the two-point statistics of the [\ion{C}{ii}] intensity field, namely its auto-correlation power spectrum $P_{\ion{C}{ii}}(k)$, which can be directly measured from TIME's data cube. Combining $P_{\ion{C}{ii}}(k)$ measured by TIME with other observations such as the CMB optical depth, we are able to constrain the global history of reionization. 

TIME will also measure the CO and [\ion{C}{i}] emission from galaxy populations from intermediate redshifts ($0.5 \la z \la 2$).  These signals are strong and will be interlopers from the standpoint of the extraction of the [\ion{C}{ii}] signal, but they are interesting in their own right as a constraint on the evolving molecular gas content in galaxies. Without relying on external data, we can distinguish these foreground lines from the [\ion{C}{ii}] signal by cross-correlating pairs of TIME bands that correspond to frequencies of two lines emitted from the same redshift (and thus tracing the same underlying LSS). In this case, [\ion{C}{ii}] emission only contributes to the uncertainty rather than the signal of the cross-correlation power spectra (see Section~\ref{sec:survey-sensitivity}). 

\subsection{Observables Requiring Ancillary Data}

In addition to observables that can be directly measured from TIME datasets, we also consider joint analysis with ancillary data, in particular cross-correlations with external tracers of the LSS at both low and high redshifts. Based on surveys of available LSS tracers, we investigate the prospects for (1) measuring the angular correlation function, $\omega_{\ion{C}{ii} \times \mathrm{LAE}}$, between [\ion{C}{ii}] intensity and Ly$\alpha$ emitters (LAEs) identified from narrowband data at $z\sim6$, and (2) measuring the cross-power spectra, $P_{\rm CO \times gal}$, between foreground CO lines and near-IR selected galaxies at the same redshifts. These cross-correlation analyses will not only help us better distinguish the low-$z$ and high-$z$ signals, but also shed light onto physical conditions of the overlapping galaxy population traced by these emission lines.


\section{Models} \label{sec:model}

Following the introduction of observables for TIME, in this section we first describe our models for tracers of the LSS (Section~\ref{sec:model:lss}), including [\ion{C}{ii}] emission from the EoR and foreground CO/[\ion{C}{i}] lines internal to TIME data sets, and external tracers like low-$z$ galaxies and high-$z$ LAEs to be cross-correlated with TIME data sets. We then present models for how these tracers can reveal about (1) the molecular gas content of galaxies near cosmic noon (Section~\ref{sec:model-mh2}), and (2) the SFH of EoR galaxies at $z\ga6$ and its implications for the EoR history as the primary goal of TIME experiment (Section~\ref{sec:model-eor}).

\subsection{Tracers of Large-Scale Structure} \label{sec:model:lss}

Our modeling framework of LSS tracers captures the two major line signals TIME will directly measure, namely the target [\ion{C}{ii}] line from the EoR and foreground CO lines from cosmic noon. It also predicts the statistics of high-redshift Ly$\alpha$ emitters (LAEs), whose spatial distribution can be cross-correlated with [\ion{C}{ii}] intensity maps to serve as an independent validation of the auto-correlation analysis, which is subject to more complicated foreground contamination. Because observational constraints on the mean emissivity of [\ion{C}{ii}]/CO emitters and their luminosity distributions are still limited, we adopt a phenomenological approach by connecting the [\ion{C}{ii}] and CO line intensities to the observed cosmic infrared background (CIB) and UV LFs, respectively, such that the model can be readily constrained by existing measurements while being flexible enough to explore the possible deviations from the fiducial case. Table~\ref{tb:lines_time} lists the emission lines observable to TIME, including their rest-frame wavelengths, mean intensities, together with their observable redshift and scale ranges (see Section~\ref{sec:survey-strategy} for details about the Fourier space that TIME measures). 


\subsubsection{Carbon Monoxide and Neutral Carbon Near Cosmic Noon} \label{sec:model:CO}

As summarized in Table~\ref{tb:lines_time}, several low-redshift foreground emission lines are brighter than the EoR [\ion{C}{ii}] line, and can be blended with the [\ion{C}{ii}] signal in an auto-correlation analysis. On the contrary, in-band cross-correlations measure (the product of) two line intensities tracing the same LSS distribution at a given redshift. Because low-$J$ CO line ratios are well known and CO correlates with molecular hydrogen, these population-averaged line strengths provide valuable insights into physical conditions of molecular gas clouds from which they originate. 

To model the emission lines near cosmic noon, we first take a CIB model of the infrared luminosity, $L_{\rm IR}$, of galaxies as a function of their host halo mass and redshift. In short, we fit a halo model \citep{Cooray_Sheth_2002} that describes the clustering of galaxies, whose SEDs are assumed to resemble a modified black-body spectrum, to the CIB anisotropy observed in different FIR bands. The resulting best-fit model is characterized by spectral indices of modified black-body spectrum, the dust temperature, and factors of mass and redshift dependence. Given that it is a well-established model whose variations have been applied to numerous studies of the CIB \cite[e.g.,][]{Shang_2012, WD_2017a, WD_2017b}, the CMB \cite[e.g.,][]{Desjacques_2015, Shirasaki_2019}, and line intensity mapping \cite[e.g.,][]{Cheng_2016, Serra_2016, Pullen_2018, Switzer_2019, Sun_2019}, we refrain from going into further details about the CIB model and refer interested readers to the aforementioned papers for more information. In this work, we adopt the CIB model described in \citet{WD_2017b} and \citet{Sun_2019}. 

Combining the total infrared luminosities derived from the CIB model and its correlation with the CO luminosity, we can express the CO luminosity as
\begin{equation}
\log \left[ \frac{L^\prime_{\mathrm{CO}(J \rightarrow J-1)}}{\mathrm{K\, km\,s^{-1}\, pc^2}} \right] = \alpha^{-1} \left[ \log \left( \frac{L_{\rm IR}}{L_{\odot}} \right) - \beta \right] + \log r_J~,
\label{eq:LCO}
\end{equation}
where we adopt $\alpha=1.27$ and $\beta=-1.00$ (see Table~\ref{tb:model_params}) as fiducial values for CO(1-0) transition \citep{Kamenetzky_2016}. Provided that the slope $\alpha$ does not evolve strongly with increasing $J$ (e.g., \citealt{CW_2013}; \citealt{Kamenetzky_2016}; but see also \citealt{Greve_2014}), higher $J$ transitions can be described by a fixed scaling factor $r_J$, whose values are determined from a recent study by \citet{Kamenetzky_2016} about the CO spectral line energy distributions (SLEDs; also known as the CO rotational ladder) based on \textit{Herschel}/SPIRE observations. Specifically, for the excitation of CO we take $r_{3}=0.73$, $r_{4}=0.57$, $r_{5}=0.32$, $r_{6}=0.19$, and terminate the $J$ ladder at $r_{7}=0.1$ \citep{Kamenetzky_2016} as the contribution from higher $J$'s becomes negligible. For simplicity, our model ignores the variation of the CO SLEDs among individual galaxies, which needs to be investigated in future work. Even though ratios of adjacent CO lines do not vary as much as the full CO rotational ladder, the diverse SLEDs observed (especially at higher $J$) will affect power spectral measurements and introduce additional systematics in the inference of molecular gas content from mid- or high-$J$ CO observations \cite[][]{CW_2013, NK_2014, MSL_2015, Mashian_2015}. As a compromise, we include a log-normal scatter, $\sigma_{\rm CO}$, to describe the level of dispersion in the strengths of all CO lines independent of $J$. As discussed in \citet{NK_2014} and \citet{MSL_2015}, such a common scatter in the CO excitation ladder might be attributed to the stochasticity in global modes of star formation, which can be characterized by the SFR surface density of galaxies. The CO luminosity can be converted from $L^\prime_{\rm CO}$ (in $\mathrm{K\,km\,s^{-1}\,pc^2}$) to $L_{\rm CO}$ (in $L_{\odot}$) by
\begin{equation}
L_{\mathrm{CO}(J \rightarrow J-1)} = 3.2 \times 10^{-11} \left[ \frac{\nu_{\mathrm{CO}(J \rightarrow J-1)}}{\mathrm{GHz}} \right]^3 L^\prime_{\mathrm{CO}(J \rightarrow J-1)}~.
\end{equation}

\begin{figure}[h!]
\centering
\includegraphics[width=0.48\textwidth]{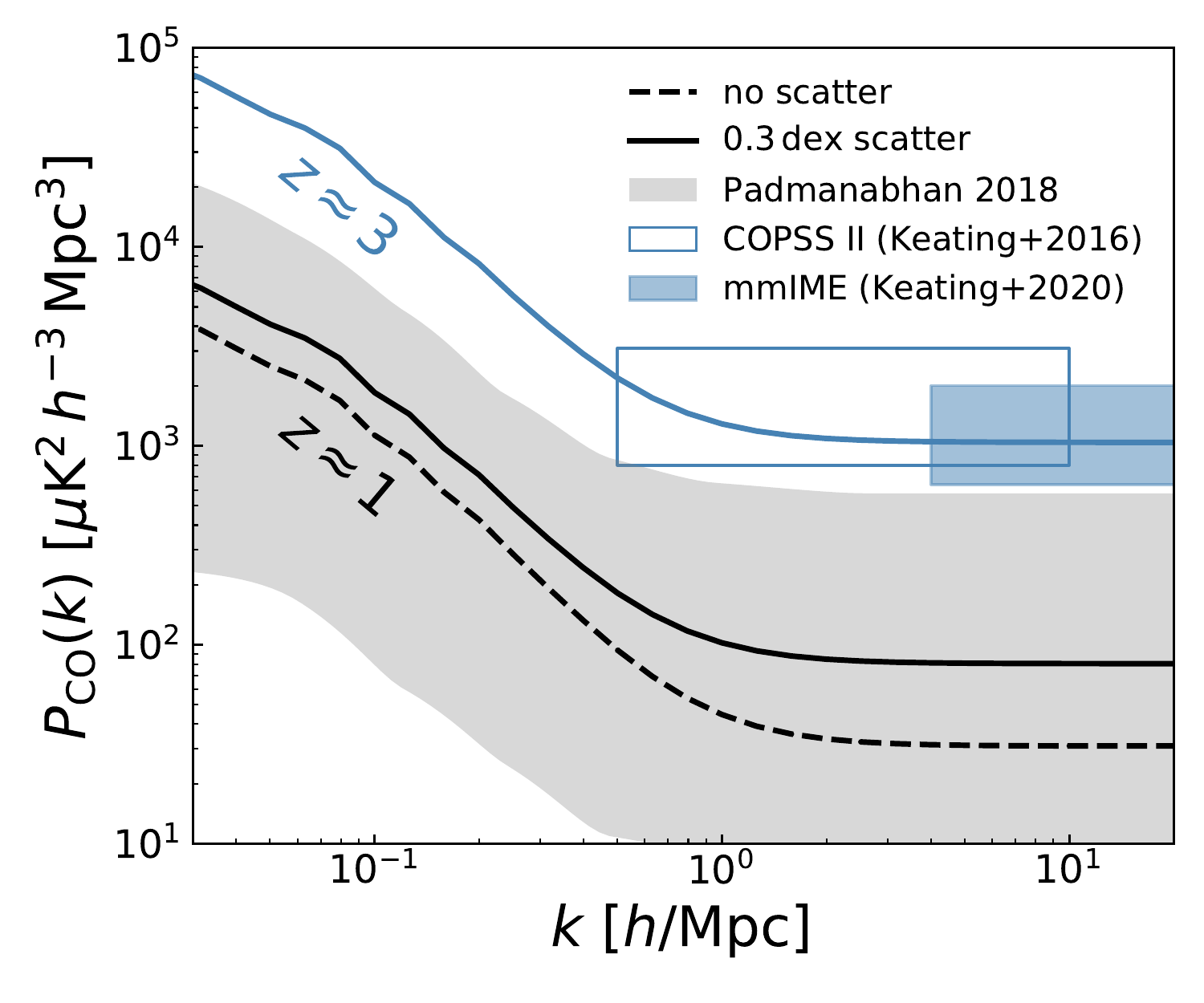}
\caption{A comparison of the CO(1-0) auto-correlation power spectra predicted by our fiducial model with results in the literature. The COPSS~II experiment \cite[][]{Keating_2016} reported a marginal detection of CO shot-noise power spectrum $2000_{-1200}^{+1100}\,\mu \mathrm{K}^2 h^{-3} \mathrm{Mpc}^3$ at $z\sim3$ (from a refined analysis by \citealt{Keating_2020}). Also shown is the independently measured shot-noise power $1140_{-500}^{+870}\,\mu \mathrm{K}^2 h^{-3} \mathrm{Mpc}^3$ at $z\sim3$ from mmIME \citep{Keating_2020}. \citet{Padmanabhan_2018} fits an empirical model to a compilation of available observational constraints on CO line emissivities at different redshifts. The solid and dashed curves represent the power spectra with and without including a 0.3\,dex lognormal scatter in the $L_{\rm CO}$--$L_{\rm IR}$ relation, respectively.}
\label{fig:coaps_z1p0}
\end{figure}

The fluctuations of CO emission can be written as the sum of a clustering term proportional to the power spectrum $P_{\delta\delta}$ of the underlying dark matter density fluctuations, and a scale-independent shot-noise term,\,\footnote{For clarity, $J$ is dropped in the expressions of CO power spectrum.}
\begin{equation}
P_{\rm CO}(k, z) = \bar{I}^2_{\rm CO}(z) \bar{b}^2_{\rm CO}(z) P_{\delta \delta}(k, z) + P^{\rm shot}_{\rm CO}(z)~.
\end{equation}
The mean CO intensity is defined as
\begin{equation}
\bar{I}_{\rm CO}(z) = \int \mathrm d M \frac{\mathrm d n}{\mathrm d M} \frac{L_{\rm CO}[L_{\rm IR}(M, z)]}{4\pi D^2_L} y(z) D^2_A~,
\end{equation}
where the integration has a lower bound of $10^{10}\,M_\odot$ \cite[][]{WD_2017b}, below which the contribution to the total CO line intensity is expected to be negligible according to the CIB model, and an upper bound of $10^{15}\,M_\odot$. $\mathrm d n / \mathrm d M$ is the dark matter halo mass function (HMF), which is defined for the virial mass $M_{\rm vir}$ following \citet{Tinker_2008} throughout this work. $D_L$ and $D_A$ are the luminosity and comoving angular diameter distances, respectively, and $y(z) \equiv d \chi / d \nu = \lambda_{\rm rf}(1+z)^2/H(z)$ maps the frequency into the line-of-sight (LOS) distance, where $\lambda_{\rm rf}$ is the rest-frame wavelength of the emission line. $\bar{b}_{\rm CO}(z)$ denotes the luminosity-averaged halo bias factor of CO as a tracer of the underlying dark matter density field, namely
\begin{equation}
\bar{b}_{\rm CO}(z) = \frac{\int \mathrm d M (\mathrm d n/\mathrm d M) b(M,z) L_{\rm CO}[L_{\rm IR}(M, z)]}{\int \mathrm d M (\mathrm d n/\mathrm d M) L_{\rm CO}[L_{\rm IR}(M, z)]}~.
\end{equation}
The shot-noise term is defined as
\begin{equation}
P^{\rm shot}_{\rm CO}(z) = \int \mathrm d M \frac{\mathrm d n}{\mathrm d M} \left\{ \frac{L_{\rm CO}[L_{\rm IR}(M, z)]}{4\pi D^2_L} y(z) D^2_A \right\}^2~.
\end{equation}
For simplicity, we neglect effects on the intensity fluctuations due to sub-halo structures such as satellite galaxies, which could be non-trivial at the redshifts from which CO lines are emitted. Nonetheless, a halo occupation distribution (HOD) formalism can be readily introduced in order to take into account such effects \cite[][]{Serra_2016, Sun_2019}. We also note that the presence of the scatter $\sigma_{\rm CO}$ in $L_{\rm CO}$ for a given $L_{\rm IR}$ affects the clustering and shot-noise components differently. To account for such an effect, we adopt the same multiplicative factors $\mathcal{S}_I$ and $\mathcal{S}_{SN}$ ($\log \mathcal{S}_I=0.5 \sigma^2_{\rm CO} \ln 10$ for the mean intensity and $\log \mathcal{S}_{SN}=2 \sigma^2_{\rm CO} \ln 10$ for the shot-noise power, respectively) as presented in \citet{Sun_2019} to scale the two components and obtain the correct form of power spectrum in the presence of scatter. Figure~\ref{fig:coaps_z1p0} shows how our model predictions with and without including a scatter of $\sigma_{\rm CO}=0.3$\,dex compare to the constraints on CO(1-0) power spectrum at $z\sim1$ derived from a compilation of observations by \cite{Padmanabhan_2018}. Also shown in blue is a comparison between our model prediction and the 68\% confidence intervals on CO(1-0) shot-noise power at $z\approx3$ from a revised analysis of COPSS~II \citep{Keating_2016} data, as well as a recent, independent measurement from the Millimeter-wave Intensity Mapping Experiment (mmIME) at 100\,GHz by \citet{Keating_2020}. 

Due to the resemblance in critical density, fine-structure lines of neutral carbon (\ion{C}{i}) tightly correlate with CO lines independent of environment, as demonstrated by observations of molecular clouds in galaxies over a wide range of redshifts. The observed correlation and coexistence of \ion{C}{i} and CO in molecular clouds can be explained by modern PDR models more sophisticated than simple, plane-parallel models \cite[e.g.,][]{BPV_2015, Glover_2015}. \ion{C}{i} has therefore been recognized as a promising tracer of molecular gas in galaxies at both low and high redshift \citep{Israel_2015, Jiao_2017, Valentino_2018, Nesvadba_2019}. Both fine-structure transitions of \ion{C}{i} at 492\,GHz and 809\,GHz are in principle detectable by TIME, but because the latter is from a much higher redshift and in fact spectrally blended with CO(7-6) transition, we will only consider [\ion{C}{i}] ${^{3}P_{1}} \rightarrow {^{3}P_{0}}$ transition at 492\,GHz (609\,$\mu$m) in this work and refer to it as the [\ion{C}{i}] line henceforth for brevity. We also choose to not include CO(7-6) line (and higher-$J$ transitions) in our subsequent analysis. Recent far-infrared observations suggest an almost linear correlation between [\ion{C}{i}] and CO(1-0) luminosities \cite[e.g.,][]{Jiao_2017}, so we empirically model the [\ion{C}{i}] line luminosity by
\begin{equation}
\log \left[ \frac{L^\prime_{\ion{C}{i}}}{\mathrm{K\, km\,s^{-1}\, pc^2}} \right] = \alpha^{-1} \left[ \log \left( \frac{L_{\rm IR}}{L_{\odot}} \right) - \beta \right] + \log r_{\ion{C}{i}}~,
\label{eq:LCI}
\end{equation}
where $\alpha$ and $\beta$ are set to the same values as in the CO case, while $r_{\ion{C}{i}}=0.18$. Equation~(\ref{eq:LCI}) provides a good fit to the observed $L_{\ion{C}{i}}$--$L_{\rm IR}$ relation covering a wide range of galaxy types and redshifts \citep{Valentino_2018, Nesvadba_2019}. 

\begin{table}[h!]
\centering
\caption{Fiducial model parameters for sensitivity analysis}
\label{tb:model_params}
\begin{tabular}{cccc}
\toprule
\toprule
Parameter & Description & Value & Prior \\
\hline
$\alpha$ & $L_{\rm CO}$--$L_{\rm IR}$ relation & $1.27$ & $[0.5, 2]$ \\
$\beta$ & $L_{\rm CO}$--$L_{\rm IR}$ relation & $-1.00$ & $[-2, 0]$ \\
$\sigma_{\rm CO}$ & scatter in $L_{\rm CO}(L_{\rm IR})$ & $0.3$\,dex & $[0, 1]$ \\
$a$ & $L_{\ion{C}{ii}}$--$L_{\rm UV}$ relation & $1.0$ & $[0.5, 2]$ \\
$b$ & $L_{\ion{C}{ii}}$--$L_{\rm UV}$ relation & $-20.6$ & $[-21.5, -19.5]$ \\
$\sigma_\ion{C}{ii}$ & scatter in $L_{\ion{C}{ii}}(L_{\rm UV})$ & $0.2$\,dex & $[0, 1]$ \\
$\xi$ & SFE in low-mass halos & 0 & $[-0.5, 0.5]$ \\
$f_{\rm esc}$ & escape fraction & 0.1 & $[0, 1]$ \\
\bottomrule
\end{tabular}
\end{table}


\subsubsection{Low-$z$ NIR-Selected Galaxies} \label{sec:model:gal}

Cross-correlating intensity fluctuations of aforementioned, low-redshift target lines for TIME with external tracers, such as galaxy samples, provides an independent measure of the line interlopers blended with the EoR [\ion{C}{ii}] signal. Therefore, we present an analytical description here to estimate how well TIME will be able to detect the cross-correlation between CO intensity maps and the distribution of near-IR (NIR) selected galaxies, whose redshifts are available from either spectroscopy ($\sigma_z/(1+z) \gtrsim 0.001$) or high-quality photometry ($\sigma_z/(1+z) \gtrsim 0.01$), such as those from the COSMOS/UVISTA survey \citep{Laigle_2016}. As discussed in \citet{Sun_2018}, the same galaxy samples can be utilized to clean foreground CO lines following a targeted masking strategy. 

\begin{figure}[h!]
\centering
\includegraphics[width=0.48\textwidth]{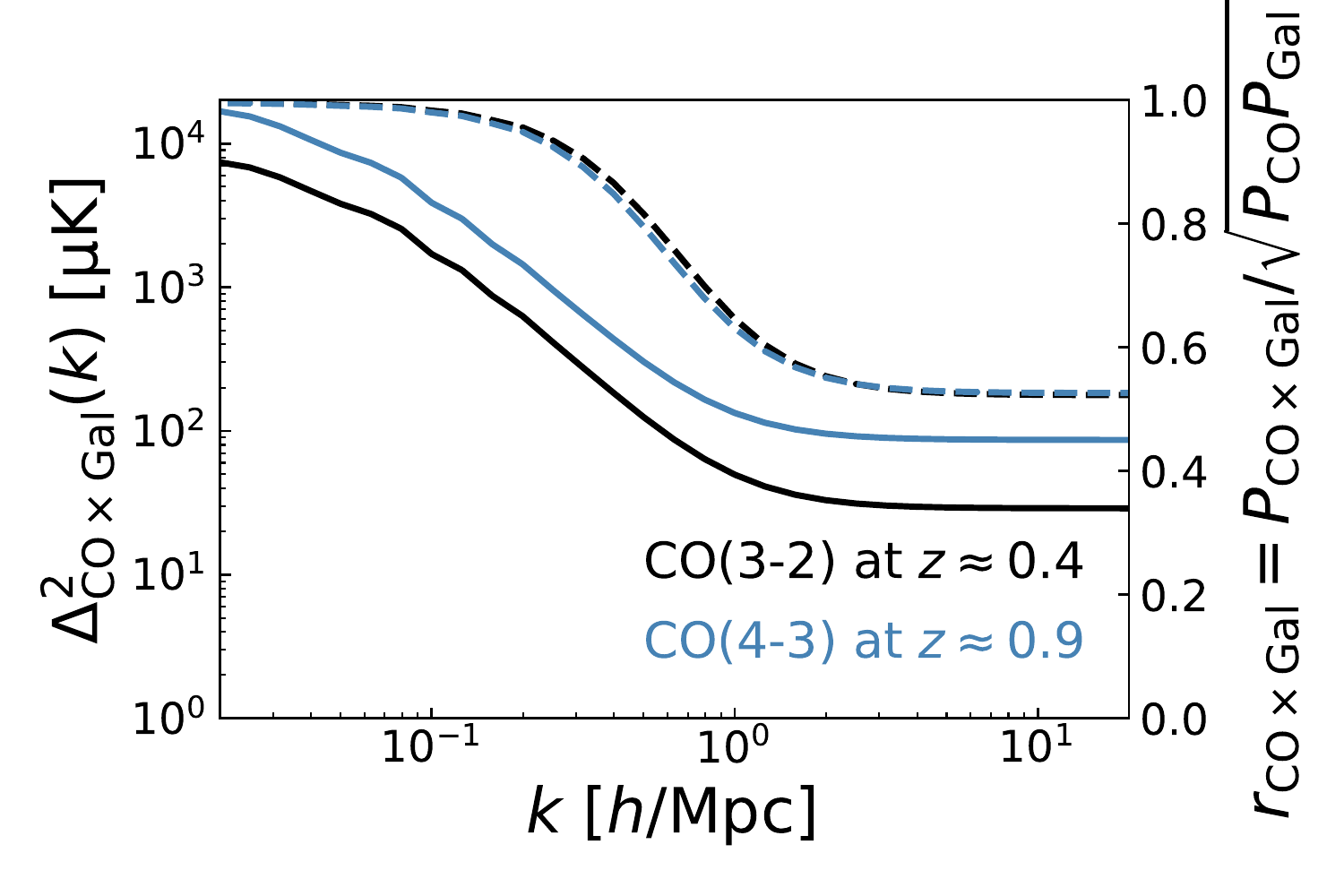}
\caption{Predicted cross-power spectrum $P_\mathrm{CO \times gal}$ and  cross-correlation coefficient $r_\mathrm{CO \times gal}(k)$ of CO(3-2) and CO(4-3) lines with galaxy distributions at $z\approx0.4$ and $z\approx0.9$, respectively. The partial correlation at large $k$ is because the cross shot-noise term only probes CO emitters overlapped with the galaxy samples.}
\label{fig:cogalccc}
\end{figure}

Specifically, the total power spectrum of the galaxy density field is the sum of a clustering term and a shot-noise term
\begin{equation}
P_{\rm gal}(k,z) = P_{\rm gal}^{\rm clust}(k,z) + P_{\rm gal}^{\rm shot}(z) = \bar{b}^2_{\rm gal}(z) P_{\delta \delta}(k, z) + \frac{1}{n_{\rm gal}}~.
\end{equation}
The bias factor of galaxies can be derived from the halo bias via
\begin{equation}
\bar{b}_{\rm gal}(z) = \frac{\int_{>M_{\rm crit}} \mathrm d M (\mathrm d n / \mathrm d M) b(M, z) \left[ N_{\rm cen} + N_{\rm sat}(M,z) \right]}{\int_{>M_{\rm crit}} \mathrm d M (\mathrm d n / \mathrm d M)}~,
\end{equation}
where $M_{\rm crit}$ is the halo mass corresponding to the critical stellar mass used for galaxy selection. $N_{\rm cen}$ and $N_{\rm sat}$ give the halo occupation statistics, namely the numbers of central galaxy and satellite galaxies per halo. For simplicity, we set $N_{\rm cen}$ to 1 for $M>10^{10}\,M_\odot$ and zero otherwise, and ignore the presence of satellite galaxies by setting $N_{\rm sat}=0$. Note that the denominator is simply the galaxy number density $n_{\rm gal}$. The cross-power spectrum between the galaxy density and the CO intensity fields is therefore
\begin{equation}
P_{\rm CO \times gal}(k,z) = \bar{b}_{\rm gal}(z) \bar{b}_{\rm CO}(z) \bar{I}_{\rm CO}(z) P_{\delta \delta}(k, z) + \frac{\bar{I}_{\rm CO, gal}(z)}{n_{\rm gal}(z)}~,
\label{eq:COxGal}
\end{equation}
where $\bar{I}_{\rm CO, gal}$ represents the mean intensity of a given CO line attributed to the selected galaxy samples with halo mass $M>M_\mathrm{crit}$, which is an important quantity extractable from the cross shot-noise power as discussed in \citet{Wolz_2017b}. In the shot-noise regime, the cross-power spectrum effectively probes the mean CO line luminosity $\langle L_{\rm CO} \rangle_\mathrm{g}$ of individual galaxy samples, given prior information of their redshifts. The subscript $\mathrm{g}$ indicates the mean CO luminosity of the galaxy sample only. Figure~\ref{fig:cogalccc} shows the cross-power spectrum together with the cross-correlation coefficient $r_{\rm CO \times gal}(k) = P_{\rm CO \times gal}(k) / \sqrt{P_{\rm CO}(k) P_{\rm gal}(k)}$ between the CO intensity maps TIME measures and galaxy distributions at $z\approx0.4$ and $z\approx0.9$.

Both photometric redshift $z_{\rm phot}$ and spectroscopic redshift $z_{\rm spec}$ can be considered, as long as the corresponding $n_{\rm gal}$ allows a sufficiently large statistical sample to be selected. For photometric data, we examine two examples where galaxies are cross-correlated with CO(3-2) line and CO(4-3) line at $z\approx0.4$ and $z\approx0.9$, respectively. We set $M_{\rm crit}=5\times10^{11}\,M_\odot$, which corresponds to a stellar mass of $M_* \gtrsim 2\times10^9\,M_\odot$ at $z\sim1$ \cite[][]{Sun_2018}, comparable to the completeness limit of deep, near-IR selected catalogs like the COSMOS/UltraVISTA \citep{Laigle_2016}. This implies a galaxy bias factor $\bar{b}_\mathrm{gal}$ of 1 (1.3) and a galaxy number density $n_\mathrm{gal}$ of $0.004\,\mathrm{Mpc^{-3}}$ ($0.003\,\mathrm{Mpc^{-3}}$) at $z\approx0.4$ (0.9), corresponding to a total of approximately 50 (200) galaxies within TIME's survey volume. Alternatively, TIME CO maps may also be cross-correlated with spectroscopic galaxies such as samples from the DEEP2 survey \citep{Mostek_2013}. Due to the limited survey area and spectral resolving power of TIME, it will not be a lot more beneficial to use spectroscopic galaxies, which have a significant lower number density. We therefore focus on the cross-correlation with photometric galaxies henceforth. 

We follow \citet{Chung_2019} to estimate the extent by which the redshift error de-correlates the cross-correlation signal. For a gaussian error $\sigma_z$ around $z_{\rm phot}$, the attenuation effect on the true power spectrum can be described by the filtering function
\begin{align}
\mathcal{F}_z(k, z) & = \int_{0}^{1} \mathrm d \mu \exp \left[- \frac{c^2 k^2 \mu^2 \tilde{\sigma}^2_z}{H^2(z)} \right] \\ \nonumber
& = \frac{\sqrt{\pi} H(z)}{2 c k \tilde{\sigma}_{z}} \mathrm{erf} \left( \frac{c k \tilde{\sigma}_{z}}{H(z)} \right)~,
\label{eq:sigmaz}
\end{align}
where $\tilde{\sigma}_z = \sigma_z$ and $\sigma_z/\sqrt{2}$ for the galaxy auto and CO--galaxy cross-power spectra, respectively, and $\mu=k_{\parallel}/k$ is the cosine of the $k$-space polar angle. We note that $\mathcal{F}_z(k, z)$ is introduced here for illustrative purpose only. Because of the anisotropic Fourier space that TIME measures (to be discussed in Section~\ref{sec:survey}), when estimating the observed 2D power spectrum we first account for the attenuation effect due to $\sigma_z$ in the LOS direction, and then average the resulting power over the Fourier space sampled. Compared with TIME's modest spectral resolution, de-correlation is negligible on clustering scales for galaxies with spectroscopic redshifts, but has some effect for high-accuracy photometric redshifts.


\subsubsection{Ionized Carbon During the EoR} \label{sec:model:CII}

A number of previous works have exploited galaxy evolution models derived from infrared observations to predict the strength of [\ion{C}{ii}] emission from the EoR \cite[e.g.,][]{Silva_2015, Cheng_2016, Serra_2016}. However, tensions often exist between the modeled SFH and that inferred from deep, UV observations after correcting for dust attenuation. Such a discrepancy is not surprising, considering that FIR observations of EoR galaxies are still lacking and a fair comparison between the SFHs extrapolated from IR-based models and UV observations at $z \ga 5$ is not necessarily guaranteed. In order to avoid such problems, here and in Section~\ref{sec:model:lae}, we adopt an alternative approach based on UV observations to model the high-redshift [\ion{C}{ii}] and Ly$\alpha$ signals that TIME will directly measure in auto- and cross-correlations. 

\begin{figure}[h!]
\centering
\includegraphics[width=0.48\textwidth]{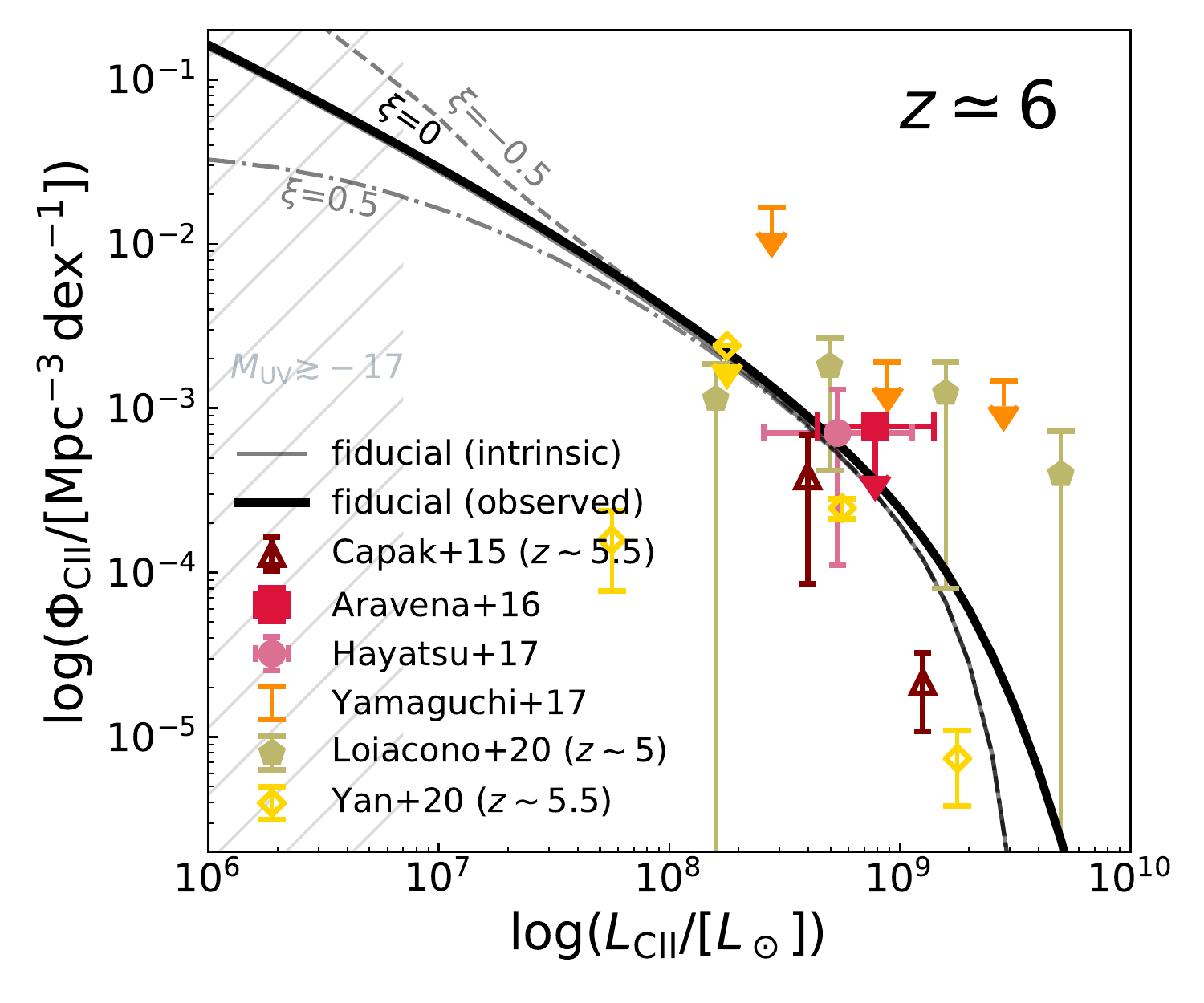}
\caption{A comparison of our modeled [\ion{C}{ii}] luminosity function, $\Phi_\ion{C}{ii}$, with constraints from ALMA observations at $z\sim6$, including both blind surveys \citep{Aravena_2016, Hayatsu_2017, Yamaguchi_2017, Loiacono_2020} and that based on UV-selected samples \citep{Capak_2015, Yan_2020}, which will always underestimate $\Phi_\ion{C}{ii}$. The black solid curve shows the observed luminosity function predicted by our fiducial [\ion{C}{ii}] model, which is related to the intrinsic one (gray solid curve) by the convolution described in Equation~(\ref{eq:ciilf_convolve}) assuming a scatter of $\sigma_\ion{C}{ii}=0.2$\,dex. The dashed and dash-dotted curves in gray deviating at the faint end illustrate the dependence on the extrapolation of the star formation efficiency $f_*(M)$ at its low-mass end, as specified by the $\xi$ parameter (see Appendix~\ref{sec:f_star}). The hatched region on the left shows the regime where in our model galaxies are fainter than $M_{\mathrm{UV}}=-17$, below current detection limit.}
\label{fig:ciilf_model}
\end{figure}

Our phenomenological model of [\ion{C}{ii}] emission assumes a correlation between the UV 1500--2800\,\AA\ continuum luminosity $L_{\rm UV}$ and the [\ion{C}{ii}] line luminosity $L_{\ion{C}{ii}}$. As will be discussed below, $L_{\rm UV}$ is used only as a proxy for the SFR of galaxies. We choose to connect $L_{\ion{C}{ii}}$ with $L_{\rm UV}$ instead of the SFR directly in order to model (1) the luminosity distribution of [\ion{C}{ii}] emitters and (2) their underlying SFH calibrated to the observed UV luminosity function of galaxies during reionization. The correlation can be parameterized as
\begin{equation}
\log \left( \frac{L_{\ion{C}{ii}}}{L_\odot} \right) = a \log \left( \frac{L_{\rm UV}}{\mathrm{erg\,s^{-1}\,Hz^{-1}}} \right) + b~, 
\label{eq:lcii_luv}
\end{equation}
where $a = 1$ and $b = -20.6$ as listed in Table~\ref{tb:model_params} are fiducial values that predict a reasonable [\ion{C}{ii}] luminosity function at $z\simeq6$ consistent with existing observational constraints based on identified high-redshift [\ion{C}{ii}] emitters. We also consider a non-trivial scatter $\sigma_{\ion{C}{ii}}=0.2\,$dex which specifies a log-normal distribution of $L_{\ion{C}{ii}}$ as a function of $L_{\rm UV}$
\begin{equation}
P_{\rm s}(x) \mathrm d x = \frac{1}{\sqrt{2\pi} \sigma_{\ion{C}{ii}}} \exp \left[ - \frac{x^2}{2 \sigma_{\ion{C}{ii}}^2} \right] \mathrm d x~,
\end{equation}
where $x = \log L_{\ion{C}{ii}} - \mu$ and $\mu = a \log L_{\rm UV} + b$. 
Under the assumption that a one-to-one correspondence exists between [\ion{C}{ii}]-emitting galaxies and their host dark matter halos, the intrinsic [\ion{C}{ii}] luminosity function can be simply obtained from the halo mass function $\mathrm d n / \mathrm d M$, connected via the UV luminosity, as
\begin{equation}
\Phi_{\ion{C}{ii}}(L_{\ion{C}{ii}}) = \frac{\mathrm d n}{\mathrm d \log M} \frac{\mathrm d \log M}{\mathrm d \log L_{\rm UV}} \frac{\mathrm d \log L_{\rm UV}}{\mathrm d \log L_\ion{C}{ii}} = \frac{\mathrm d n}{a \mathrm d \log L_{\rm UV}}~.
\end{equation}
Following \cite{Behroozi_2010}, the observed luminosity function after accounting for the scatter is given by the convolution
\begin{equation}
\Phi_{\ion{C}{ii}}^{\rm obs}(L_{\ion{C}{ii}}) = \int_{-\infty}^{\infty} \Phi_{\ion{C}{ii}}(10^x) P_{\rm s}(x - \log L_{\ion{C}{ii}}) \mathrm d x~,
\label{eq:ciilf_convolve}
\end{equation}
which effectively flattens the bright end of the luminosity function, since there are more faint sources being up-scattered than bright sources being down-scattered. Figure~\ref{fig:ciilf_model} shows a comparison between the [\ion{C}{ii}] luminosity function predicted by our fiducial model (as well as its variations) and constraints from a few recent high-redshift [\ion{C}{ii}] surveys with ALMA, based on either serendipitous (i.e., blindly detected) [\ion{C}{ii}] emitters (ASPECS, \citealt{Aravena_2016}; \citealt{Hayatsu_2017}; \citealt{Yamaguchi_2017}; ALPINE, \citealt{Loiacono_2020}) or observations of UV-selected targets (\citealt{Capak_2015}; ALPINE, \citealt{Yan_2020}), which are strictly speaking lower limits because [\ion{C}{ii}]-bright but UV-faint galaxies are potentially missing. We note that [\ion{C}{ii}] luminosity is known to be affected by physical conditions of the PDR in numerous ways \cite[][]{Ferrara_2019}. Theoretical models \cite[e.g.,][]{Lagache_2018} are in slight tension with existing constraints on the [\ion{C}{ii}] luminosity function. This may indicate problems with assumptions made about the PDR model, or failure to properly account for cosmic variance in estimates of the luminosity function (see e.g., \citealt{KMK_2020}, \citealt{TF_2020} and references therein, for recent studies about the impact of cosmic variance on high-redshift galaxy surveys and intensity mapping measurements).

The UV continuum luminosity is correlated with the SFR as
\begin{equation}
\dot{M}_{*} = \mathcal{K}_{\rm UV} L_{\rm UV}
\end{equation}
where the conversion factor is taken to be $\mathcal{K_{\rm UV}} = 1.15 \times 10^{-28}\,M_\odot \mathrm{yr^{-1} / erg\,s^{-1}\,Hz^{-1}}$, which is valid for stellar populations with a Salpeter IMF \citep{Salpeter_1955} and a metallicity $Z \sim 0.05 Z_\odot$ during the EoR following \citet{SF_2016}. The SFRD informed by UV data can then be expressed as
\begin{equation}
\dot{\rho}_*(z) = \int^{M_{\rm max}}_{M_{\rm min}} \mathrm d M \frac{\mathrm d n}{\mathrm d M} \dot{M}_{*}(M, z)~,
\label{eq:SFRD}
\end{equation}
where we choose $M_{\rm min} = 10^8\,M_\odot$, corresponding to the minimum halo mass for star formation implied by the atomic cooling threshold, and $M_{\rm max}=10^{15}\,M_\odot$. As will be discussed in Section~\ref{sec:model-eor}, the SFR, $\dot{M}_*$, as a function of halo mass and redshift can be specified by the star formation efficiency (SFE) and the rate at which halo mass grows. The shapes of both [\ion{C}{ii}] luminosity function and power spectrum are therefore affected by the halo mass dependence of these factors. Since the reionization history is irrelevant to star formation after reionization was complete, we do not match the SFRD inferred from UV observations to that obtained by extrapolating the CIB model to $z \gtrsim 5$, which is itself highly uncertain. 

The spatial fluctuations of [\ion{C}{ii}] emission can be described by the [\ion{C}{ii}] auto-correlation power spectrum
\begin{equation}
P_{\ion{C}{ii}}(k, z) = \bar{I}^2_{\ion{C}{ii}}(z) \bar{b}^2_{\ion{C}{ii}}(z) P_{\delta \delta}(k, z) + P^{\rm shot}_{\ion{C}{ii}}(z)~.
\label{eq:ciips}
\end{equation}
The mean [\ion{C}{ii}] intensity is
\begin{equation}
\bar{I}_{\ion{C}{ii}}(z) = \int^{M_{\rm max}}_{M_{\rm min}} \mathrm d M \frac{\mathrm d n}{\mathrm d M} \frac{L_{\ion{C}{ii}}[L_{\rm UV}(M,z)]}{4\pi D^2_L} y(z) D^2_A~,
\end{equation}
and $\bar{b}_{\ion{C}{ii}}(z)$ is the [\ion{C}{ii}] luminosity-averaged halo bias factor defined as
\begin{equation}
\bar{b}_{\ion{C}{ii}}(z) = \frac{\int^{M_{\rm max}}_{M_{\rm min}} \mathrm d M (\mathrm d n/\mathrm d M) b(M,z) L_\ion{C}{ii}[L_{\rm UV}(M)]}{\int^{M_{\rm max}}_{M_{\rm min}} \mathrm d M (\mathrm d n/\mathrm d M) L_\ion{C}{ii}[L_{\rm UV}(M)]}~.
\end{equation}
The shot-noise term is
\begin{equation}
P^{\rm shot}_{\ion{C}{ii}}(z) = \int^{M_{\rm max}}_{M_{\rm min}} \mathrm d M \frac{\mathrm d n}{\mathrm d M} \left\{ \frac{L_{\ion{C}{ii}}[L_{\rm UV}(M,z)]}{4\pi D^2_L} y(z) D^2_A \right\}^2~.
\end{equation}
Similar to the CO case, we use the scaling factors given in \citet{Sun_2019} to account for the effects of $\sigma_{\ion{C}{ii}}$ on the [\ion{C}{ii}] power spectrum. 


\subsubsection{High-$z$ LAEs} \label{sec:model:lae}

\begin{figure}[h!]
\centering
\includegraphics[width=0.48\textwidth]{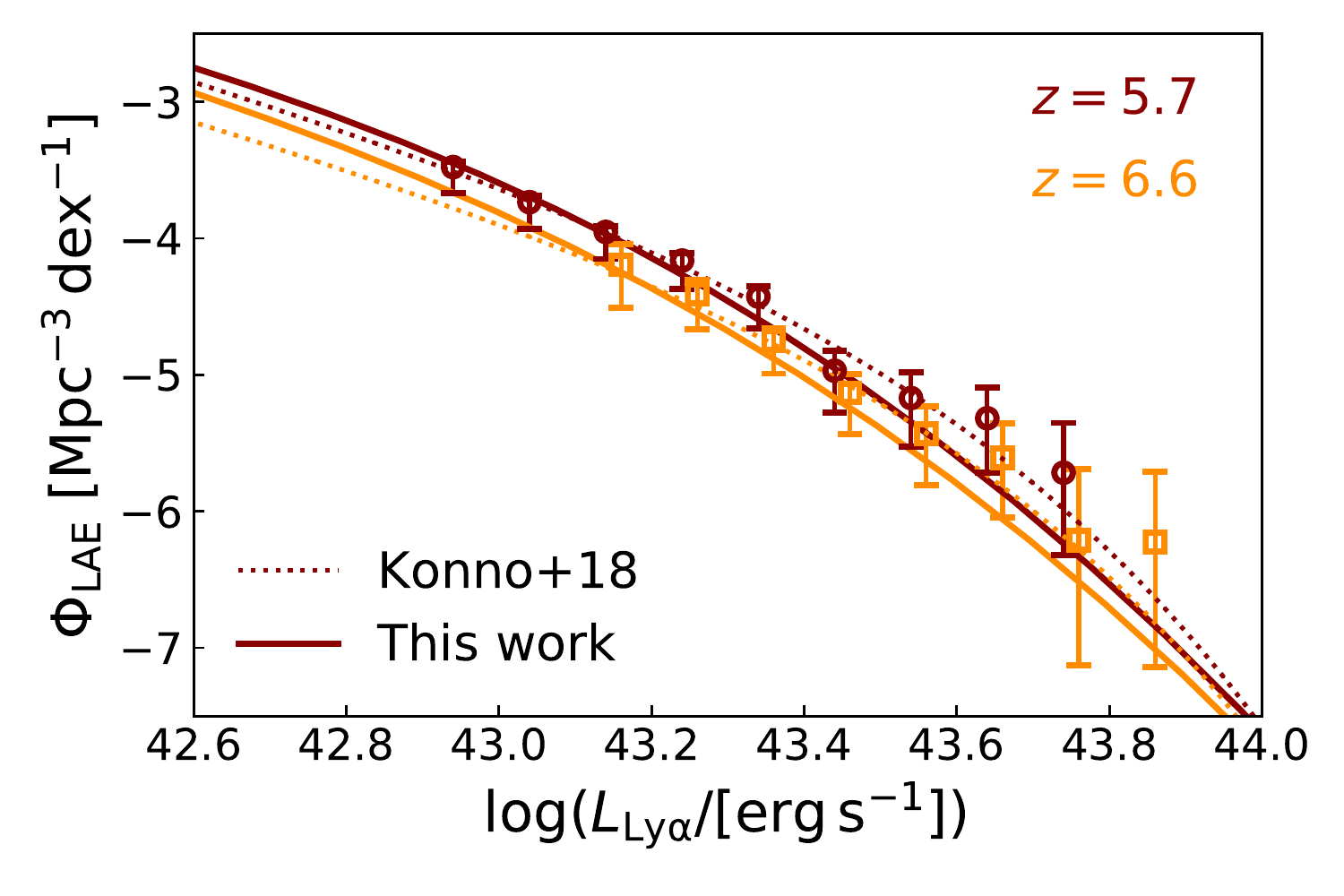}
\caption{Comparison between the observed LAE luminosity functions at $z=5.7$ and $z=6.6$ predicted by our analytical model (solid curves) and taken from \citet{Konno_2018} (data points and dotted curves). }
\label{fig:laelf_model}
\end{figure}

In order to estimate TIME's sensitivity to the cross-correlation between high-redshift [\ion{C}{ii}] emission and LAEs, we adopt a semi-analytical approach to paint [\ion{C}{ii}] and Ly$\alpha$ emission onto the halo catalogs from the Simulated Infrared Dusty Extragalactic Sky \cite[SIDES,][]{Bethermin_2017} simulation. Analytic models have been widely used to investigate physical properties of high-redshift LAEs \cite[e.g.,][]{SSS_2009, Jose_2013, LMR_2016, LMR_2017a, LMR_2017b, Sarkar_2019}. Here, to model Ly$\alpha$ luminosity of LAEs, we assume that Ly$\alpha$ photons are solely produced by recombinations under ionization equilibrium. As a result, for a given halo mass and redshift, it can be approximately related to the SFR by
\begin{equation}
L_{\rm Ly\alpha} = \frac{f_{\gamma}\dot{M}_*(M, z)/\eta}{m_{\rm p}/(1-Y)} (1 - f_{\rm esc}) f^{\rm Ly\alpha}_{\rm esc} f_{\rm Ly\alpha} E_{\rm Ly\alpha}~,
\end{equation}
where $m_\mathrm{p}$ is the mass of hydrogen atom. The ionizing photon produced per stellar baryon $f_{\gamma}$, the escape fraction of ionizing photons $f_{\rm esc}$, the fraction of recombinations ending up as Ly$\alpha$ emission $f_{\rm Ly\alpha}$ and the helium mass fraction $Y$ are taken to be $f_{\gamma}=4000$ (typical for low-metallicity Pop~II stars with a Salpeter initial mass function), $f_{\rm esc}=0.1$, $f_{\rm Ly\alpha}=0.67$ and $Y=0.24$, respectively. The factors $(1 - f_{\rm esc})$ and $f^{\rm Ly\alpha}_{\rm esc}$ account for the fraction of ionizing photons failing to escape (and thus leading to recombinations) and the fraction of Ly$\alpha$ photons emitted that eventually reach the observer. Because the production of Ly$\alpha$ emission is also subject to local dust extinction, a scale factor $\log \eta = \langle A_{\rm UV} \rangle/2.5$, whose value is specified by the dust correction formalism described in Appendix~\ref{sec:f_star}, is included here to obtain the obscured star formation rate. As in cases of [\ion{C}{ii}] and CO emission, we consider a log-normal scatter $\sigma_{\rm Ly\alpha}$ around the mean $L_{\rm Ly\alpha}$-$M$ relation above, which makes the observed LAE luminosity function a convolution of the intrinsic function with the log-normal distribution. In our model, we take $f^{\rm Ly\alpha}_{\rm esc}=0.6$ and $\sigma_{\rm Ly\alpha} = 0.3$\,dex, consistent with the observationally determined Ly$\alpha$ escape fraction \citep{Jose_2013} and the dispersion about the luminosity-halo mass relation \citep{More_2009}, to obtain reasonably good fits to the luminosity functions measured by \citet{Konno_2018}, as shown in Figure~\ref{fig:laelf_model}. The luminosity-halo mass relation is then used to paint both [\ion{C}{ii}] and Ly$\alpha$ emission onto dark matter halos catalogued to obtain maps of LAE spatial distribution and [\ion{C}{ii}] intensity fluctuations. 

The limiting magnitude $m_{\rm lim}^{\rm AB}$ of LAE surveys can be related to the line luminosity $L_{\rm Ly\alpha}$ of LAEs by $L_{\rm Ly\alpha} = 4 \pi D_L^2 F_{\rm Ly\alpha}$ and
\begin{equation}
F_{\rm Ly\alpha} = 3\times10^{-5} \times \frac{10^{(8.90 - m_{\rm lim}^{\rm AB})/2.5} \Delta \lambda} {\lambda^2}\,\mathrm{erg\,s^{-1}\,cm^{-2}}~, \nonumber
\end{equation}
where we take $\Delta \lambda = 131$\,\AA\ and $\lambda = 8170$\,\AA\ for $z=5.7$ and $\Delta \lambda = 120$\,\AA\ and $\lambda = 9210$\,\AA\ for $z=6.6$ as specified in \citet{Konno_2018}. Meanwhile, to generate mock LAE catalogs we consider limiting magnitudes of the planned, ultra-deep (UD) survey of the HSC, namely $m_{\rm lim}^{\rm AB} = 26.5$ and 26.2 at $z=5.7$ and 6.6, respectively, which correspond to minimum Ly$\alpha$ luminosities of $\log (L_{\rm Ly\alpha}/ \mathrm{erg\,s^{-1}})$ = 42.3 and 42.4. For such survey depths, we predict the comoving number density of LAEs to be $n^{z=5.7}_{\rm LAE} = 1.4 \times 10^{-3}\,\mathrm{Mpc}^{-3}$ and $n^{z=6.6}_{\rm LAE} = 5.7 \times 10^{-4}\,\mathrm{Mpc}^{-3}$ by integrating the LAE luminosity functions our model implies. As a result, no more than a few LAEs are expected to exist in the survey volume of TIME due to its limited survey area of about 0.01\,deg$^2$. One caveat to our LAE model is that we ignore the impact of patchy reionization on the spatial distribution of LAEs through the Ly$\alpha$ transmission fraction, which is affected by, and thus informs, the growth of ionized bubbles around LAEs \cite[e.g.,][]{SSM_2016}. We note, though, that for estimating the [\ion{C}{ii}]--LAE cross-correlation TIME will measure, our simple model calibrated against the LAE luminosity functions from the SILVERRUSH survey should suffice. In fact, thanks to the large survey areas covered ($14$ and $21\,\mathrm{deg}^2$ at $z=5.7$ and $6.6$, respectively), the patchiness effect is already captured, at least in part, by the observed LAE statistics. To fully address the suppression on the LAE number density due to patchy reionization, both numerical \cite[e.g.,][]{McQuinn_2007} and semi-analytical \cite[e.g.,][]{DFG_2008} methods can be applied. We will explore how such effects may be probed by the [\ion{C}{ii}]--LAE cross-correlation in future work. 

Therefore, we consider the measurement of two-point correlation function, instead of power spectrum, to maximally extract the information about large-scale correlation between distributions of LAEs and [\ion{C}{ii}] intensity. In general, for a given normalized selection function $\mathcal{N}(z)$, the angular correlation function is related to the spatial correlation function by the Limber equation
\begin{equation}
\omega(\theta, z) = \int \mathrm d z' \mathcal{N}(z') \int \mathrm d z'' \mathcal{N}(z'') \xi \left[ r(\theta, z', z''), z \right]~,
\label{eq:tpcf_to_acf}
\end{equation}
where we approximate $\mathcal{N}(z)$ by top-hat functions over $z=5.67$--5.77 and $z=6.52$--6.63 corresponding to the bandwidths of narrow-band filters used in the SILVERRUSH survey \citep{Ouchi_2018, Konno_2018}. Specifically, the angular cross-correlation function between the [\ion{C}{ii}] intensity map measured by TIME and the LAE distribution is (in units of Jy/sr)
\begin{equation}
\omega_{\rm \ion{C}{ii}\times LAE}(\theta) \equiv \frac{\sum\limits_i^{N(\theta)} \Delta I_{\ion{C}{ii}}^i (\theta)}{N(\theta)} \approx b_{\rm LAE} \bar{b}_{\ion{C}{ii}} \bar{I}_{\ion{C}{ii}} \omega_{\rm DM}(\theta)~,
\label{eq:acf}
\end{equation}
where for the bin $\theta$, $\Delta I_{\ion{C}{ii}}^i (\theta) = I_{\ion{C}{ii}}^i (\theta) - \bar{I}_{\ion{C}{ii}}$ denotes the [\ion{C}{ii}] intensity fluctuation at pixel $i$, whereas $N(\theta)$ denotes the total number of LAE-pixel pairs. Determined from the LAE distributions generated with our semi-analytical approach, the LAE bias $b_\mathrm{LAE} \approx 6$ at both $z=5.7$ and 6.6 is consistent with the upper limits on $b_\mathrm{LAE}$ estimated from the SILVERRUSH survey. The approximation is valid on large scales where the clustering of LAEs and [\ion{C}{ii}] emission are linearly biased tracers of the dark matter density field. The dark matter angular correlation function $\omega_{\rm DM}$ is derived using Equation~(\ref{eq:tpcf_to_acf}) from the spatial correlation function
\begin{equation}
\xi_{\rm DM}(r, z) = \frac{1}{2\pi^2} \int \mathrm d k\, k^2 P_{\delta \delta}(k, z) \frac{\sin(kr)}{k r}~.
\end{equation}


\subsection{Molecular Gas Content} \label{sec:model-mh2}

Over $0.5 \la z \la 2$, TIME can detect more than one CO rotational line over its 183--326\,GHz bandwidth (see Table~\ref{tb:inst_specs}). Section~\ref{sec:model:CO}. By cross-correlating a pair of adjacent CO lines emitted from galaxies at the same redshift, we are able to simultaneously constrain $\alpha$, $\beta$, and $\sigma_{\rm CO}$ as defined in Eq.~\ref{eq:LCO}. As already mentioned in Section~\ref{sec:model:CO}, provided that the CO SLED is known and does not appreciably vary over the galaxy population, we can place sensitive constraints on the luminosity density of CO(1-0) line using the intensity fluctuations of the higher-$J$ CO transitions in TIME's spectral range. The cosmic molecular gas density can be consequently derived from the CO(1-0) line luminosity density as
\begin{equation}
\rho_{\rm H_2}(z) = \alpha_{\rm CO} \rho_{L^\prime_{\rm CO}}(z) = \alpha_{\rm CO} \int \mathrm d M \frac{\mathrm d n}{\mathrm d M} L^\prime_{\rm CO}[L_{\rm IR}(M, z)]~,
\end{equation}
where we adopt a universal CO-to-$\rm H_2$ conversion factor $\alpha_{\rm CO} = 4.3\,M_\odot(\mathrm{K\,km\,s^{-1}\,pc^2})^{-1}$ for Milky Way-like environments, as given by \cite{BWL_2013}. One important caveat is that our model assumes the ratios of CO lines with different $J$'s, as given by the excitation state of CO, are well-known. This is of course an oversimplification given the complexity of physical processes driving variations in the CO SLEDs in galaxies \citep{NK_2014}, even though the variation in line ratios for adjacent CO lines tends to be small \cite[e.g.,][]{CW_2013, Casey_2014PhR}. The variation of $\alpha_{\rm CO}$ serves as another source of uncertainty, but we note that it is a systematic uncertainty intrinsic to the usage of CO as tracer affecting nearly all measurements of the molecular gas content and a topic of extensive investigation at different redshifts \citep{BWL_2013, Amorin_2016, Gong_2018}.


\subsection{Reionization History} \label{sec:model-eor}

We embed our model of [\ion{C}{ii}] emission presented in Section~\ref{sec:model:CII} into a simple picture of reionization to demonstrate how TIME can probe the EoR. Our methods to model the production of [\ion{C}{ii}] emission and the progress of reionization are related to the cosmic SFH (see equations~\ref{eq:SFRD} and~\ref{eq:source}). TIME data constrain the SFRD during reionization, despite the uncertainty in the conversion from [\ion{C}{ii}] luminosity to star formation rate. In addition, if analyzed jointly with other observational constraints that probe different aspects of the EoR, such as quasar absorption spectra and the CMB optical depth, TIME observations can further improve our knowledge of key EoR parameters, including the escape fraction of hydrogen-ionizing photons $f_{\rm esc}$.

Following \cite{SF_2016} and \cite{Mirocha_2017}, in this work we adopt a commonly-used, two-zone model of the IGM \cite[][]{Furlanetto_2006, PL_2010, LF_2013} where the reionization history is characterized by the following set of differential equations that describe the redshift evolution of the \ion{H}{ii}-region filling factor $Q_{\ion{H}{ii}}$ and the electron fraction $x_{\rm e}$ outside \ion{H}{ii} regions\,\footnote{It is assumed that only X-ray photons can ionize the ``cavities'' of neutral gas between \ion{H}{ii} regions. }, 
\begin{equation}
\frac{\mathrm d Q_{\ion{H}{ii}}}{\mathrm d z} = \zeta \frac{\mathrm d f_{\rm coll}}{\mathrm d z} + \frac{C(z)\alpha_{\rm B}(T_{\rm e})}{H(z)}(1+z)^2 \bar{n}^0_{\rm H} Q_{\ion{H}{ii}}
\label{eq:ode_Q}
\end{equation}
and
\begin{equation}
\frac{\mathrm d x_{\rm e}}{\mathrm d z} = \mathcal{C}_{\rm ion} f_{X, \mathrm{ion}}(x_{\rm e}) \frac{\mathrm d f_{\rm coll}}{\mathrm d z} \approx 50.2 f_X f_{X, \mathrm{ion}}(x_{\rm e}) \frac{\mathrm d f_{\rm coll}}{\mathrm d z}~,
\label{eq:ode_xe}
\end{equation}
where $\bar{n}^{0}_{\rm H}$ is mean (comoving) number density of hydrogen. $C(z) \equiv \langle n_{\rm e}^2 \rangle / \langle n_{\rm e} \rangle^2$ defines the clumping factor of the IGM, whose globally-averaged value is approximately 3 as suggested by numerical simulations \cite[][]{Pawlik_2009, Shull_2012, DAloisio_2020}. $\alpha_{\rm B}(T_{\rm e})$ is the case-B recombination coefficient, and we take $T_{\rm e} \sim 2 \times 10^4$\,K valid for freshly reionized gas \citep{HH_2003, KFG_2012}. The overall ionizing efficiency, $\zeta$, is defined as the product of the star formation efficiency (SFE) $f_*$, the escape fraction of ionizing photons $f_{\rm esc}$, the average number of ionizing photons produced per stellar baryon $f_\gamma=4000$ and a correction factor $A_{\rm He} = 4/(4-3Y) = 1.22$ for the presence of helium, namely $\zeta = A_{\rm He} f_* f_{\rm esc} f_\gamma$. In our fiducial model, we set $f_{\rm esc}=0.1$, which leads to a reionization history consistent with current observational constraints (see Figure~\ref{fig:eor}). For simplicity, we only consider a population-averaged and redshift-independent escape fraction in this work, even though in practice it may evolve with halo mass and redshift \cite[e.g.,][]{Naidu_2020}. In Equation~(\ref{eq:ode_xe}), $f_{X, \mathrm{ion}}$ denotes the fractions of X-ray energy going to ionization, whose value is estimated by \cite{FS_2010}, and $f_X$ is a free, renormalization parameter for the efficiency of X-ray production, which is set to 1 in our model. In order to solve Equations~(\ref{eq:ode_Q}) and (\ref{eq:ode_xe}), we use {\footnotesize COSMOREC} \citep{CT_2011} to generate the initial conditions at $z=30$. 

The two differential equations above are closely associated with the redshift derivative of the collapse fraction of dark matter halos, $\mathrm d f_{\rm coll} / \mathrm d z$, which is always \textit{negative} by definition \cite[][]{Furlanetto_2017}
\begin{equation}
\bar{\rho} \frac{\mathrm d f_{\rm coll}}{\mathrm d z} \frac{\mathrm d z}{\mathrm d t} = \int^{M_{\rm max}}_{M_{\rm min}} \mathrm d M \frac{\mathrm d n}{\mathrm d M} \dot{M} + \left( \dot{M} M \frac{\mathrm d n}{\mathrm d M} \right)\Big|_{M_{\rm min}}~, \label{eq:dfcoll}
\end{equation}
where $\bar{\rho}$ is the mean matter density and the second term of Equation~(\ref{eq:dfcoll}) describing the evolution due to mass growth at the boundary is subdominant at the redshifts of interest. Following Equations~ (\ref{eq:ode_Q}) and (\ref{eq:dfcoll}), the total ionization rate $\zeta \mathrm d f_{\rm coll}/\mathrm d z$ is related to the cosmic star formation rate density $\dot{\rho}_*(z)$ by
\begin{align}
\zeta \frac{\mathrm d f_{\rm coll}}{\mathrm d z} & = \frac{A_{\rm He} f_{\rm esc} f_\gamma}{\bar{\rho}} \int^{M_{\rm max}}_{M_{\rm min}} \mathrm d M \frac{\mathrm d n}{\mathrm d M} \frac{\mathrm d t}{\mathrm d z} f_*(M,z) \dot{M} \nonumber \\
& = \frac{A_{\rm He} f_{\rm esc} f_\gamma \Omega_{\rm m}}{\bar{\rho} \Omega_{\rm b}} \times \dot{\rho}_*(z) \times \frac{\mathrm d t}{\mathrm d z}~,
\label{eq:source}
\end{align}
where the star formation rate of a given dark matter halo is $\dot{M}_*(M,z) = f_*(M,z) \Omega_{\rm b}/\Omega_{\rm m} \dot{M}(M,z)$. In order to find the SFE $f_*$ and the growth rate of halo mass $\dot{M}$, we perform the halo abundance matching technique to the UV LF and halo mass function respectively, following \cite{Mirocha_2017}. In particular, the potential redshift evolution of $f_*$, likely driven by feedback processes such as supernova explosions, is assumed to be negligible so that it can be described by a modified double-power law in $M$. The dust correction uses the observed UV continuum slope (see Appendix~\ref{sec:f_star} for details), although observed LFs are probably only modestly affected by dust extinction \citep{Capak_2015}. As also elaborated in Appendix~\ref{sec:f_star}, to characterize the degeneracy between the abundance of faint sources and the minimum halo mass, we allow the low-mass end of $f_*(M)$ to deviate from a perfect power law, as shown by Equation~(\ref{eq:defn_xi}). A modulation factor $\xi$ is introduced to make $f_*(M)$ either asymptote to a constant floor value when $\xi<0$ or decay exponentially when $\xi>0$. As listed in Table~\ref{tb:model_params}, we set the fiducial value of $\xi$ to 0 such that the low-mass end of $f_*(M)$ follows a power law implied by observed UV luminosity functions at $z\gtrsim6$ \cite[][]{Mirocha_2017}.

Once the redshift evolutions of $Q_{\ion{H}{ii}}$ and $x_{\rm e}$ have been solved, we can calculate the Thomson scattering optical depth for CMB photons as \cite[][]{Robertson_2015, SF_2016}
\begin{equation}
\tau_{\rm es}(z) = \frac{3 H_0 \Omega_{\rm b} c \sigma_{\rm T}}{8 \pi G m_{\rm p}} \int_0^z \mathrm d z' \frac{\bar{x}_i(z')(1+z')^2(1-Y+\frac{N_{\rm He}Y}{4})}{\sqrt{\Omega_{\rm m}(1+z')^3 + 1 - \Omega_{\rm m}}}~,
\end{equation}
where $\sigma_{\rm T} = 6.65\times10^{-25}\,\mathrm{cm^2}$ is the cross section of Thomson scattering, and $\bar{x}_i(z) = Q_{\ion{H}{ii}}(z) + [1 - Q_{\ion{H}{ii}}(z)]x_{\rm e}(z)$ is the overall ionized fraction. For simplicity, we further set $N_{\rm He}$ to 2 for $z < 3$ and 1 otherwise (i.e., instantaneous helium reionization at $z=3$) to account for the degree of helium ionization \cite[][]{FO_2008}. As will be discussed in Section~\ref{sec:results:eor}, with $\bar{x}_i$ and $\tau_\mathrm{es}$ in hand, we can constrain our model by combining the [\ion{C}{ii}] power spectra TIME measures with independent constraints on the IGM neutrality and CMB optical depth inferred from observations.

\begin{figure*}[t]
    \centering
    \begin{minipage}{.48\linewidth}
        \includegraphics[width=\textwidth]{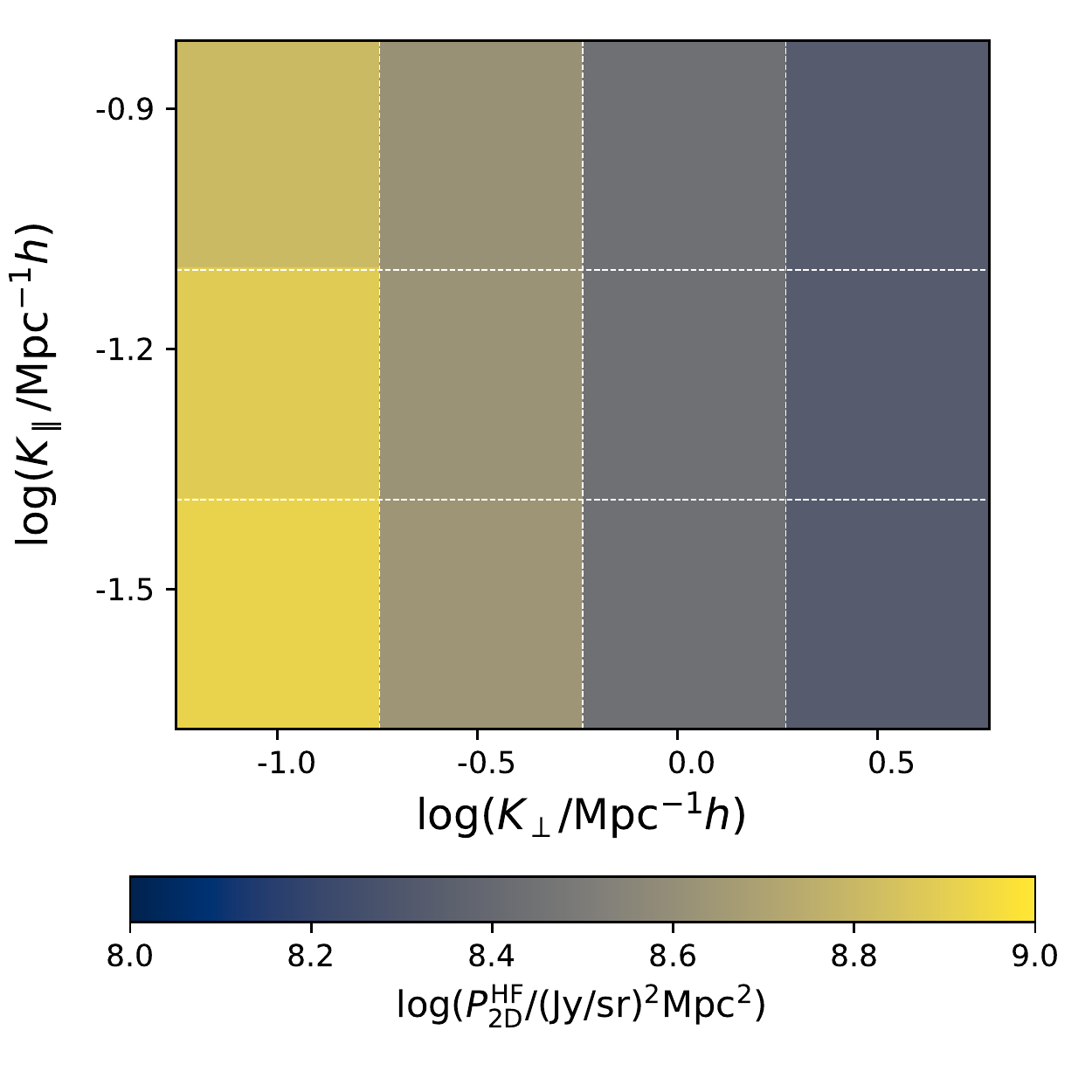}
    \end{minipage}
    \begin{minipage}{.48\linewidth}
        \includegraphics[width=\textwidth]{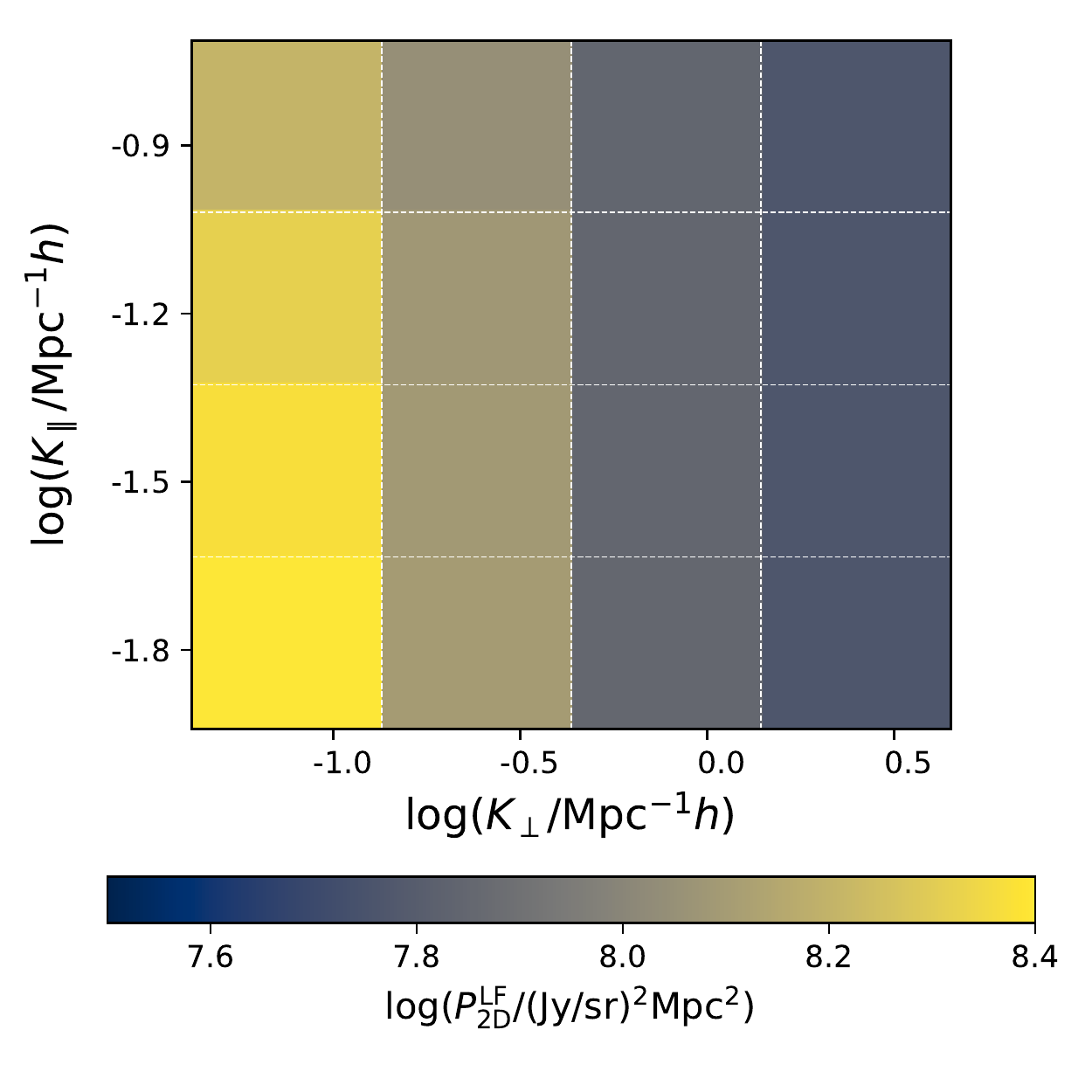}
    \end{minipage}
    \caption{The 2D binned [\ion{C}{ii}] auto-power spectrum measured in TIME low-$z$/HF (left) and high-$z$/LF (right) sub-bands and binned in $K_\perp$ (perpendicular to the LOS) versus $K_{\parallel}$ (parallel to the LOS) space. The scale change between $K_\perp$ and $K_{\parallel}$ reflects the anisotropic Fourier space that TIME measures.}
    \label{fig:2DPS}
\end{figure*}


\section{Mock Observations} \label{sec:survey}
Based on the survey strategy and sensitivity analysis to be described in the following sub-sections, we estimate TIME measurements in auto- and cross-correlations from the instrument parameters listed in Table~\ref{tb:inst_specs}, and use them to forecast constraints on physical quantities of interest relevant to the EoR and galaxy evolution (Section~\ref{sec:results}).


\subsection{Survey Strategy} \label{sec:survey-strategy}
With a line scan design, TIME will directly observe a two-dimensional map of intensity fluctuations in instrument coordinates, namely a spatial coordinate defined by the angular position and spectral frequency. As a result, the two-point statistics are described by a 2D power spectrum defined in the observed comoving frame of the instrument, which relates to the theoretical 3D power spectrum defined in Equation~(\ref{eq:ciips}) by the survey window function.

From the definition of window function $W_{ii} \left( k, \vec{K}_i \right)$ discussed in Appendix~\ref{sec:wf}, we obtain an integral equation that maps the true 3D power spectrum $P(k)$ of a sky mode $k$ to the observed 2D power spectrum $\mathcal{P}(K)$ of an instrument mode $K$
\begin{align}
\mathcal{P}\left( \vec{K}_i \right) & = L_x L_z \int_{-\infty}^{\infty} \mathrm d \ln k \Delta^2(k) W_{ii} \left( k, \vec{K}_i \right)~, 
\label{eq:p3d_to_p2d}
\end{align}
where $L_x$ and $L_z$ measure dimensions of survey volume perpendicular and parallel to the LOS direction, respectively, and $\Delta^2(k) \equiv k^3 P(k) / 2\pi^2$ is the dimensionless spatial power spectrum containing both the clustering and shot-noise terms. The window function $W_{ii}$ describes the relationship between $K$ and $k$, thereby acting as a kernel that projects the spatial power spectrum $P(k)$ into $\mathcal{P}(K)$ measured in the observing frame of TIME. A given instrument mode $K$ can be further decomposed into two components parallel ($K_{\parallel}$) and perpendicular ($K_{\perp}$) to the LOS, respectively, with $K = \sqrt{K^2_{\parallel}+K^2_{\perp}}$. In particular, the minimum accessible scales are defined by the survey size and bandwidth, whereas the maximum accessible scales are defined by the beam size and spectral resolution \cite[][]{Uzgil_2019}. By discretizing the linear integral equation above with a trapezoidal-rule sum, we can arrive at a simple matrix representation of Equation~(\ref{eq:p3d_to_p2d})
\begin{equation}
\vec{\mathcal{P}} = \mathbf{A} \vec{P}~, 
\label{eq:mat_eqn}
\end{equation}
where $\mathbf{A}$ is an $m \times n$ transfer matrix, with each row summing up to unity, that converts a column vector $\vec{P}$, which represents the true power spectra $P(k)$ binned into $n$ bins of $k$, into another column vector $\vec{\mathcal{P}}$, which represents the 2D power spectrum $\mathcal{P}(K)$ measured in $m$ bins of $K$. 

In practice, though, various foreground cleaning techniques such as voxel masking \citep{BKK_2015, Sun_2018} may be applied in order to remove contamination due to both continuum foregrounds (e.g., atmosphere, the CMB, etc.) and line interlopers. As a result, it is unlikely that the window function will have a simple analytic form. Therefore, it must be calculated numerically to account for the loss of survey volume and/or accessible $k$ space due to foreground cleaning. 

\begin{table}[h!]
\centering
\caption{Experimental parameters for TIME and TIME-EXT}
\hspace*{-1.5em}
\begin{threeparttable}[b]
\setlength{\tabcolsep}{2pt}
\makebox[0.95\linewidth]{\footnotesize
\begin{tabular}{ccc}
\toprule
\toprule
Parameter & TIME & TIME-EXT \\
\hline
Number of spectrometers ($N_{\rm feed}$) & 32 & 32 \\
Dish size ($D_{\rm ap}$) & 12\,m & 10\,m \\
Beam size ($\theta_{\rm FWHM}$)\tnote{a} & 0.43\,arcmin & 0.52\,arcmin \\
Spectral range ($\nu_{\rm min}$, $\nu_{\rm max}$)\tnote{b} & 183--326\,GHz & 183--326\,GHz \\
\multirow{2}{*}{Spectral bands} & LF: 200--265\,GHz & LF: 200--265\,GHz \\
 & HF: 265--300\,GHz & HF: 265--300\,GHz \\
Resolving power ($R$) & 90--120 & 90--120 \\
Observing site & ARO & LCT \\
Noise equivalent intensity (NEI) & $5\,\mathrm{MJy\,sr^{-1}\,s^{1/2}}$ & $2.5\,\mathrm{MJy\,sr^{-1}\,s^{1/2}}$ \\
Total integration time ($t_{\rm obs}$) & 1000\,hours & 3000\,hours \\
Survey power\tnote{c} & 1 & 12 \\
\bottomrule
\end{tabular}}
\begin{tablenotes}
\item [a] {\footnotesize $\theta_{\rm FWHM}$ is evaluated at 237\,GHz, corresponding to $z_\ion{C}{ii} = 7$. }
\item [b] {\footnotesize TIME has 44 (30+14 in LF and HF sub-bands, respectively) scientific spectral channels over 200--300\,GHz, and 16 additional channels monitoring atmospheric water vapor.}
\item [c] {\footnotesize The survey power is defined to scale as $N_{\rm feed} t_{\rm obs} / \mathrm{NEI}^2$}. 
\end{tablenotes}
\end{threeparttable}
\label{tb:inst_specs}
\end{table}

Using the mode counting method to be described in Section~\ref{sec:survey-sensitivity}, we aim to determine a survey strategy that optimizes our [\ion{C}{ii}] auto-correlation measurements, while ensuring a reasonable chance for successfully detecting the cross-correlation signals. In particular, we consider two defining factors of the survey, namely its geometry (i.e., aspect ratio of the survey area) and depth. We find that while the scale-independent shot-noise component dominating the total S/N of the power spectrum is not sensitive to survey geometry, a line scan offers the most economical way to overlap large-scale $K_\parallel$ modes with $K_\perp$ modes---a desirable property that allows cross-check of systematics that manifest themselves differently in $K_\parallel$ and $K_\perp$ dimensions. It is also a favorable geometry of TIME, which has an instantaneous field of view (FOV) of $32\times1$ beams due to the arrangement of the grating spectrometer array in the focal plane. The length of the line scan, on the other hand, is set by the trade-off between accessing large scales  (small $K_\parallel$) and maintaining a survey depth that ensures a robust [\ion{C}{ii}] detection. The resulting survey strategy after optimization is a line scan with $180\times1$ beams across, covering a total survey area of approximately $1.3 \times 0.007\,\mathrm{deg}^2$, which applies to all the analysis in the remainder of the paper. 

Figure~\ref{fig:2DPS} shows explicitly the Fourier space that TIME will sample via the line scan in its two sub-bands, a low-$z$/high-frequency (HF)
sub-band with bandwidth 265--300\,GHz ($5.3 < z_\ion{C}{ii} < 6.2$), and a high-$z$/low-frequency
(LF) sub-band with bandwidth 200--265\,GHz ($6.2 < z_\ion{C}{ii} < 8.5$). The 2D binned [\ion{C}{ii}] power spectrum is shown for each individual bin in $K_\parallel$ versus $K_\perp$ space. The line scan can access modes at scales $K_\parallel \sim K_\perp \sim 0.1\,h/\mathrm{Mpc}$, a regime where the power is dominated by the clustering component.


\subsection{Sensitivity Analysis} \label{sec:survey-sensitivity}

As discussed in the previous section, the effect of the window function is non-trivial for the clustering signal, so it is most reasonable to estimate the measurement uncertainty first in the observing frame (i.e., instrument space) and then propagate it to obtain the uncertainty on the true power spectrum. Here, we follow \citet{Gong_2012} to provide an overview of the sensitivity analysis based on the mode counting method. 

Table~\ref{tb:inst_specs} summarizes the instrument specifications for TIME and an extended version of the experiment, TIME-EXT, which may offer more than an order of magnitude improvement in survey power by combining (1) lower photon noise offered by a better-sited telescope with fewer mirrors like the LCT (S.~Golwala, private communication)\footnote{See slides from the Infrared Science Interest Group (IR SIG) seminar given by Sunil Golwala, available at the time of writing at \url{https://fir-sig.ipac.caltech.edu/system/media_files/binaries/29/original/190115GolwalaLCTIRSIGWeb.pdf}} and (2) longer integration time. For the observed [\ion{C}{ii}] auto power spectrum after binning in $K$ space, the uncertainty can be expressed as
\begin{equation}
\delta \mathcal{P}_{\ion{C}{ii}} (K) = \frac{ \mathcal{P}_{\ion{C}{ii}}(K) + \mathcal{P}_{\ion{C}{ii}}^\mathrm{n} }{\sqrt{N_{\rm m}(K)}}~, 
\label{eq:cii_unc}
\end{equation}
where the noise power $\mathcal{P}^{\rm n}$ is related to the noise equivalent intensity (NEI), angular sizes of the beam ($\Omega_{\rm beam}$) and the survey ($\Omega_{\rm survey}$), number of spectrometers $N_{\rm feed}$, total observing time $t_{\rm obs}$, and voxel volume $V_{\rm vox}$ by
\begin{equation}
\mathcal{P}^{\rm n} = \sigma^2_{\rm n} V_{\rm vox} = \frac{(\mathrm{NEI})^2 V_{\rm vox}}{N_{\rm feed} (\Omega_{\rm beam}  / \Omega_{\rm survey}) t_{\rm obs}}~.
\end{equation}
For TIME, the NEI values assumed are $5\,\mathrm{MJy\, sr^{-1}\,s^{1/2}}$ and $10\,\mathrm{MJy\, sr^{-1}\,s^{1/2}}$ for the high-$z$/LF and low-$z$/HF sub-bands, respectively, which are estimated assuming operation at ARO with 3\,mm perceptible water vapor (PWV) content. These numbers are assumed to be a factor of 2 smaller for TIME-EXT, since the LCT is better-sited and requires fewer number of coupling mirrors \cite[][]{Hunacek_2020PhDT}. $N_{\rm m}(K)$ is the total number of independent Fourier modes accessible to the instrument, determined by both how the Fourier space is sampled by the instrument and the loss due to e.g., foreground cleaning. We conservatively assume the lowest $K_\parallel$ and $K_\perp$ modes are contaminated by scan-synchronous systematics, so they are rejected from our mode counting, which in turn affects the accessible $K$ range for a given survey. It is also important to note that, due to the survey geometry of TIME, Fourier space is not uniformly sampled. Consequently, instead of managing to derive an analytical expression for $N_{\rm m}(K)$, we simply count the number of independent $K$ modes in a discrete manner for any given binning scheme (see also \citealt{Chung_2020}). 

For the CO cross-power spectrum, the uncertainty can be similarly expressed as
\begin{equation}
\delta \mathcal{P}_{J \times J'} = \frac{\left[ \mathcal{P}^2_{J \times J'} + \delta \mathcal{P}_J \delta \mathcal{P}_{J'} \right]^{1/2}}{\sqrt{2 N_{\rm m}}}~, 
\end{equation}
where $\delta \mathcal{P}_J(K) = \mathcal{P}_J(K) + \mathcal{P}_J^\mathrm{n}$. When evaluating $\delta \mathcal{P}_J$, we also include the expected [\ion{C}{ii}] auto power at the corresponding redshift and wavenumber\footnote{Following assumptions made in \cite{Sun_2018}, we use the approximation $ k_{\ion{C}{ii}} \approx \sqrt{3} k_{\rm CO} / \sqrt{2(\chi_{\ion{C}{ii}}/\chi_{\rm CO})^2+(y_{\ion{C}{ii}}/y_{\rm CO})^2} $ and the rescaling factor $P_{\ion{C}{ii}}(k_{\rm CO})/P_{\ion{C}{ii}}(k_{\ion{C}{ii}}) = (\chi_{\rm CO}/\chi_{\ion{C}{ii}})^2 y_{\rm CO}/y_{\ion{C}{ii}} $ to project the [\ion{C}{ii}] power spectrum into the observing frame of CO.} as an additional source of uncertainty for CO cross-correlation measurements that would not be removed by simple continuum subtraction. A clarification of the factor of 2 in the denominator is provided in Appendix~\ref{sec:unc_ps}. Similarly, the uncertainty on the CO--galaxy cross-power spectrum is
\begin{equation}
\delta \mathcal{P}_{\rm CO \times gal} = \frac{\left[ \mathcal{P}^2_{\rm CO \times gal} + \left( \mathcal{P}_{\rm CO} + \mathcal{P}^{\rm n}_{\rm CO} \right) \left( \mathcal{P}^{\rm clust}_{\rm gal} + n^{-1}_{\rm gal} \right) \right]^{1/2}}{\sqrt{2 N_{\rm m}}}
\end{equation}

We note that the finite spatial and spectral resolutions of the instrument will also affect the minimum physical scales, or equivalently $K_{\rm \perp, max}$ and $K_{\rm \parallel, max}$, that can be probed. In order to account for the reduction of sensitivity due to this effect, for $K_{\perp}$ and $K_{\parallel}$ modes we divide the thermal noise part of the uncertainty by scaling factors
\begin{equation}
\mathcal{R_{\perp}}(K_{\perp}) = e^{-K^2_{\perp}/K^2_{\rm \perp, max}}
\end{equation}
and
\begin{equation}
\mathcal{R_{\parallel}}(K_{\parallel}) = e^{-K^2_{\parallel}/K^2_{\rm \parallel, max}}~,
\end{equation}
respectively, where $K_{\rm \perp, max} \approx 2\pi \left( \chi \Omega^{1/2}_{\rm beam} \right)^{-1}$ and $K_{\rm \parallel, max} \approx 2\pi \left( \mathrm \delta \nu d \chi / \mathrm d \nu \right)^{-1}$ are characterized by the comoving radial distance $\chi$, the angular size of the beam $\Omega_\mathrm{beam}$, and the spectral resolution $\delta \nu$. 

These estimated uncertainties are combined with observables predicted by our fiducial model to generate mock data and allow parameter inference, which will be presented in the next section.


\section{Results} \label{sec:results}

Assuming a line scan optimized for reliably detecting the [\ion{C}{ii}] intensity fluctuations from the EoR as described in Section~\ref{sec:survey}, we adopt the fiducial model parameters given in Table~\ref{tb:model_params} and use the mode counting method discussed to create mock signals of the [\ion{C}{ii}], CO, and [\ion{C}{i}] power spectra TIME will measure. We then implement a Bayesian analysis framework for parameter estimation and solve it with the affine-invariant Markov Chain Monte Carlo (MCMC) code {\footnotesize EMCEE} \citep{FM_2013PASP}. For the inference of [\ion{C}{ii}], the calibration dataset for parameter fitting is taken to be the mock auto power spectra measured in two redshift bins by TIME, to be combined with independent constraints on the EoR history such as $\tau_{\rm es}$. For adjacent pairs of CO transitions and [\ion{C}{i}], the calibration dataset is taken to be the mock cross-power spectra. The likelihood function for fitting mock observations can be expressed as
\begin{equation}
l \left( \hat{x} \big| \hat{\theta} \right) = \prod^{N_{z}}_{i=0} \prod^{N_{K}}_{j=0} p_{ij}(K, z)~,
\end{equation}
where $N_K$ ($N_z$) denotes the number of $K$ (redshift) bins in which auto- or cross-power spectra are measured. The probability of the data vector $\hat{x}$ for a given set of model parameters $\hat{\theta}$ is assumed to be described by a normal distribution
\begin{equation}
p_{ij}= \frac{1}{\sqrt{2\pi} \sigma_{ij}(K, z)} \exp \left\{ - \frac{\left[ \mathcal{P}(K, z) - \mathcal{P}(K, z | \hat{\theta}) \right]^2}{2 \sigma^2_{ij}(K, z)} \right\}~,
\end{equation}
where $\sigma_{ij}$ represents the gaussian error associated with the measurement. As specified in Table~\ref{tb:model_params}, broad, uniform priors on the model parameters are used. The bounds are chosen to ensure that parameter values suggested by observations in literature fall well within the prior ranges.

The predicted detectability of various target signals of TIME and TIME-EXT, together with the constraints to be placed on the key astrophysical parameters involved in our models, are summarized in Table~\ref{tb:constraints}. We note that for brevity TIME-EXT forecasts will be shown for [\ion{C}{ii}] measurements only. The detectability of low-$z$ CO and [\ion{C}{i}] lines with cross-correlation will also be improved, though by a significantly smaller amount, as these measurements are dominated by sample variance rather than instrument noise --- the latter in general contributes less than half of the total power spectrum uncertainty in these cases.


\subsection{Constraints on [\ion{C}{ii}] Intensity}

\begin{figure}[h!]
\centering
\includegraphics[width=0.48\textwidth]{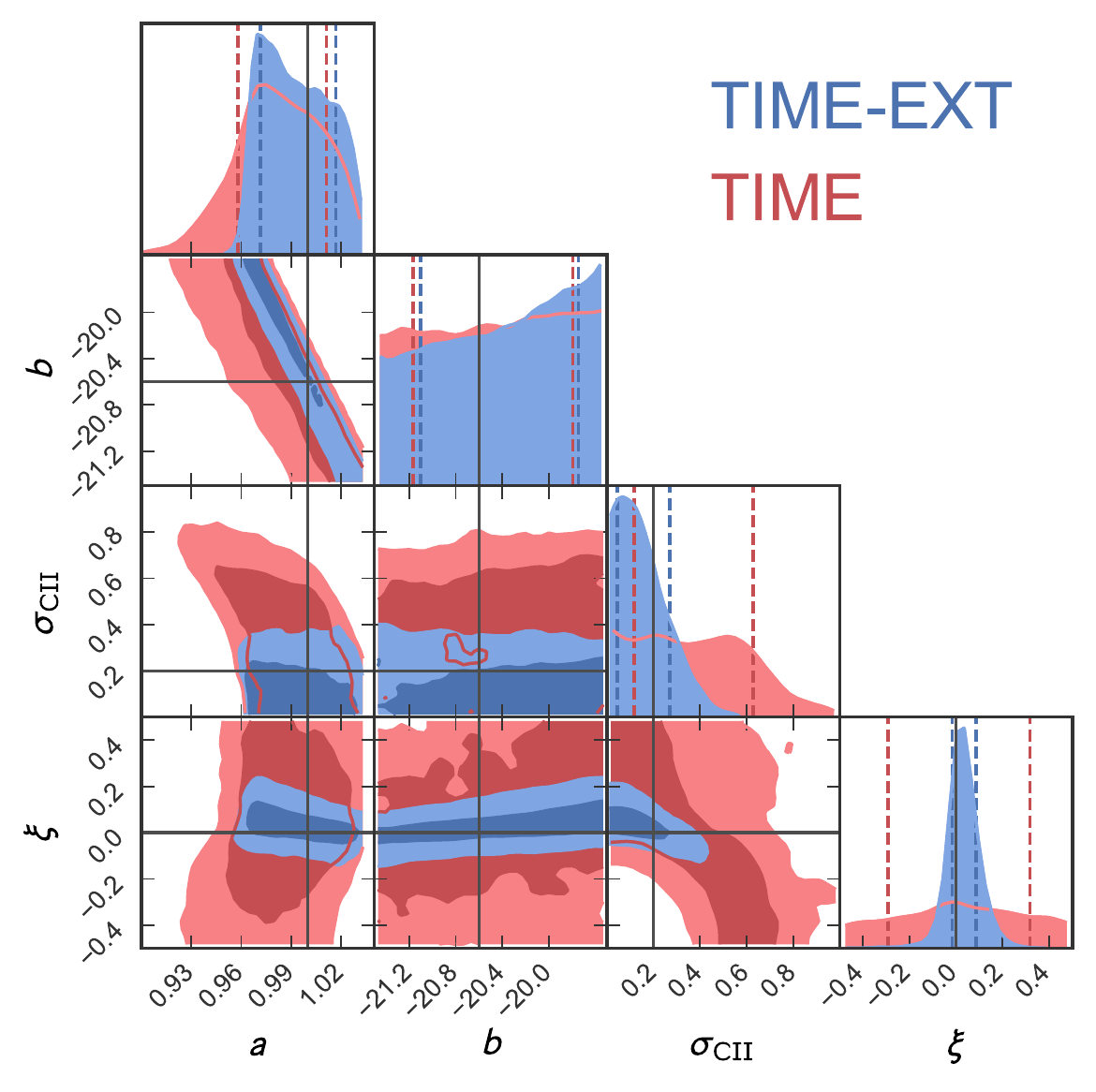}
\includegraphics[width=0.48\textwidth]{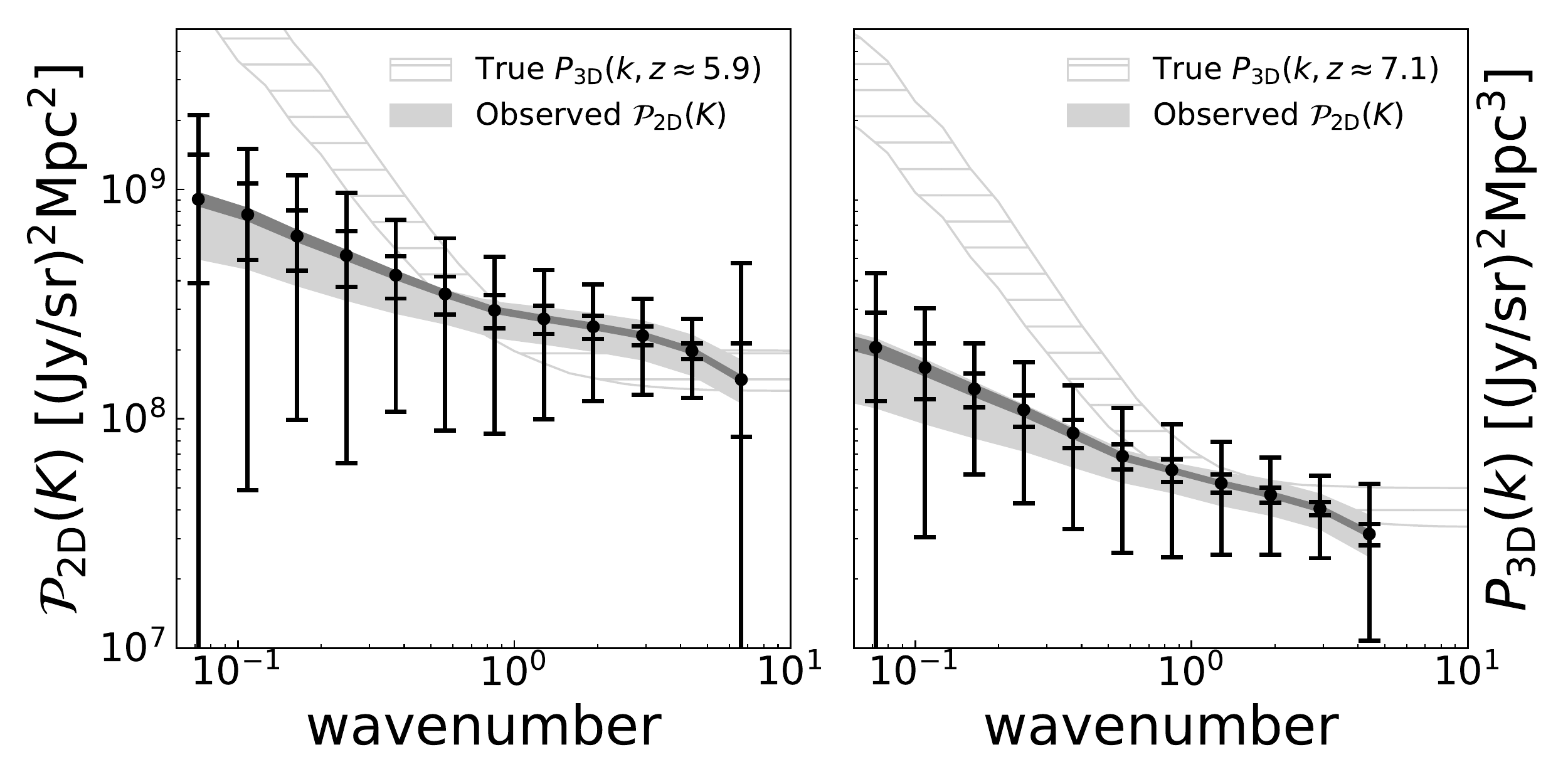}
\caption{Top: the joint posterior distribution of \{$a$, $b$, $\sigma_{\ion{C}{ii}}$, $\xi$\} constrained by TIME (red) and TIME-EXT (blue). True values of parameters in our fiducial model are indicated by the solid lines in gray, whereas the 68\% confidence intervals of marginalized distributions are shown by the vertical dashed lines. Bottom: constraining power of TIME's HF (low-$z$) and LF (high-$z$) bands on the [\ion{C}{ii}] power spectrum from a $1.3 \times 0.007\,\mathrm{deg}^2$ line scan. The data points denote TIME (outer) and TIME-EXT (inner) sensitivities to the binned, observed 2D power spectra $\mathcal{P}(K)$, estimated using the mode counting method described in Section~\ref{sec:survey-sensitivity}. The light and dark shaded bands represent the 68\% confidence intervals of the observed power spectra, inferred from the posterior distribution constrained by TIME and TIME-EXT, respectively. For reference, horizontally-hatched regions show the true, 68\% confidence intervals of 3D power spectra $P(k)$ constrained by TIME.} 
\label{fig:ciionlyps}
\end{figure}

\begin{figure}[h!]
\centering
\includegraphics[width=0.48\textwidth]{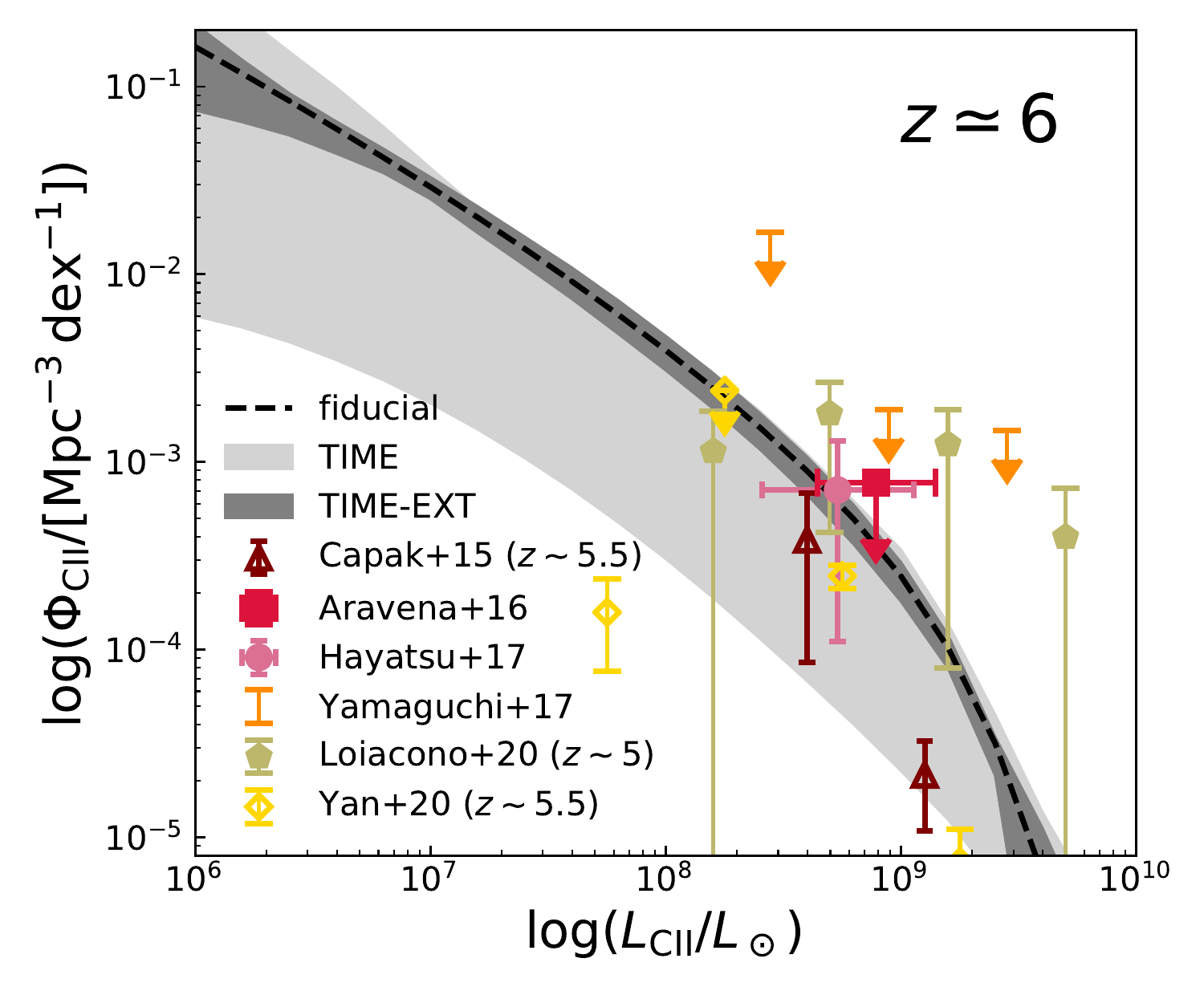}
\caption{Same as Figure~\ref{fig:ciilf_model}, but with the light and dark shaded regions indicating the 68\% confidence interval reconstructed from the joint posterior distribution of \{$a$, $b$, $\sigma_{\ion{C}{ii}}$, $\xi$\} constrained by TIME and TIME-EXT, respectively.}
\label{fig:ciilf_posterior}
\end{figure}

Using the measured [\ion{C}{ii}] auto-correlation power spectra, we can quantify the strength of [\ion{C}{ii}] emission by simultaneously constraining parameters $a$, $b$, $\sigma_{\ion{C}{ii}}$, and $\xi$ related to the [\ion{C}{ii}] power spectrum in our model (see Section~\ref{sec:model:CII}). Figure~\ref{fig:ciionlyps} shows the posterior distributions from the MCMC analysis, in which power spectrum templates specified by \{$a$, $b$, $\sigma_{\ion{C}{ii}}$, $\xi$\} are first projected into observing frame by the window function and then fit to the mock, observed 2D power spectra in the two sub-bands of TIME, which have a total S/N of 5.3 (HF) and 5.8 (LF), respectively. These numbers increase to 23 (HF) and 30 (LF) for TIME-EXT because of its enhanced survey power, as summarized in Table~\ref{tb:inst_specs}. Among the four parameters, constraining power is observed for $a$, $\sigma_{\ion{C}{ii}}$, and $\xi$ that affect (and therefore benefit from having access to) the full shape of the power spectrum, whereas $b$ controls only the normalization of the power spectrum and is prior dominated. In particular, a clear anti-correlation between $\sigma_{\ion{C}{ii}}$ and $\xi$ exists because they have similar effects on the power spectrum shape --- increasing $\sigma_{\ion{C}{ii}}$ elevates the shot-noise power (2nd moment of luminosity function), while increasing $\xi$ suppresses the star formation rate and [\ion{C}{ii}] emissivity of faint galaxies and therefore reduces the clustering power. The shot-noise power, on the other hand, is dominated by bright sources and thus not much affected by the faint-end behavior controlled by $\xi$. Such a degeneracy can be greatly reduced by TIME-EXT thanks to its increased constraining power on $\xi$, which is more than a factor of 5 better than TIME. The weak anti-correlation between $a$ and $\sigma_{\ion{C}{ii}}$ or $\xi$ (not obvious for TIME due to its low S/N) has a similar origin, since a steeper slope $a$ also gives rise to a flatter [\ion{C}{ii}] power spectrum with fractionally higher shot-noise power. 

\begin{figure}[h!]
\centering
\includegraphics[width=0.48\textwidth]{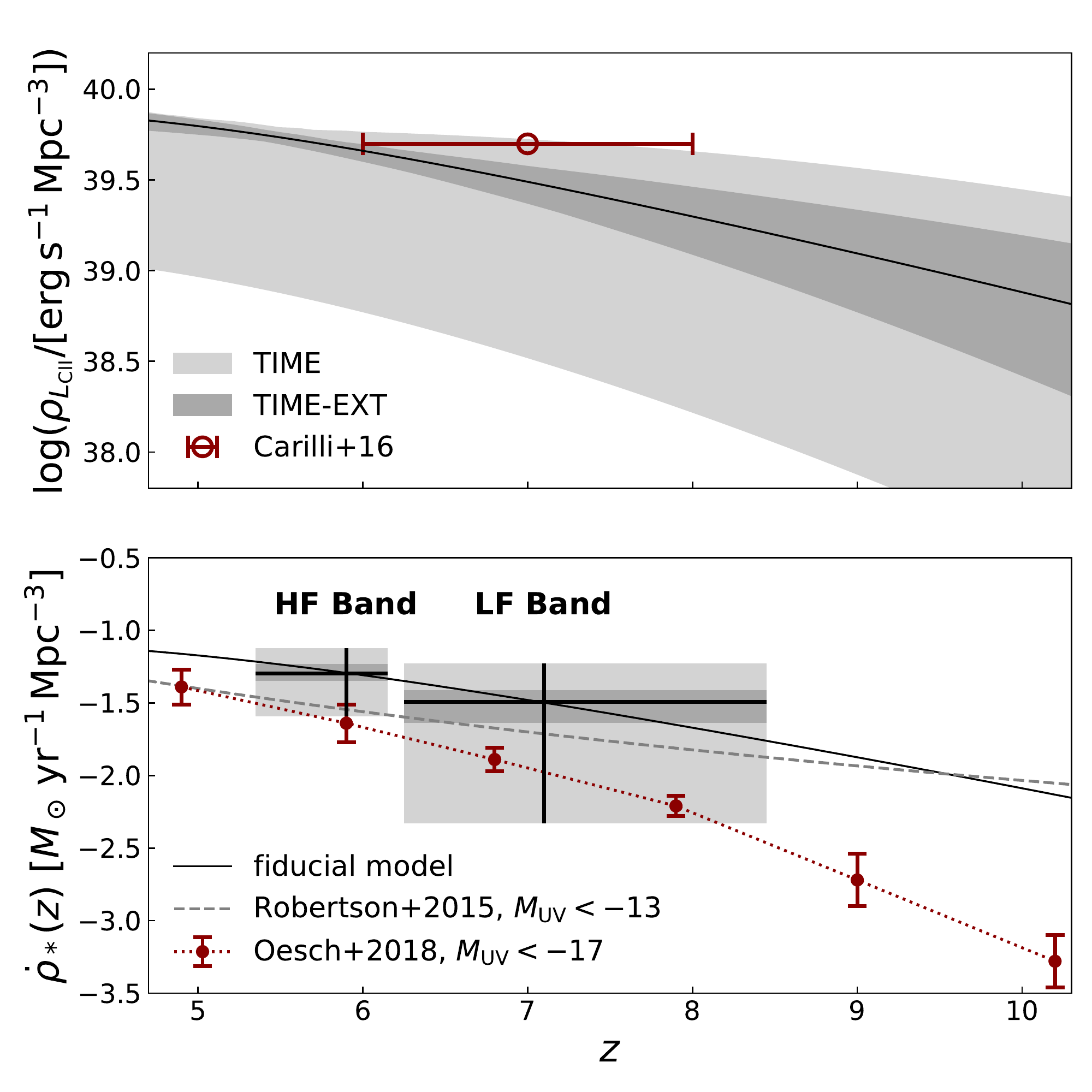}
\caption{Top: constraints (68\% confidence interval) on the [\ion{C}{ii}] luminosity density, calculated from TIME and TIME-EXT measurements, compared against the mean line brightness temperature at 242\,GHz measured from the ASPECS blind survey \citep{Carilli_2016}. Bottom: constraints (68\% confidence interval) on the cosmic SFRD provided by the low-$z$/HF and high-$z$/LF sub-bands of TIME and TIME-EXT. The solid curve shows our fiducial SFH assuming $\xi=0$ and a minimum halo mass of $M_{\rm min} = 10^8\,M_\odot$. For comparison, the dashed line shows the best-fit cosmic SFRD integrated down to $0.001\,L_{\star}$ ($M_{\rm UV}<-13$ at $z\sim7$) from \citet{Robertson_2015}, whereas the data points in red represent the observed SFRD from \citet{Oesch_2018} after the dust correction and with a limiting magnitude of $M_{\rm UV}<-17$.}
\label{fig:lcii_sfrd}
\end{figure}

From the joint posterior distribution, we are able to infer how accurately the [\ion{C}{ii}] luminosity function can be constrained by the measured power spectrum. As shown in Figure~\ref{fig:ciilf_posterior}, the integral constraints from [\ion{C}{ii}] power spectrum allow us to determine the [\ion{C}{ii}] luminosity function to within a factor of a few for TIME and smaller than 50\% for TIME-EXT. Even though the detailed shape determined from integral constraints is model dependent, such measurements provide unique information of the aggregate [\ion{C}{ii}] emission from galaxies, including the faintest [\ion{C}{ii}] emitters cannot be accessed by even the deepest galaxy observation to date. We can also determine the [\ion{C}{ii}] luminosity density evolution during the EoR. Figure~\ref{fig:lcii_sfrd} shows the level of constraint TIME is expected to provide on the [\ion{C}{ii}] luminosity density over $5 < z < 10$ assuming our fiducial [\ion{C}{ii}] model. We note that overall our fiducial model predicts lower [\ion{C}{ii}] luminosity density compared with the mean line brightness temperature in ALMA 242\,GHz band measured by \cite{Carilli_2016}. The apparent discrepancy between the measurement and our model may be understood in two ways. First, the ALMA observation based on individual, blindly-detected line emitters shall be interpreted as a lower limit because contribution from galaxies too faint to be blindly detected is not included. That said, it may include a substantial contribution from emission lines such as CO and [\ion{C}{i}] at lower redshifts, which typically requires near-IR counterparts to characterize \cite[see also][]{Decarli_2020}. 

Combined with improved measurements of the total SFR based on both optical/near-IR and mm-wave data, TIME's measurements of the distribution and overall density of [\ion{C}{ii}] emission help narrow down the uncertainty exists in the connection between [\ion{C}{ii}] line luminosity and the SFR, particularly at high $z$. Physical processes that determine [\ion{C}{ii}] luminosity and its scatter in EoR galaxies, including the ISM properties (e.g., metallicity and the interstellar radiation field), feedback, as well as the impact of stochasticity, can be consequently studied. 

\subsection{EoR Constraints Inferred From [\ion{C}{ii}] Measurements} \label{sec:results:eor}

\begin{figure}[h!]
\centering
\includegraphics[width=0.49\textwidth]{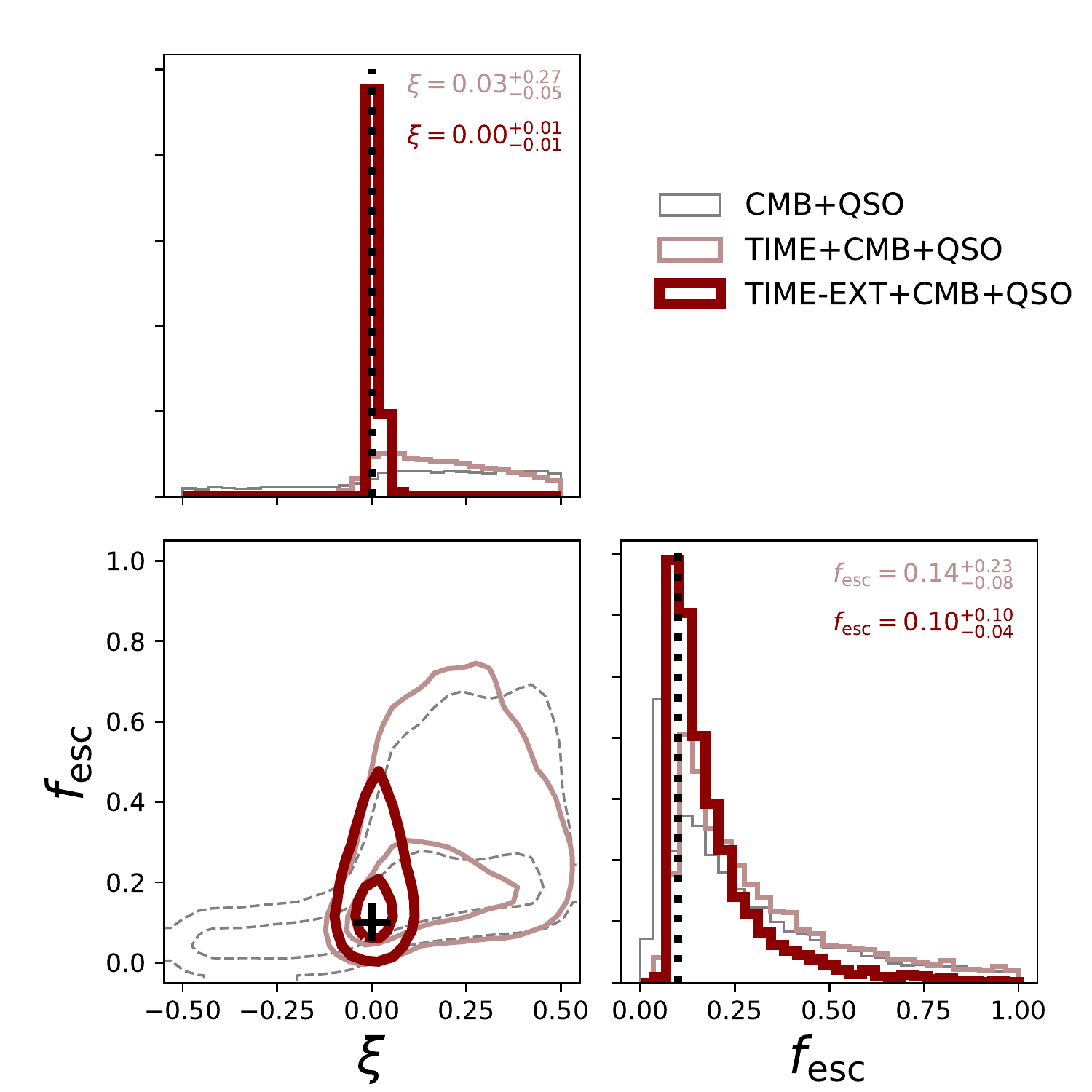}
\caption{The joint posterior distribution of the parameter $\xi$ measuring the contribution to reionization from faint galaxies and the escape fraction of ionizing photons $f_{\rm esc}$. The black cross and dotted lines indicate the fiducial values. The comparison among contours and histograms of different colors illustrates the improvement thanks to the addition of TIME and TIME-EXT measurements to constraints from the CMB optical depth and quasar absorption spectra. The 68\% confidence intervals (based on the highest posterior density) estimated from the marginalized distributions are quoted.}
\label{fig:corner_eor}
\end{figure}

 To illustrate the information TIME adds to our understanding of the EoR history, we consider two contrasting cases, namely whether or not to combine TIME data with other EoR constraints, including the integral constraint from Thomson scattering optical depth of CMB photons and constraints on the end of the EoR from quasar absorption spectra. Specifically, to include these observations as independent constraints in the MCMC analysis, we compare predictions of our reionization model (assuming Gaussian statistics) to $\tau_{\rm es} = 0.055\pm0.009$ \citep{Planck_XLVI} and $1-\bar{x}_i(z=5.5) < 0.1$ that represents an up-to-date, though conservative, constraint on the IGM neutrality near the end of reionization from quasar observations at $z \lesssim 6$ \cite[e.g.,][]{Fan_2006, McGreer_2015, Davies_2018}. 

Using these combined datasets, we simultaneously fit two EoR parameters of our model, namely the modulation factor $\xi$ controlling the contribution from the faint galaxy population and the population-averaged escape fraction of ionizing photons $f_{\rm esc}$, using the MCMC method. Values of [\ion{C}{ii}] parameters ($a$, $b$, and $\sigma_{\ion{C}{ii}}$) are fixed to their fiducial values in this exercise in order to better demonstrate the information contributed by a [\ion{C}{ii}] intensity mapping experiment. While fixing [\ion{C}{ii}] parameters is likely an oversimplified assumption given uncertainties associated with how well [\ion{C}{ii}] traces the SFR of EoR galaxies, future galaxy and LIM observations at mm/sub-mm wavelengths are expected to greatly improve the prior on the conversion from [\ion{C}{ii}] luminosity to the SFR. We therefore consider an idealized case of constraining $f_{\rm esc}$ with TIME/TIME-EXT when this conversion is perfectly known, similar to what is routinely done when inferring $f_{\rm esc}$ from rest-frame UV observations of EoR galaxies \cite[][]{Robertson_2015, Mason_2015, SF_2016, Yue_2018, Naidu_2020}. As a final note, we also verify that with the 5-parameter fitting the distributions of $\xi$ and $f_{\rm esc}$ do not deteriorate catastrophically. The resulting posterior distributions of the parameters are shown in Figure~\ref{fig:corner_eor}, where cases combing both TIME and external data from the CMB and quasars are compared against the case without TIME shown in gray. From the marginalized distributions, we find an average escape fraction of ionizing photons $f_{\rm esc} = 0.14^{+0.23}_{-0.08}$ ($f_{\rm esc} = 0.10^{+0.10}_{-0.04}$) and a faint-end modulation factor $\xi = 0.03_{-0.05}^{+0.27}$ ($\xi = 0.00_{-0.01}^{+0.01}$) for TIME (TIME-EXT), where the uncertainties are quoted for a 68\% confidence interval derived from the highest posterior density (HPD). By imposing a tight constraint on the faint-end slope of galaxy LF parametrized by $\xi$, TIME(-EXT) reduces the degeneracy between it and the escape fraction. An accurate measurement of $\xi$ also informs how the ionization background built up during the EoR may have suppressed star formation in galaxies hosted by low-mass halos. Effects of stellar and reionization feedbacks on the faint-end of galaxy LF provides important information about the interplay between reionization and its driving forces \citep{Furlanetto_2017, Yue_2018}.

\begin{figure}[h!]
\centering
\includegraphics[width=0.48\textwidth]{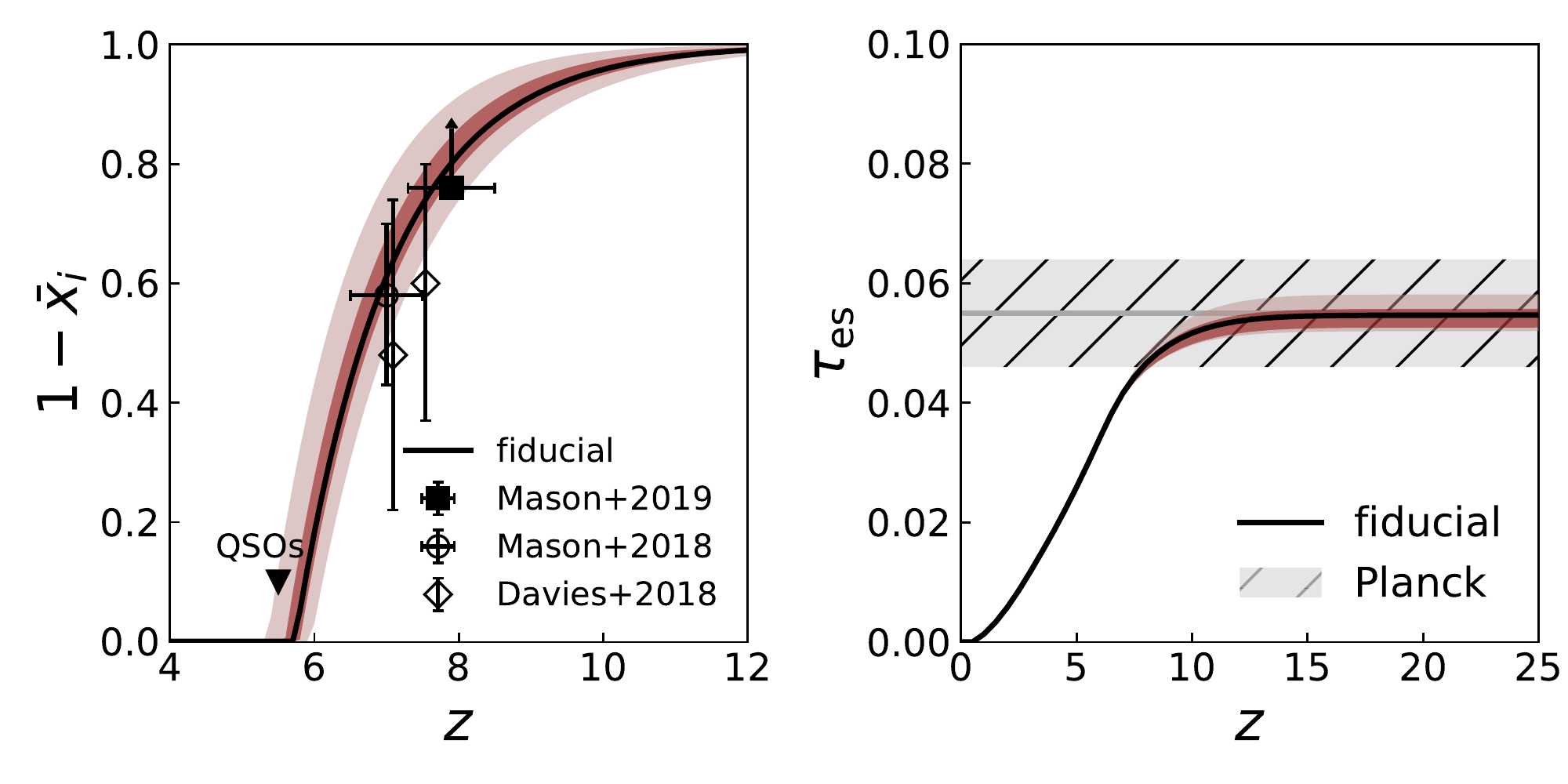}
\caption{Left: the redshift evolution of the average IGM neutrality $1-\bar{x}_i$ compared with the reionization timeline constraints from recent observations of LBGs and IGM damping wings of quasars at $z\ga7$. The dark (light) shaded region denotes the 68\% confidence level inferred from the SFRD constrained by TIME (TIME-EXT), when $f_\mathrm{esc}$ is held fixed at 0.1. Right: the CMB electron-scattering optical depth inferred from TIME and TIME-EXT measurements compared with constraints from Planck \citep{Planck_XLVI}. }
\label{fig:eor}
\end{figure}

TIME also sheds light onto the global history of reionization by constraining the cosmic SFR with integrated [\ion{C}{ii}] emission. Figure~\ref{fig:eor} shows the reionization timeline inferred from a joint analysis of TIME, the CMB optical depth, and quasar absorption spectra. The left panel shows the constraints on the evolution of the mean IGM neutrality $1 - \bar{x}_{i}$, compared with estimates 
based on Ly$\alpha$ emission from Lyman Break galaxies (LBGs) \cite[][]{Mason_2018, Mason_2019} and damping wing signatures of quasars \cite[][]{Davies_2018} at $z\ga7$. The reionization history implied by our fiducial model agrees reasonably well with the independent Ly$\alpha$ and quasar observations, which suggest that the IGM is about half ionized at $z\simeq7$. The right panel shows the inferred Thomson scattering optical depth of CMB photons. We note that because TIME only directly constrains the SFRD, $1-\bar{x}_i$ inferred this way is also subject to the uncertainty in $\tau_{\rm es}$, which is non-trivial compared with the fraction to be explained by hydrogen reionization at $z\ga6$ ($\Delta \tau_{\rm es} \approx 0.02)$. Nevertheless, the constraints from TIME are less susceptible to sample variance, and, in contrast to analyses of UV galaxies \cite[e.g.,][]{Robertson_2015, Mason_2015, SF_2016}, immune to the uncertainty associated with faint-end extrapolation. 


\subsection{[\ion{C}{ii}]--LAE Cross-Correlation}

As discussed in Section~\ref{sec:survey-strategy}, the survey strategy of TIME optimizes the detectability of large-scale modes. A line scan, however, limits the spatial overlap between [\ion{C}{ii}] data and LAEs available for a cross-power spectral analysis. Because the two-point correlation function in this case is computed as a function of angular distance, we can include LAEs that do not fall exactly along the scan path, thereby increasing the number of LAE--voxel pairs available for constraining [\ion{C}{ii}]-LAE angular clustering. Using Equation~(\ref{eq:acf}), we compute the angular correlation function between LAEs and the [\ion{C}{ii}] line intensity measured by TIME. To estimate the detectability of the cross-correlation signal, we first extract mock [\ion{C}{ii}] data in the TIME spectral channel corresponding to the redshift of LAEs identified by the Subaru HSC narrow-band filter. The [\ion{C}{ii}] data from a line scan of 180 beams wide is then cross-correlated with angular positions of LAEs simulated in a $1.4\times1.4=2\,\mathrm{deg}^2$ field. 

\begin{figure}[h!]
\centering
\includegraphics[width=0.48\textwidth]{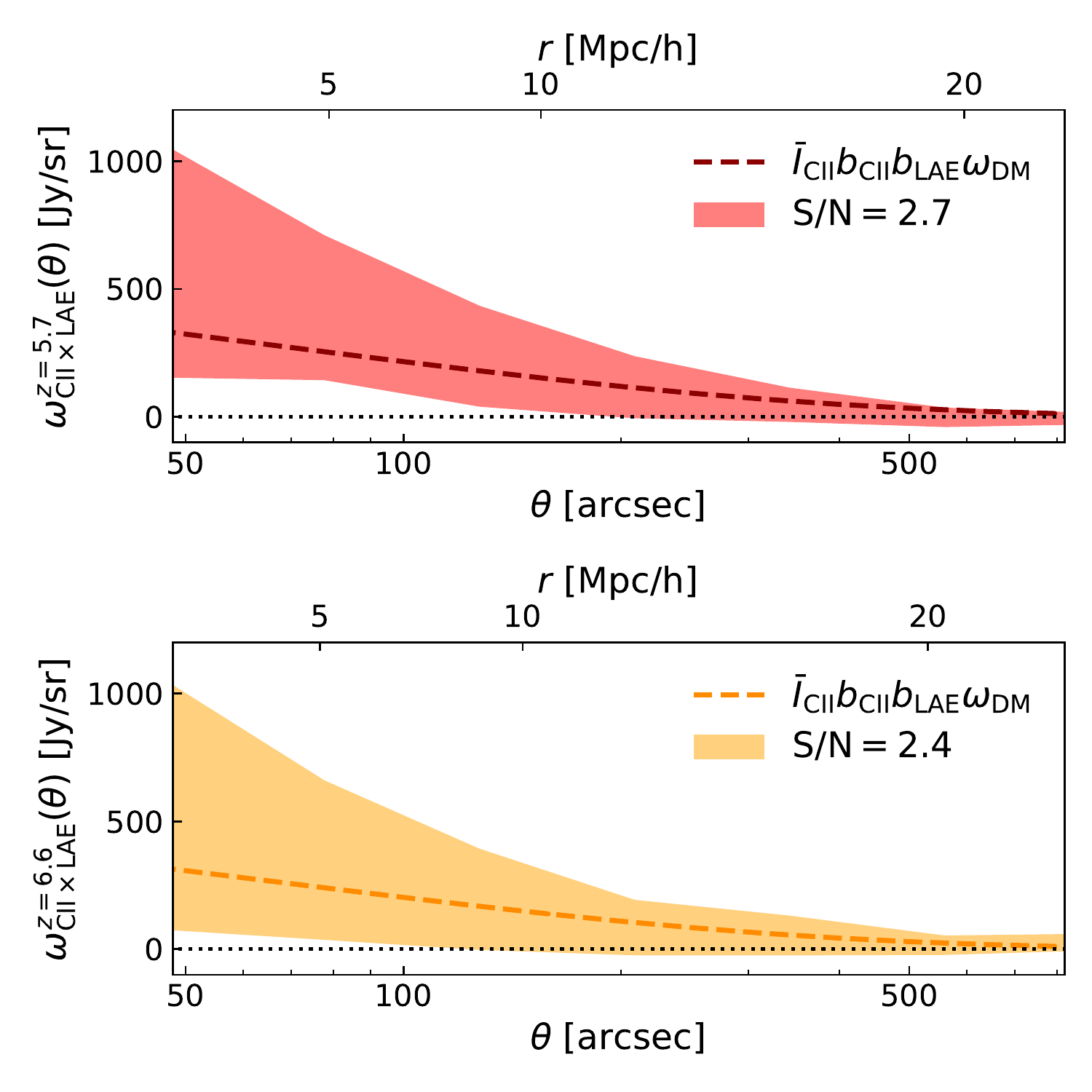}
\caption{Sensitivity to the angular cross-correlation function between the [\ion{C}{ii}] line intensity and LAEs surveyed by Subaru HSC at $z=5.7$ and 6.6. The dashed lines show analytical approximations $\omega_{\ion{C}{ii} \times \mathrm{LAE}} \approx b_{\rm LAE} \bar{b}_{\ion{C}{ii}} \bar{I}_{\ion{C}{ii}} \omega_{\rm DM}$ scaled from the angular correlation function of dark matter. The shaded region indicates the 68\% confidence interval from measurements of TIME, estimated by bootstrapping 1000 randomized LAE catalogs.}
\label{fig:ciilae_acf}
\end{figure}

Figure~\ref{fig:ciilae_acf} shows the sensitivity of [\ion{C}{ii}]--LAE angular correlation function $\omega_{\ion{C}{ii} \times \mathrm{LAE}}$ at $z=5.7$ and 6.6, as predicted by our semi-analytical approach. For comparison, we also show the angular correlation function of dark matter (from linear theory) scaled by $b^2_{\rm LAE}$. While only marginal detections of the angular correlation function are expected due to the limited survey size, upper limits inferred from this cross-correlation provide a valuable independent check against our [\ion{C}{ii}] auto-correlation analysis. Taking $b_{\rm LAE} \sim 6$ inferred from our simulated LAE distributions, which is broadly consistent with measurements from \citet{Ouchi_2018}, and restricting the fitting to linear scales with $r > 10\,\mathrm{Mpc}$, we obtain $b_{\ion{C}{ii}} \bar{I}_{\ion{C}{ii}} = 2700\pm3200\,\mathrm{Jy/sr}$ at $z=5.7$ and $2600\pm2900\,\mathrm{Jy/sr}$ at $z=6.6$, respectively. Because of the restricted number of LAE--voxel pairs given TIME's small survey area, sample variance contributes a significant fraction ($>60\%$) of the uncertainty in $\omega_{\ion{C}{ii} \times \mathrm{LAE}}$ measurements predicted above, which is estimated by bootstrapping 1000 randomized LAE catalogs. Thus, with the same survey area as TIME but lower instrument noise, TIME-EXT only slightly improves the detectability of $\omega_{\ion{C}{ii} \times \mathrm{LAE}}$. Nevertheless, as will be discussed in Section~\ref{sec:discuss:ng}, precise measurements of $\omega_{\ion{C}{ii} \times \mathrm{LAE}}$ during the EoR will be one of the major targets for next-generation [\ion{C}{ii}] LIM experiments covering $\sim10\,\mathrm{deg}^2$ of sky.

\subsection{Probing Physics of Molecular Gas Growth with CO and [\ion{C}{i}] Intensities}

Here, we consider two potential applications of in-band cross-correlation to measure the strengths of CO and [\ion{C}{i}] lines from $0.5 \la z \la 2$. The mean intensities of these lines extracted from cross-power spectra reveal physical information about molecular gas in galaxies near cosmic noon. In the first scenario, we assume a fixed CO rotational ladder, with the line ratios to CO(1-0) specified by the scaling factors provided in Section~\ref{sec:model:CO}, and constrain the molecular gas density evolution by converting luminosities of higher-$J$ CO lines into CO(1-0) luminosity (see Section~ \ref{sec:model-mh2}). We relax our assumption that the CO rotational ladder is known in the second scenario, and use the cross-correlations of three pairs of CO and [\ion{C}{i}] lines to determine their individual strengths.

\subsubsection{Cross-correlating high-$J$ CO lines at $0.5 \lesssim z \lesssim 2$}

\begin{figure}[h!]
\centering
\includegraphics[width=0.48\textwidth]{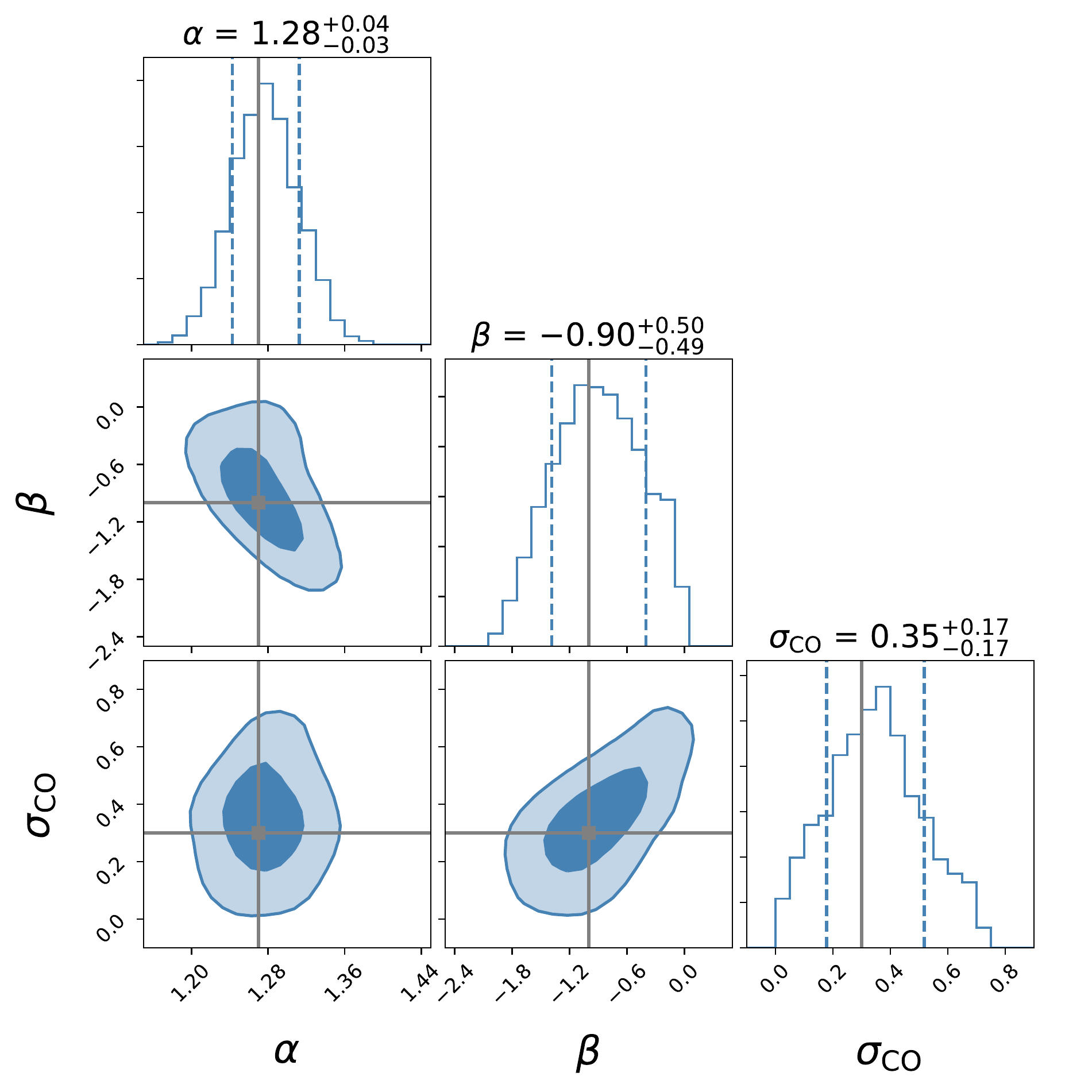}
\includegraphics[width=0.48\textwidth]{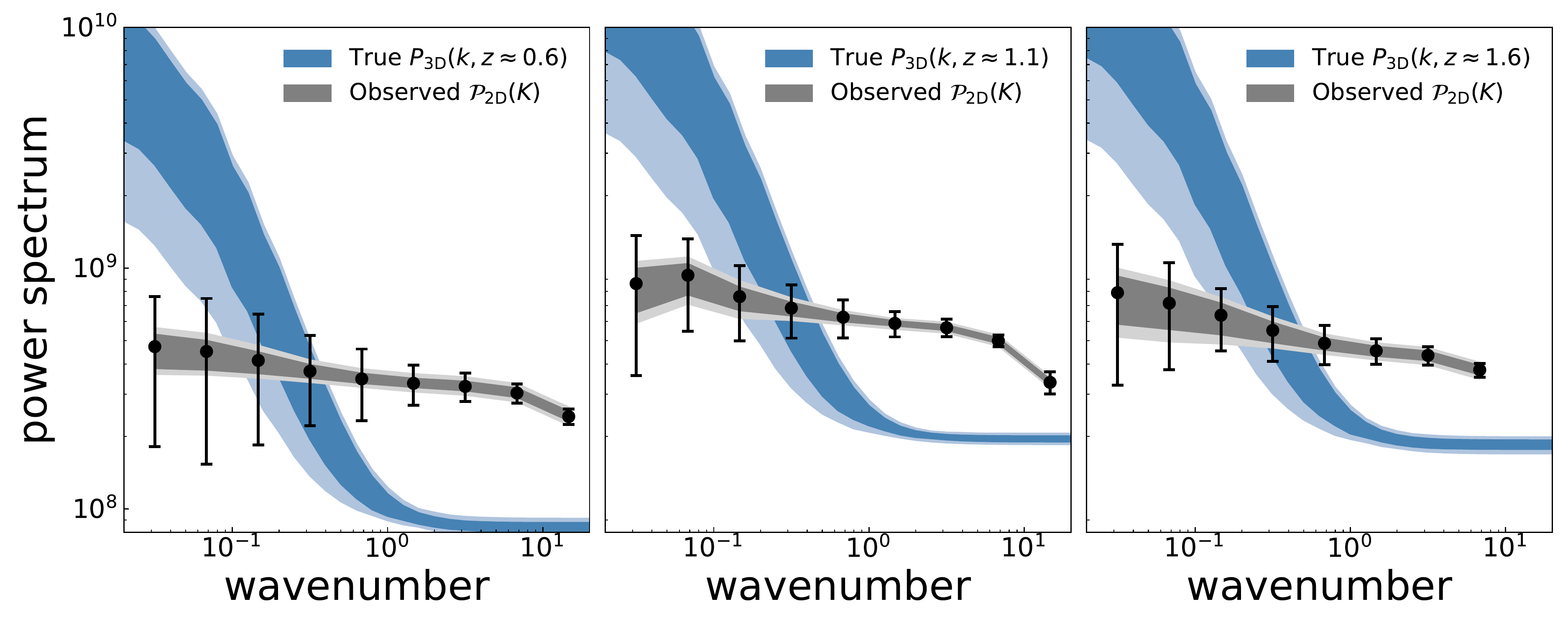}
\caption{Top: joint posterior distributions of the free parameters of our CO model inferred from cross-correlating pairs of adjacent CO rotational lines over $0.5 \la z \la 2$ observable by TIME. Bottom: TIME constraints on the cross-correlation power spectra of CO(3--2)$\times$CO(4--3) at $z\approx0.6$, CO(4--3)$\times$CO(5--4) at $z\approx1.1$ and CO(5--4)$\times$CO(6--5) at $z\approx1.6$. 68\% and 95\% confidence intervals of the cross-power spectra, derived from 1000 random samples of the posterior distribution, are shown. }
\label{fig:cops}
\end{figure}

By cross-correlating intensity maps measured at frequencies corresponding to a pair of adjacent CO lines emitted at the same redshift, TIME can measure the intensity product of two CO lines. Thanks to the wide bandwidth of TIME, we are able to detect multiple CO transitions over $0 < z < 2$ and thereby determine the evolution of molecular gas content. In order to quantify how well TIME constrains $\rho_{\rm H_2}$, we create mock CO data with our fiducial model outlined in Section~\ref{sec:model:CO} and apply an MCMC analysis in a similar manner to the [\ion{C}{ii}] case. Specifically, we consider measuring the cross-power spectra of CO(3--2)$\times$CO(4--3) at $z \approx 0.6$ ($0.53<z<0.73$), CO(4--3)$\times$CO(5--4) at $z \approx 1.1$ ($0.90<z<1.31$), and CO(5--4)$\times$CO(6--5) at $z \approx 1.6$ ($1.29<z<1.88$), which end up using 13, 21, and 25 TIME spectral channels, respectively. 

\begin{figure}[h!]
\centering
\includegraphics[width=0.48\textwidth]{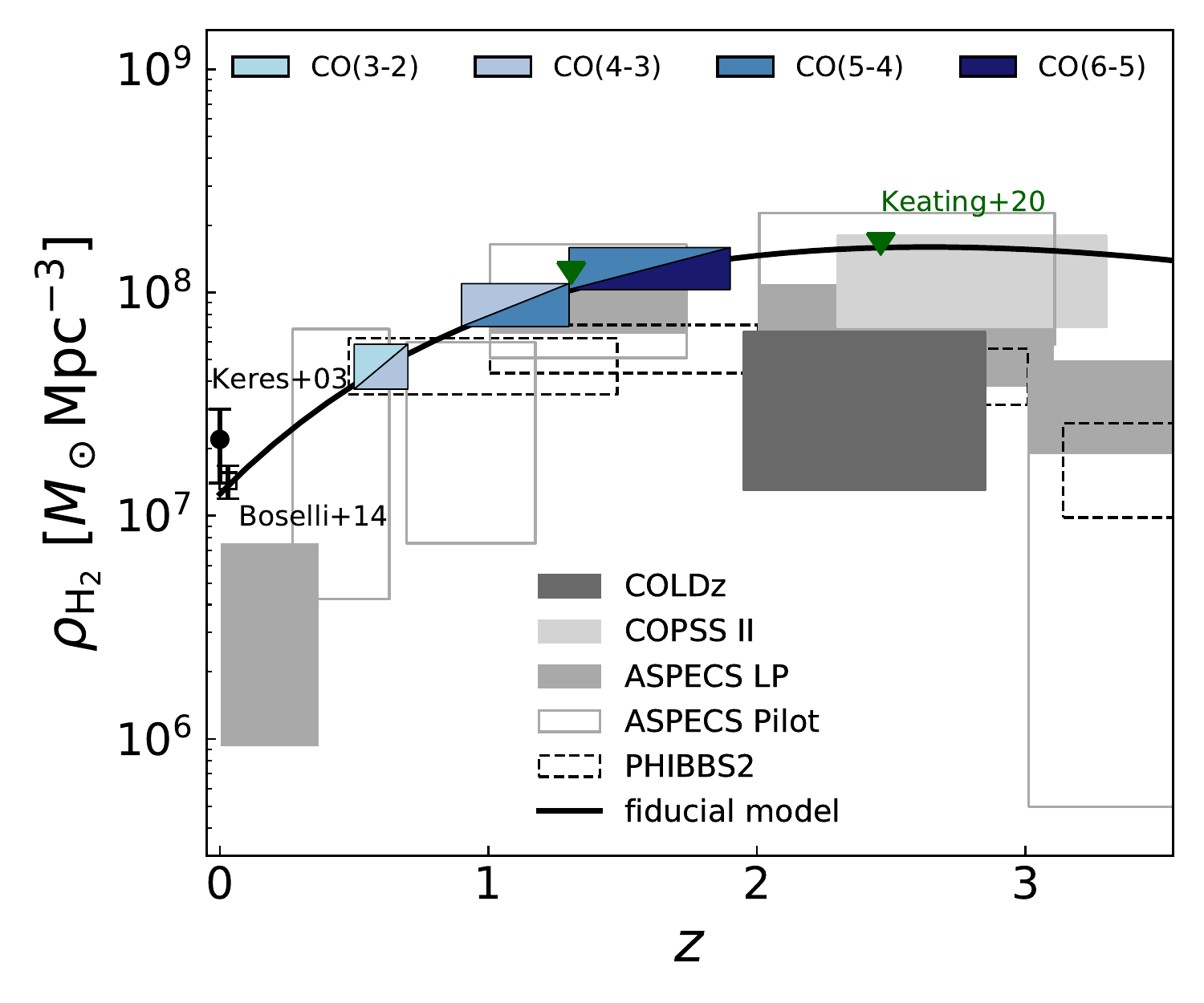}
\caption{The evolution of molecular gas density over $0<z<3.5$, compared with several molecular line observations showing a wide range of variation, as indicated by the data from COLDz \citep{Riechers_2019}, COPSS~II \citep{Keating_2016}, mmIME \citep{Keating_2020}, ASPECS Pilot \citep{Decarli_2016}, ASPECS large program \citep{Decarli_2019}, PHIBBS2 \citep{Lenkic_2020}, and near $z=0$ by \cite{Keres_2003} (filled circle) and \cite{Boselli_2014} (open square). Boxes in blue denote the constraints TIME is expected to provide by cross-correlating pairs of adjacent CO lines emitted from the same redshift, assuming our fiducial CO model (black curve).}
\label{fig:rhoH2}
\end{figure}

The resulting posterior distributions of our CO model parameters $\{ \alpha, \beta, \sigma_{\rm CO} \}$ are shown in Figure~\ref{fig:cops}, in tandem with the reproduced cross-power spectra of CO(3--2)$\times$CO(4--3), CO(4--3)$\times$CO(5-4) and CO(6--5)$\times$CO(5--4). As is the case of [\ion{C}{ii}], the slope $\alpha$ anti-correlates with the intercept $\beta$ of the log-log relation specifying the line luminosity. However, an anti-correlation between the slope $\alpha$ and scatter $\sigma_{\rm CO}$ is not observed, as the CO slope is always greater than unity and thus only weakly affects the shape of the power spectrum. The constraints on the cosmic evolution of molecular gas density $\rho_{\rm H_2}$ inferred from MCMC analysis of the cross-power spectra are shown in Figure~\ref{fig:rhoH2} as boxes in various shades of blue, indicating the pairs of CO lines being cross-correlated at different redshift intervals. These constraints are competitive compared with a collection of estimates from existing molecular line observations \cite[][]{Keres_2003, Boselli_2014, Keating_2016, Keating_2020, Riechers_2019, Decarli_2016, Decarli_2019}. While interpreting the CO signal requires common assumptions about the CO excitation and the relation between CO luminosity and $\rm H_2$ mass as discussed in Section~\ref{sec:model-mh2}, overall TIME complements other CO surveys by providing high-significance ($\mathrm{S/N} \ga 5$ in each redshift bin) constraints on the cosmic molecular gas density evolution near cosmic noon. 

By quantifying the volume-averaged depletion timescale of cold molecular gas, which can be defined as $\langle t_{\rm depl} \rangle = \rho_{\rm H_2}/\dot{\rho}_*$, these $\rho_{\rm H_2}$ measurements from TIME provide important information for understanding the nearly factor-of-10 decline of cosmic SFRD over this redshift range \cite[see also][]{Walter_2020arXiv, Decarli_2020}. 

\subsubsection{Cross-correlating CO and [\ion{C}{i}] lines at $z\approx1.1$}

\begin{figure}[h!]
\centering
\includegraphics[width=0.48\textwidth]{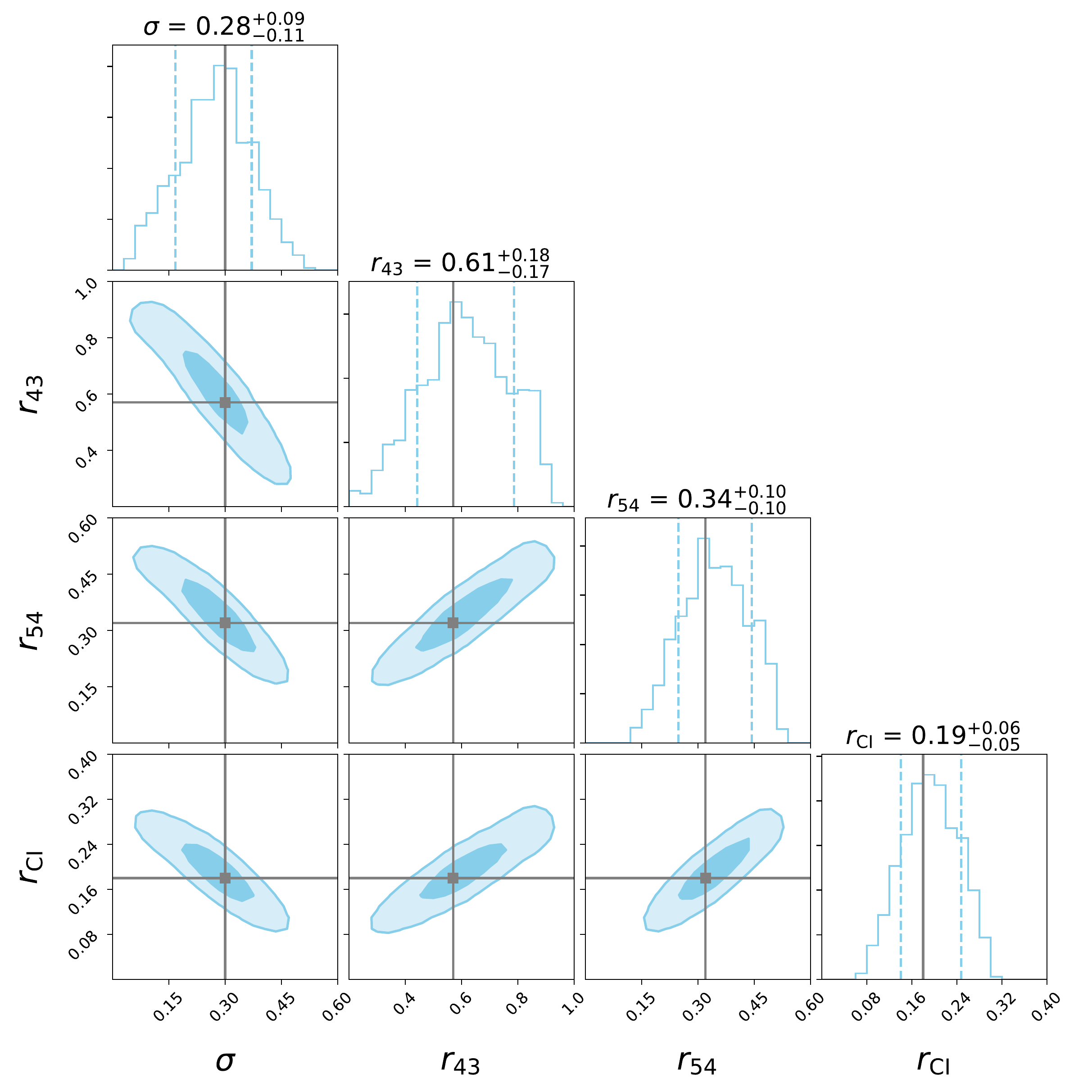}
\includegraphics[width=0.48\textwidth]{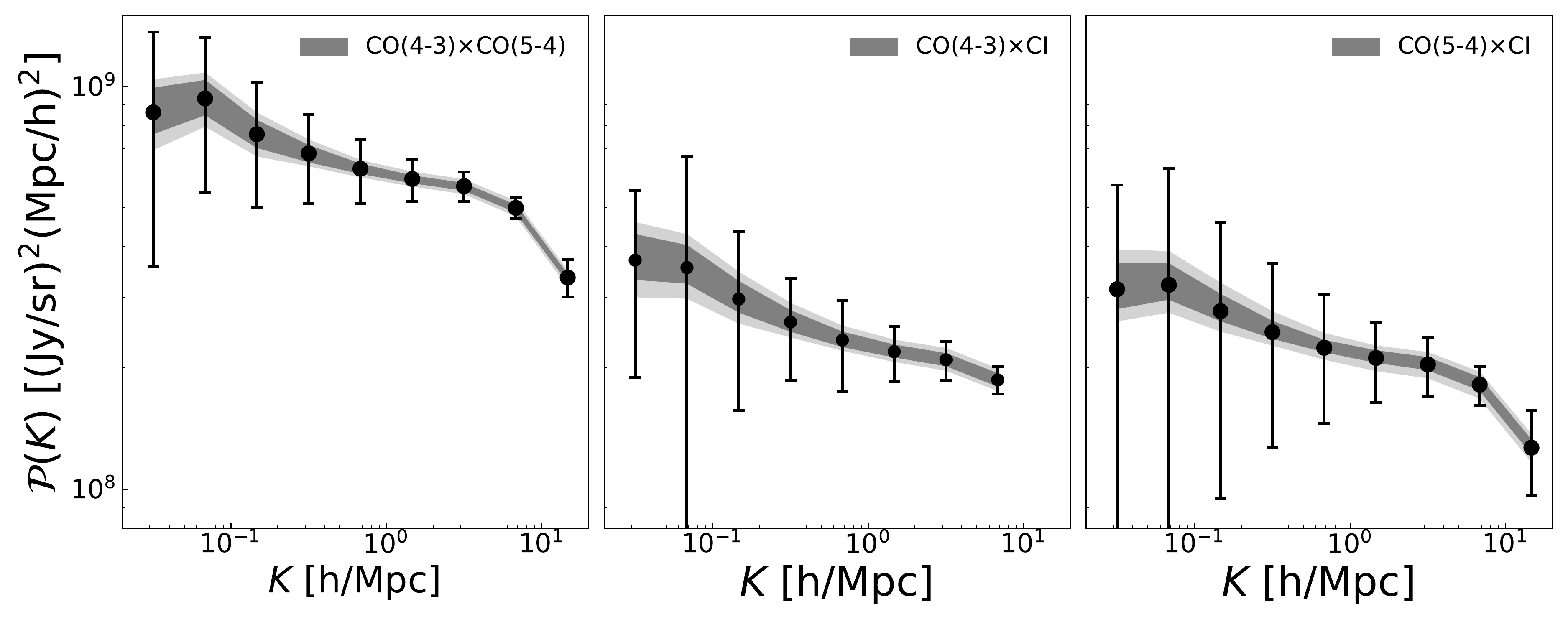}
\caption{Top: joint posterior distributions of the free parameters inferred from cross-correlating the CO(4--3), CO(5--4), and [\ion{C}{i}] lines at $z\sim1.1$. Assuming all three lines are proportional to CO(1-0) line by some unknown scaling factors and a common log scatter $\sigma$, values of $r_{43}$, $r_{54}$, and $r_{\ion{C}{i}}$ and $\sigma$ can be determined at 3-sigma level from the mutual cross-correlations. Bottom: the constraining power of TIME on the mutual cross-power spectra. 68\% and 95\% confidence intervals of the cross-power spectra, derived from 1000 random samples of the posterior distribution, are shown. }
\label{fig:coci_cps}
\end{figure}

In addition to measuring pairs of adjacent CO transitions at different redshift, TIME can simultaneously observe the CO(4--3), CO(5--4) and [\ion{C}{i}] lines emitted by the same sources at $0.9 \lesssim z \lesssim 1.3$. Provided that these three lines are perfectly correlated (as assumed in our model), their mutual cross-correlations can uniquely determine the mean emission from each individual line, while being immune to bright line interlopers \citep{Serra_2016, BVNL_2019}. We note that although CO and [\ion{C}{i}] lines can be similarly related to the total infrared luminosity by empirical scaling relations described in Section~\ref{sec:model:CO}, in practice [\ion{C}{i}] is likely not perfectly correlated with CO(4--3) or CO(5--4) line due to source-to-source variations such as the gas excitation state, galaxy type, [\ion{C}{i}] abundance, and so forth. The resulting de-correlation, quantifiable by measuring the [\ion{C}{i}] auto-power spectrum, needs to be taken into account in actual data analysis, but is ignored here for simplicity.

To illustrate the constraining power TIME will offer on these line strengths, we fit templates of cross-power spectrum specified by a set of four free parameters $\{\sigma, r_{43}, r_{54}, r_{\ion{C}{i}} \}$, to mock observations created assuming their fiducial values as specified in Section~\ref{sec:model:CO} and Table~\ref{tb:model_params} with the MCMC technique. Uninformative priors over [0,1] are assumed. Figure~\ref{fig:coci_cps} shows the posterior distributions of the free parameters constrained by the mock observed cross-power spectra $\mathcal{P}_{\rm CO(4-3) \times CO(5-4)}$, $\mathcal{P}_{\rm CO(4-3) \times \ion{C}{i}}$, and $\mathcal{P}_{\rm CO(5-4) \times \ion{C}{i}}$, which are measured at $\mathrm{S/N}=26$, 18, and 13, respectively. Under the assumption that all these lines are proportional to CO(1-0) line and share the same log scatter $\sigma$, values of $r_{43}$, $r_{54}$, and $r_{\ion{C}{i}}$ and $\sigma$ can be determined at 3-sigma level from the mutual cross-correlations. The anti-correlation between the scaling factors and $\sigma$ is because increasing $\sigma$ will increase the overall amplitudes of all the cross-power spectra.

These measurements of [\ion{C}{i}]-to-CO line ratios provide direct constraints on the mass fraction of neutral atomic carbon $f_{\ion{C}{i}}=M_{\ion{C}{i}}/M_{\rm C}$ across the entire galaxy population at $z\sim1$, which can be compared against the values ($\la 10\%$) derived from ALMA observations of individual main-sequence galaxies at similar redshift \cite[e.g.,][]{Valentino_2018} to better understand how different phases of carbon co-exist in PDRs and molecular clouds.


\begin{figure}[h!]
\centering
\includegraphics[width=0.48\textwidth]{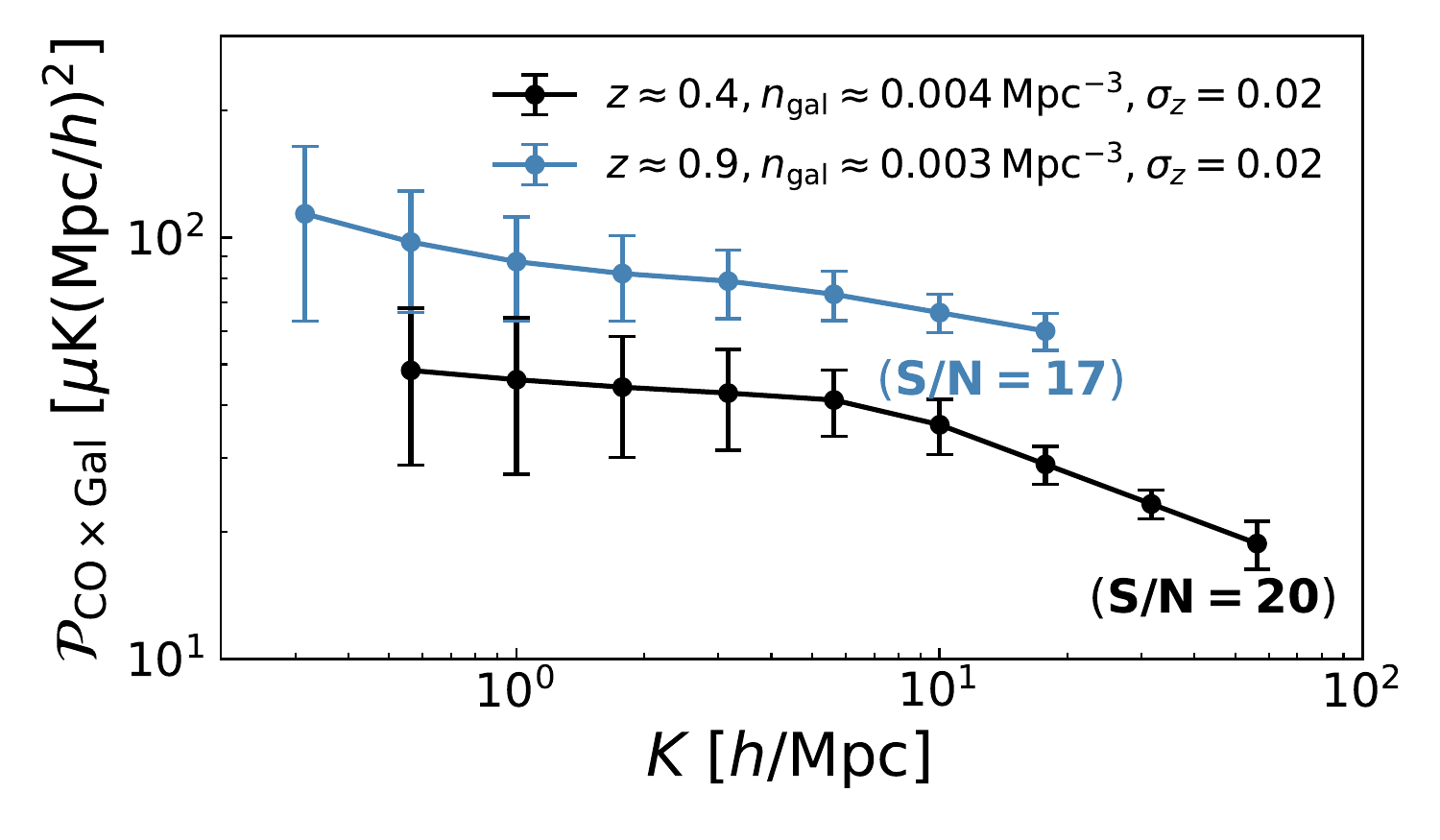}
\includegraphics[width=0.48\textwidth]{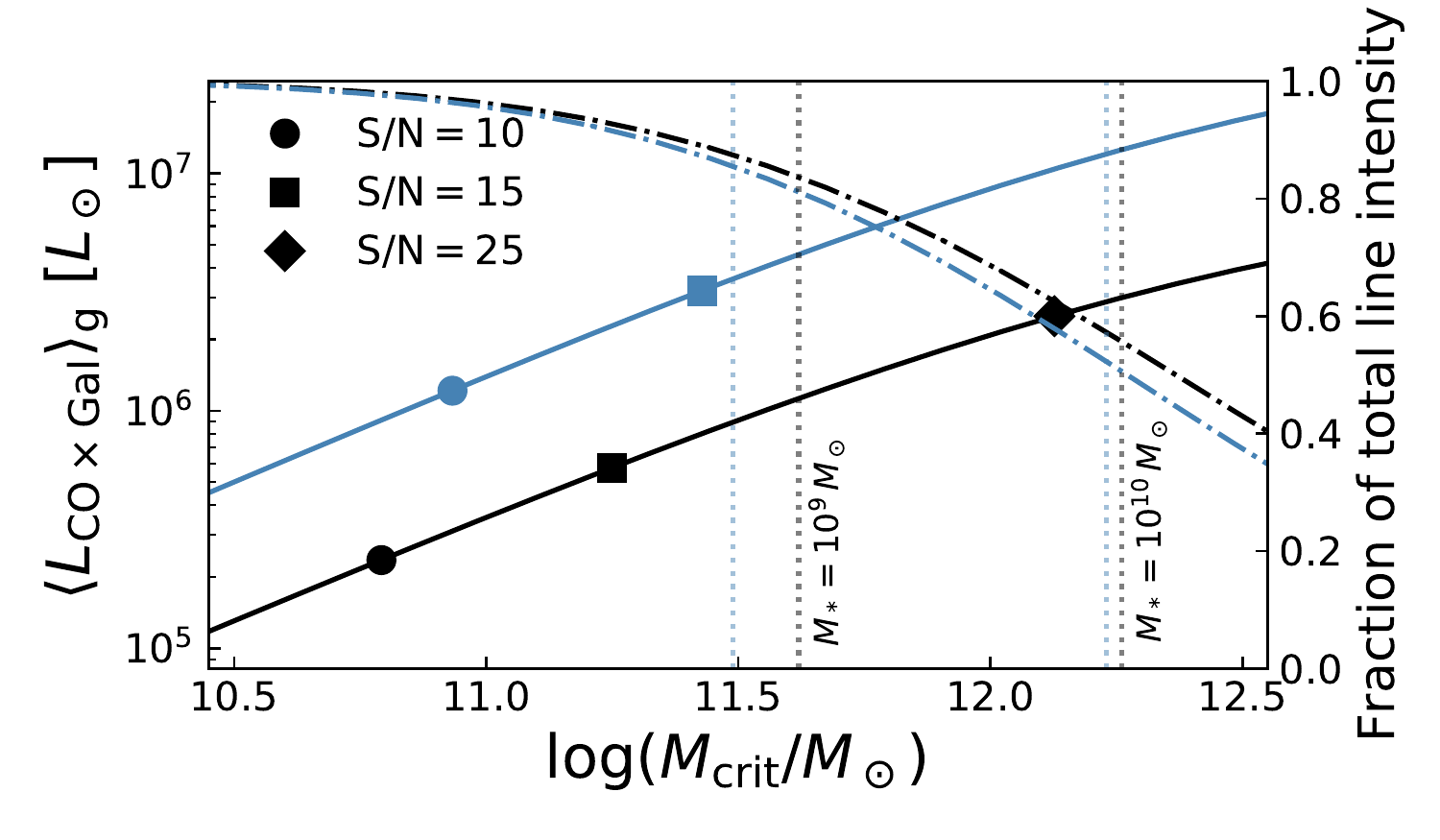}
\caption{Top: predicted TIME sensitivities to the CO--galaxy cross-power spectra at $z \approx 0.4$ and 0.9 for CO(3-2) and CO(4-3) lines, respectively. Bottom: mean CO line luminosity of individual galaxy samples measurable from the cross shot power (left axis) and the fraction of total CO line intensity consisting of the galaxy samples (right axis) as a function of selection threshold, measured in critical halo mass $M_{\rm crit}$ or stellar mass $M_{*}$ (dotted lines). }
\label{fig:COxGal}
\end{figure}

\subsection{CO--Galaxy Cross-Correlation}

As discussed in Section~\ref{sec:model:gal}, the cross-correlation between TIME maps of low-redshift CO lines with the galaxy density field serves as an important check for separating line foregrounds and a useful probe of CO emission associated with the selected galaxy population. Despite the small survey volume of TIME, a significant number of photometric/spectroscopic galaxies are still expected to be incorporated, allowing a meaningful measurement of the cross-power spectrum at $z\sim1$. The top panel of Figure~\ref{fig:COxGal} shows the predicted detectability of CO--galaxy cross-power spectrum by TIME, as described in Equation~(\ref{eq:COxGal}) and projected into the observing frame of TIME by the corresponding 
window function. We consider two examples in which TIME maps of CO(3-2) and CO(4-3) lines are cross-correlated with photometric galaxies ($\sigma^{\rm phot}_z \approx 0.02$) at $z\approx0.4$ and 0.9, respectively. As summarized in Table~\ref{tb:constraints}, these cross-power measurements allow us to infer the mean CO intensity $\bar{I}_{\rm CO,gal}$ attributed to the overlapped galaxy sample from the shot-noise component, which dominates the total power spectrum. The product of mean CO bias and intensity $\bar{b}_{\rm CO} \bar{I}_{\rm CO}$ may also be weakly constrained by the clustering component, which is only marginally detected due to the limited survey size of TIME. 

The amplitude and detectability of the CO--galaxy shot power is sensitive to the selection threshold of galaxy samples. As demonstrated in the bottom panel of Figure~\ref{fig:COxGal}, as the selection threshold (measured in critical halo mass or stellar mass) increases, galaxies selected approach the brighter end of the CO luminosity function. This in turn enhances the overall detectability of the cross shot power, though at the expense of probing a less representative sample of CO-emitting galaxies, as indicated by the right axis, which shows the fraction of total CO line emission associated with the galaxies selected. Given prior information on galaxy redshifts, we can probe the shape of the CO luminosity function by measuring this cross-shot power for galaxy samples with varying critical mass.

\begin{table*}[hbt]
\centering
\caption{Predicted constraints on astrophysical parameters from different TIME observables}
\label{tb:constraints}
\begin{tabular}{ccccc}
\toprule
\toprule
Observable & TIME (TIME-EXT) S/N & Parameter & TIME (TIME-EXT) Constraint & Reference \\
\hline
\multirow{6}{*}{$P_{\ion{C}{ii}}$} & \multirow{6}{*}{HF: 5.3 (23.1), LF: 5.8 (29.9)} & $a$ & $0.98^{+0.03}_{-0.03}$ ($0.99^{+0.02}_{-0.02}$) & \multirow{6}{*}{Eq.~(\ref{eq:lcii_luv}, \ref{eq:sfe}, \ref{eq:ciips})/Fig.~(\ref{fig:ciionlyps})} \\
 & & $b$ & $-20.46^{+0.67}_{-0.70}$ ($-20.36^{+0.62}_{-0.74}$) & \\
 & & $\sigma_{\ion{C}{ii}}$ & $0.44^{+0.24}_{-0.27}$ ($0.14^{+0.13}_{-0.09}$) & \\
 & & $\xi$ & $-0.01^{+0.31}_{-0.30}$ ($0.03^{+0.06}_{-0.04}$) & \\
 & & \multirow{2}{*}{$\bar{b}_\ion{C}{ii} \bar{I}_\ion{C}{ii}$ [Jy/sr]} & HF: $3260^{+480}_{-850}$ ($3970^{+130}_{-200}$) & \\ 
 & & & LF: $1580^{+560}_{-390}$ ($1870^{+170}_{-110}$) & \\
 \hline
\multirow{2}{*}{$P_{\ion{C}{ii}}$ (with $\tau_{\rm es}$ and QSOs)} & \multirow{2}{*}{HF: 5.3 (23.1), LF: 5.8 (29.9)} & $\xi$ & $0.03^{+0.27}_{-0.05}$ ($0.00^{+0.01}_{-0.01}$) & \multirow{2}{*}{Eq.~(\ref{eq:sfe}, \ref{eq:ode_Q})/Fig.~(\ref{fig:corner_eor})} \\
 & & $f_{\rm esc}$ & $0.14^{+0.23}_{-0.08}$ ($0.10^{+0.10}_{-0.04}$) & \\
\hline
\multirow{2}{*}{$\omega_{\rm \ion{C}{ii} \times LAE}$} & \multirow{2}{*}{$z=5.7$: 2.7, $z=6.6$: 2.4} & $\bar{b}^{z=5.7}_{\ion{C}{ii}} \bar{I}^{z=5.7}_{\ion{C}{ii}}$ [Jy/sr] & $2700\pm3200$ & \multirow{2}{*}{Eq.~(\ref{eq:acf})/Fig.~(\ref{fig:ciilae_acf})} \\
 & & $\bar{b}^{z=6.6}_{\ion{C}{ii}} \bar{I}^{z=6.6}_{\ion{C}{ii}}$ [Jy/sr] & $2600\pm2900$ & \\
  \hline
 \multirow{3}{*}{\shortstack[l]{$P_{\rm CO(3-2) \times CO(4-3)}$ at $z\sim0.6$, \\ $P_{\rm CO(4-3) \times CO(5-4)}$ at $z\sim1.1$, \\ $P_{\rm CO(5-4) \times CO(6-5)}$ at $z\sim1.6$}} & \multirow{3}{*}{20, 26, 22} & $\alpha$ & $1.28^{+0.04}_{-0.03}$ & \multirow{3}{*}{Eq.~(\ref{eq:LCO})/Fig.~(\ref{fig:cops})} \\
 & & $\beta$ & $-0.90^{+0.50}_{-0.49}$ & \\
 & & $\sigma$ & $0.35^{+0.17}_{-0.17}$ & \\
 \hline
 \multirow{4}{*}{\shortstack[l]{$P_{\rm CO(4-3) \times \ion{C}{i}}$, $P_{\rm CO(5-4) \times \ion{C}{i}}$, \\ $P_{\rm CO(4-3) \times CO(5-4)}$ at $z\sim1.1$}} & \multirow{4}{*}{18, 13, 26} & $\sigma$ & $0.28^{+0.09}_{-0.11}$ & \multirow{4}{*}{Eq.~(\ref{eq:LCO}, \ref{eq:LCI})/Fig.~(\ref{fig:coci_cps})} \\
 & & $r_{43}$ & $0.61^{+0.18}_{-0.17}$ & \\
 & & $r_{54}$ & $0.34^{+0.10}_{-0.10}$ & \\
 & & $r_{\ion{C}{i}}$ & $0.19^{+0.06}_{-0.05}$ & \\
 \hline
 \multirow{2}{*}{\shortstack[l]{$P_{\rm CO(3-2) \times gal\ (phot)}$ at $z\sim0.4$}} & \multirow{2}{*}{20} & $\bar{b}_{\rm CO} \bar{I}_{\rm CO}$ [$\mathrm{\mu K}$] & $0.087^{+0.236}_{-0.068}$ & \multirow{2}{*}{Eq.~(\ref{eq:COxGal})/Fig.~(\ref{fig:COxGal})} \\
 & & $\bar{I}_{\rm CO, gal}$ [$\mathrm{\mu K}$] & $0.102^{+0.005}_{-0.005}$ & \\
 \hline
 \multirow{2}{*}{\shortstack[l]{$P_{\rm CO(4-3) \times gal\ (phot)}$ at $z\sim0.9$}} & \multirow{2}{*}{17} & $\bar{b}_{\rm CO} \bar{I}_{\rm CO}$ [$\mathrm{\mu K}$] & $0.129^{+0.372}_{-0.105}$ & \multirow{2}{*}{Eq.~(\ref{eq:COxGal})/Fig.~(\ref{fig:COxGal})} \\
 & & $\bar{I}_{\rm CO, gal}$ [$\mathrm{\mu K}$] & $0.229^{+0.011}_{-0.015}$ & \\
\bottomrule
\end{tabular}
\end{table*}


\section{Foreground Contamination and Mitigation Strategies} \label{sec:foreground}

To reach the full scientific potential of TIME as outlined in previous sections, systematic effects in the measurement must be carefully controlled and mitigated. Among the culprits, the astrophysical and atmospheric foreground emissions are major challenges for a line intensity mapping experiment. The astrophysical foregrounds include continuum emission such as the CMB, the CIB, and spectral line interlopers such as the low-$z$ CO rotational transition lines contaminating the high-$z$ [\ion{C}{ii}] signals. 

\subsection{Continuum Emission}

The primary and secondary CMB temperature fluctuations as well as the CIB fluctuation arising from aggregate dusty galaxy emission, are spectrally smooth with well-measured spectral characteristics \cite[][]{Planck_2020}, and are thus distinct from the spectral line fluctuations TIME aims to measure. This is analogous to the component separation problem in 21cm cosmology where the spectrally smooth synchrotron foreground emission dominates over the 21cm line fluctuation, except that the foreground-to-signal ratio for TIME is more forgiving by about one to two orders of magnitudes in intensity as a function of scales. At this level, several techniques including the principal component analysis (PCA) or the independent component analysis (ICA) have been demonstrated with data to effectively separate the continuum foreground from line emission at minimum loss of signal \cite[][]{Chang_2010, Switzer_2013, Wolz_2017a}. 

We model atmospheric emission based on data taken at Mauna Kea at 143 and 268\,GHz \cite[][]{Sayers_2010}, and scale it to the typical atmosphere opacity values for Kitt Peak. We note that the TIME spectrometer covers the full 183\,GHz to 326\,GHz band, while only the 201\,GHz to 302\,GHz sub-band is used for science.  The other channels at the high- and low-frequency edges (a total of 16) serve as atmospheric monitors \citep{Hunacek_2016JLTP}. Because they access the water lines, they combine to provide greater sensitivity to the PWV fluctuations than the combined science band channels, allowing effective tracking removal of the water vapor fluctuations to below the instrumental noise levels. Given that the PWV fluctuations amount to a time-dependent amplitude modulation of the emission constant across frequency, the same PCA-based removal techniques may be used for mitigation. 

We simulate the above astrophysical and atmospheric continuum foregrounds and add their contribution to a simulated TIME signal lightcone based on the SIDES simulation \cite[][]{Bethermin_2017} to investigate the de-contamination strategy. A detailed analysis will be described in future TIME publications, and we summarize here that with a PCA-based foreground removal technique, the CMB, CIB and atmospheic emissions can be removed to high fidelity with minimum loss of spectral line signals. As noted previously, we approximate continuum foreground removal by removing the largest spatial and spectral modes from our analysis. 

\subsection{Spectral Line Interlopers}

As noted earlier, the low-redshift CO rotational lines present a rich science opportunity to probe the molecular gas growth in the universe and to trace the LSS. They however can be confused with the high-$z$ [\ion{C}{ii}] line emission at the same observed frequencies, and present a challenge as spectral line interlopers. Several mitigation strategies have been proposed, including the usage of cross-correlation \citep{Silva_2015}, masking \citep{BKK_2015, Sun_2018}, anisotropic power spectrum effect \citep{LT_2016, Cheng_2016}, as well as map-space de-blending techniques involving deep learning \citep{Moriwaki_2020} and spectral template fitting with sparse approximation \citep{CCB_2020}. 

For TIME, for the purpose of [\ion{C}{ii}] measurement we plan to follow the targeted masking strategy laid out in \citet{Sun_2018} using an external galaxy catalog to identify and mask bright low-$z$ CO emitters. As elucidated in \citet{Sun_2018}, using the total IR luminosity as a proxy for CO emission in NIR-selected galaxies, we can clean CO interlopers to a level sufficient for a robust [\ion{C}{ii}] detection by masking no more than 10\% of the total voxels. Because of the small masking fraction required and that CO foregrounds are not spatially correlated with [\ion{C}{ii}] emission from much higher redshift, masking only causes a modest reduction of survey sensitivity\footnote{The power spectrum S/N roughly scales as the square root of survey volume via $\sqrt{N_\mathrm{m}}$, so masking $<10\%$ of voxels for removing CO interlopers only moderately changes the sensitivity.} and does not bias the [\ion{C}{ii}] measurement itself. The coupling between Fourier modes arising from the survey volume lost to extra real-space filtering (i.e., masking) can be corrected by inverting the mode-coupling matrix, $\mathbf{M}_{KK^\prime}$, which can be directly calculated from the masked data cube by generalizing the window function calculation presented in Appendix~\ref{sec:wf}. Nevertheless, CO residuals may lead to an actual loss of sensitivity, although methods such as cross-correlation can be used to quantify the residual line-interloper contamination. A detailed presentation of how to correct for the mode coupling due to foreground cleaning and estimate the level of residuals is beyond the scope of this work. We therefore postpone a more thorough analysis of these issues to future publications.


\section{Discussion} \label{sec:discuss}

\subsection{Implications and Limitations of Power Spectral Constraints from TIME and TIME-EXT} \label{sec:discuss:scale}

Due to their distinct physical origins, clustering (proportional to the square of the first luminosity moment) and shot-noise (proportional to the second luminosity moment) components of the power spectrum have different sensitivities to different populations of line emitters. For this reason, while astrophysical parameters may still be constrained by measuring only one single component such as the shot noise, it is favorable to access the full power spectrum at different scales in order to maximize the effectiveness of parameter estimation \cite[][]{YF_2019}. Given the projection effect of its line-scan geometry, TIME and TIME-EXT will directly measure a 2D power spectrum much less scale-dependent compared with the true 3D power spectrum, as shown in Figure~\ref{fig:ciionlyps}, and any particular observed mode $K$ results from a non-trivial mixing of sky modes $k$ described by the window function $W_{ii} \left( k, \vec{K}_i \right)$. The connection between parameter constraints and observed modes is therefore less straightforward. Nevertheless, although the shot noise dominates large $K \gtrsim 1\,h/\mathrm{Mpc}$ modes for both [\ion{C}{ii}] and CO signals, the access to the clustering component at smaller $K$'s helps lift the degeneracy among parameters affecting the shape of the power spectrum differently. For instance, the faint-end modulation factor $\xi$ has a minute effect on the shot-noise power dominated by bright sources. Hence, it can be most easily constrained by measuring the clustering component with higher significance, as indicated by the comparison between TIME and TIME-EXT in the top panel of Figure~\ref{fig:ciionlyps}. 

For TIME, our fiducial model predicts that the uncertainties in the [\ion{C}{ii}] auto-power spectra and CO cross-power spectra are dominated by the instrument noise and sample variance, respecitively. Therefore, with the same survey strategy but more than 10 times greater survey power, TIME-EXT is expected to outperform TIME by measuring the [\ion{C}{ii}] auto-power spectrum at a high significance of S/N$>$20. This allows the [\ion{C}{ii}] parameters to be constrained to a level limited by their intrinsic degeneracies, which have to be broken by additional observables such as the one-point statistics \cite[][]{Breysse_2017} and/or data sets such as independent constraints from galaxy detection. Further insights into [\ion{C}{ii}] line emission from the EoR can be obtained from LIM measurements beyond the auto-correlation, such as the cross-correlation of [\ion{C}{ii}] with other EoR probes, some examples of which are discussed in the next sub-section. 

\begin{figure}[t]
    \centering
    \begin{minipage}{.98\linewidth}
        \includegraphics[width=\textwidth]{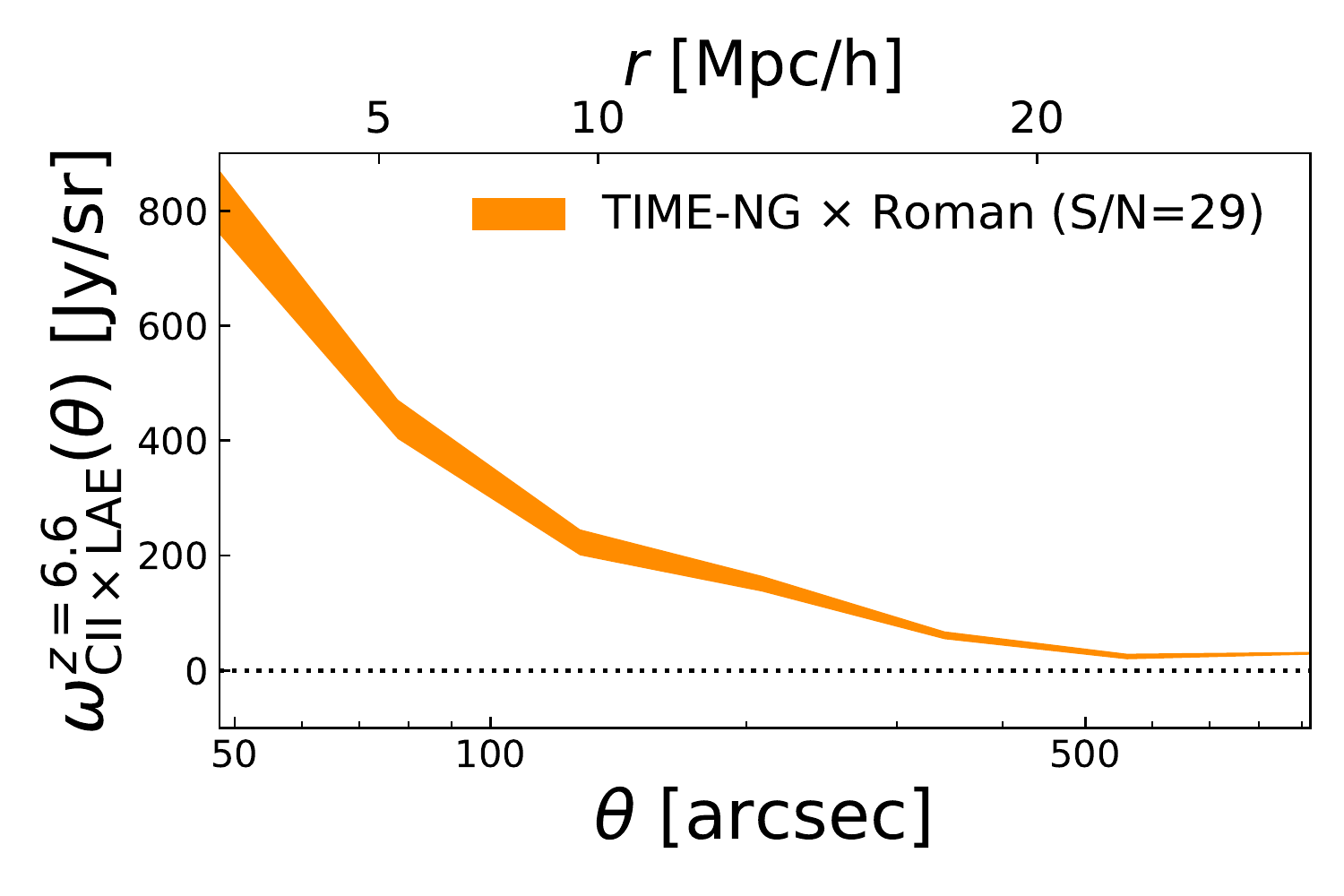}
    \end{minipage}
    \begin{minipage}{.98\linewidth}
        \includegraphics[width=\textwidth]{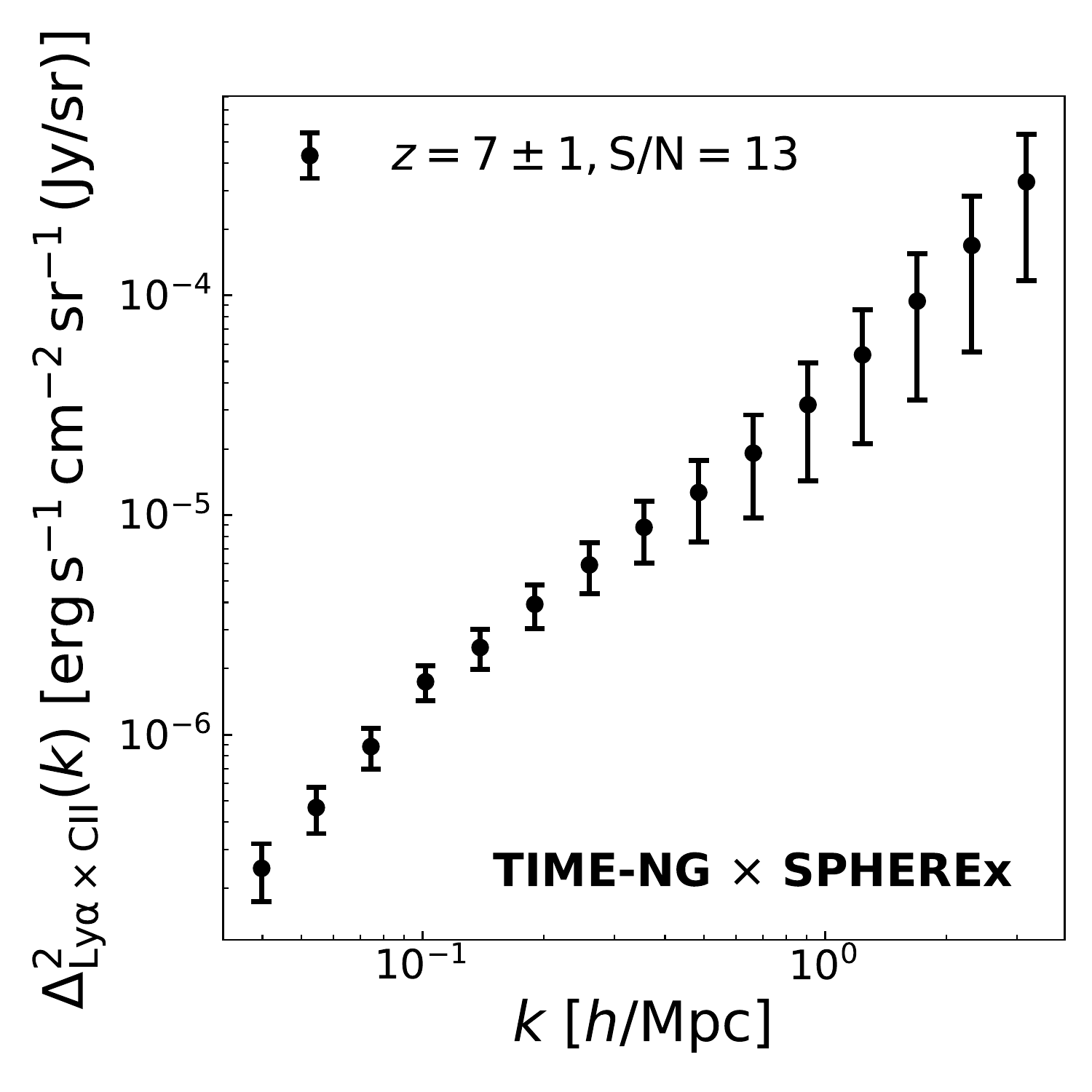}
    \end{minipage}
    \caption{Synergy between TIME-NG and observations of LAEs and Ly$\alpha$ intensity fluctuations, assuming an overlapped survey area of 10\,deg$^2$ and a factor of 120 improvement in survey power from TIME to TIME-NG. Top: predicted angular cross-correlation function at $z=6.6$ between [\ion{C}{ii}] intensity measured by TIME-NG and LAEs observed with a Roman Space Telescope GO survey reaching a minimum Ly$\alpha$ luminosity of $\log (L_{\rm Ly\alpha,min}/ \mathrm{erg\,s^{-1}})$ = 42.7 (or $m_{\rm lim}^{\rm AB}=25.5$, which implies $n_{\rm LAE} \simeq 10^{-4}\,\mathrm{Mpc}^{-3}$). Bottom: predicted dimensionless cross-power spectrum at $z\approx7$ between [\ion{C}{ii}] intensity measured by TIME-NG and Ly$\alpha$ intensity measured in SPHEREx deep field (with a 1-$\sigma$ surface brightness sensitivity level of $10^3\,\mathrm{Jy/sr}$). }
    \label{fig:TIMENG}
\end{figure}

\subsection{Next-generation [\ion{C}{ii}] LIM Experiment} \label{sec:discuss:ng}

So far, we have outlined the rich and diverse science enabled by TIME and TIME-EXT, as two distinct phases of a first-generation [\ion{C}{ii}] LIM experiment. In the future, we anticipate the sensitivity, in part limited by the number of spectrometers on TIME, can be advanced by the development of a more densely populated focal plane empowered by on-chip mm-wave spectrometers \citep{Redford_2018, Endo_2019, Karkare_2020JLTP}. A next-generation TIME experiment, TIME-NG, can have an increased number of spectrometers by at least a factor of 10. Combined with a lower atmospheric loading from a better observing site and with a dedicated telescope, TIME-NG can achieve at least three times lower NEI level compared to TIME, and expect a few thousands of hours of integration time. All these factors can result in another order of magnitude improvement in survey power to measure the [\ion{C}{ii}] power spectrum compared with TIME-EXT.

The high-significance measurements of the [\ion{C}{ii}] statistical properties will not only characterize the science cases summarized in this paper with higher precision, but more significantly, enrich the multi-tracer probe of reionization in the coming decade to further our understanding of reionization beyond presented here \cite[e.g.,][]{Chang_2019BAAS}. The improved sensitivity of TIME-NG will make possible a variety of cross-correlation analyses between [\ion{C}{ii}] and other tracers of the EoR, including LAEs and LBGs to be surveyed by the Nancy Grace Roman and Euclid Telescopes in the near future, as well as emission lines from other LIM experiments such as the Ly$\alpha$ diffuse emission from SPHEREx \citep{Dore_2014, Dore_2016}, and the 21cm emission from HERA \citep{DeBoer_2017} and the SKA \citep{Koopmans_2015}. Using models [\ion{C}{ii}] and LAEs presented in this work together with physical models of Ly$\alpha$ and 21cm line motivated by observations \cite[e.g.,][]{Gong_2012, Chang_2015, Heneka_2017, Dumitru_2019}, we estimate that, for a 10\,deg$^2$ survey and a TIME-NG-like capability with 3000 hours of integration, the [\ion{C}{ii}]-LAE cross-correlation with an anticipated Roman General Observer (GO) survey \cite[][]{Spergel_2015} of the same size and a depth of $m_{\rm lim}^{\rm AB}=25.5$ can be measured at high significance, as shown in the top panel of Figure~\ref{fig:TIMENG} and elaborated in the caption. In addition, the [\ion{C}{ii}]-Ly$\alpha$ and [\ion{C}{ii}]-21cm cross-power spectra at $z\sim7$, with SPHEREx and SKA, respectively, can both be solidly detected at an S/N $\gtrsim  5$. The bottom panel of Figure~\ref{fig:TIMENG} shows the [\ion{C}{ii}]--Ly$\alpha$ cross-power spectrum estimated based on the Ly$\alpha$ power spectrum from \citet{Heneka_2017}.  We expect a significant detection by overlapping TIME-NG with SPHEREx deep field, whose surface brightness sensitivity level is taken to be $10^3$\,Jy/sr\footnote{See the public file containing the forecasted surface brightness sensitity level of SPHEREx at \url{https://github.com/SPHEREx/Public-products/blob/master/Surface_Brightness_v28_base_cbe.txt}.}. Subsequent multi-tracer analyses, based on detecting these  cross-correlations at circum-galactic to inter-galactic scales during reionization, will provide a comprehensive view of how ionized bubbles grew out of the production and escape of ionizing photons from galaxies. 


\section{Summary} \label{sec:conclusions}

Complementary to conventional galaxy surveys, intensity mapping of the redshifted [\ion{C}{ii}] and CO lines from the reionization era and the epoch of peak star formation reliably probes aggregate line emission, offering invaluable insight into the total cosmic star formation and the evolution of the molecular gas content of galaxies during those epochs. 

We presented a modeling framework that self-consistently models the target signals of TIME and predicts its capability of constraining a series of physical quantities of interest. Using forecasts based on realistic TIME instrument specifications and our fiducial model informed by observations available to date, we identified a line-scan survey geometry optimized for measuring of [\ion{C}{ii}] intensity fluctuations 
from the EoR. 

Starting from the optimized line survey, we generate mock power spectra of our [\ion{C}{ii}] signal as well as line interlopers including rotational CO and [\ion{C}{i}] line from lower redshifts. We then analyzed results within a Bayesian inference framework to forecast parameter constraints, given the sensitivity levels of TIME and TIME-EXT (the extended TIME survey from the LCT). Based on our analysis, we expect TIME(-EXT) to measure the [\ion{C}{ii}] power spectrum during reionization with a total S/N greater than 5 (20) and thereby provide robust constraints on the [\ion{C}{ii}] luminosity density and the cosmic SFRD over $6 \lesssim z \lesssim 9$. Combining such measurements with the Thomson scattering optical depth of CMB photons and quasar absorption spectra, we also expect to constrain the population-averaged escape fraction of ionizing photons to the level of $f_{\rm esc} \approx 0.1^{+0.2}_{-0.1}$ and $f_{\rm esc} \approx 0.1^{+0.10}_{-0.05}$ respectively for the two phases of TIME experiment. Such measurements are independent of the faint-end extrapolation of galaxy LF, which will be robustly constrained by TIME-EXT.

Through in-band cross-correlations, we predict that TIME and TIME-EXT will measure the cross-power spectra of interloping CO and [\ion{C}{i}] lines at $0.5 \lesssim z \lesssim 2$ with high significance (S/N$>$10). Thanks to the wide bandwidth, these cross-correlation measurements can be used to infer the cosmic molecular gas density near cosmic noon assuming prior knowledge of the CO rotational ladder, whereas the mutual cross-correlations among CO(4-3), CO(5-4), and [\ion{C}{i}] lines at $z\sim1.1$ can extract the individual line strengths, shedding light on the excitation state of CO and the relation with neutral carbon in the ISM, averaged over the entire galaxy population. 

The synergy of TIME maps and external galaxy surveys serves as a useful sanity check of foreground removal, while also providing additional astrophysical information about the overlapping galaxy population. We therefore analyze the prospects for cross-correlating TIME [\ion{C}{ii}] maps with narrow-band selected LAEs at $z=5.7$ and $z=6.6$ from the Subaru HSC survey. Due to TIME's limited survey size, only upper limits can be extracted on $\bar{b}_{\ion{C}{ii}} \bar{I}_{\ion{C}{ii}}$ from the angular cross-correlation function of [\ion{C}{ii}] and LAEs. At lower redshifts, we expect significant detections of the cross-power spectra between TIME CO maps and galaxies with known redshifts. From these shot-noise-dominated measurements, we placed stringent constraints on the mean CO intensity attributed to the galaxy sample of interest. 

Finally, we discuss that a next-generation [\ion{C}{ii}] experiment, TIME-NG, can map [\ion{C}{ii}] intensity fluctuations during the EoR with high significance on $\sim 10\,\mathrm{deg}^2$ scales, opening exciting opportunities for multi-tracer analyses based on cross-correlating [\ion{C}{ii}] maps with other EoR probes such as LAEs, Ly$\alpha$, and the 21cm line. 

\acknowledgments
We would like to thank the anonymous referees for their comments that improved the manuscript. We are indebted to Lluis Mas-Ribas for helpful comments on an early version of the paper and Lin Yan for discussion about the [\ion{C}{ii}] luminosity function measured from the ALPINE survey. We are also grateful for Garrett (Karto) Keating for compiling the observational constraints on molecular gas density. TCC and GS acknowledge support from the JPL Strategic R\&TD awards. AC acknowledges support from NSF AST-1313319 and 2015-2016 UCI Office
of Research Seed Funding Award. DPM and RPK were supported by the National Science Foundation through CAREER grant AST-1653228. RPK was supported by the National Science Foundation through Graduate Research Fellowship grant DGE-1746060. ATC was supported by a KISS postdoctoral
fellowship and a National Science
Foundation Astronomy and Astrophysics Postdoctoral Fellowship under Grant No. 1602677. This work is supported by National Science Foundation award number 1910598. Part of the research described in this paper was carried out at the Jet Propulsion Laboratory, California Institute of Technology, under a contract with the National Aeronautics and Space Administration. 

\textit{Software:} \texttt{ares} \cite[][]{Mirocha_2017}, \texttt{corner} \cite[][]{FM_2016}, \texttt{emcee} \cite[][]{FM_2013PASP}, \texttt{matplotlib} \cite[][]{Hunter_2007}, \texttt{numpy} \cite[][]{Walt_2011}, \texttt{pygtc} \cite[][]{BC_2016}, and \texttt{scipy} \cite[][]{Jones_2001}. 

\bibliography{science}

\appendix

\section{Modeling the Star Formation Efficiency} \label{sec:f_star}

\subsection{Dust Correction}
Following \citet{SF_2016} and \citet{MF_2019}, we derive the dust extinction correction at any given magnitude $M_{\rm UV}$ by combining the \citet{Meurer_1999} relation between the dust extinction (evaluated at 1600\,\AA) $A_{\rm UV}$ and the slope of UV continuum $\beta$, 
\begin{equation}
A_{\rm UV} = 4.43 + 1.99 \beta \geq 0~,
\end{equation}
with the $\beta$--$M_{\rm UV}$ relation, which is modeled by \citet{Bouwens_2014} as a linear relation with a constant gaussian error $\sigma_\beta = 0.34$. The correction factor averaged over the $\beta$-distribution, $\langle A_{\rm UV} \rangle$, is then applied to obtain the dust-corrected UVLFs from the observed ones, and consequently define the SFRs with and without the dust correction. 

\subsection{The Star Formation Efficiency as a Function of Halo Mass}
\begin{figure}[h!]
\centering
\includegraphics[width=0.6\textwidth]{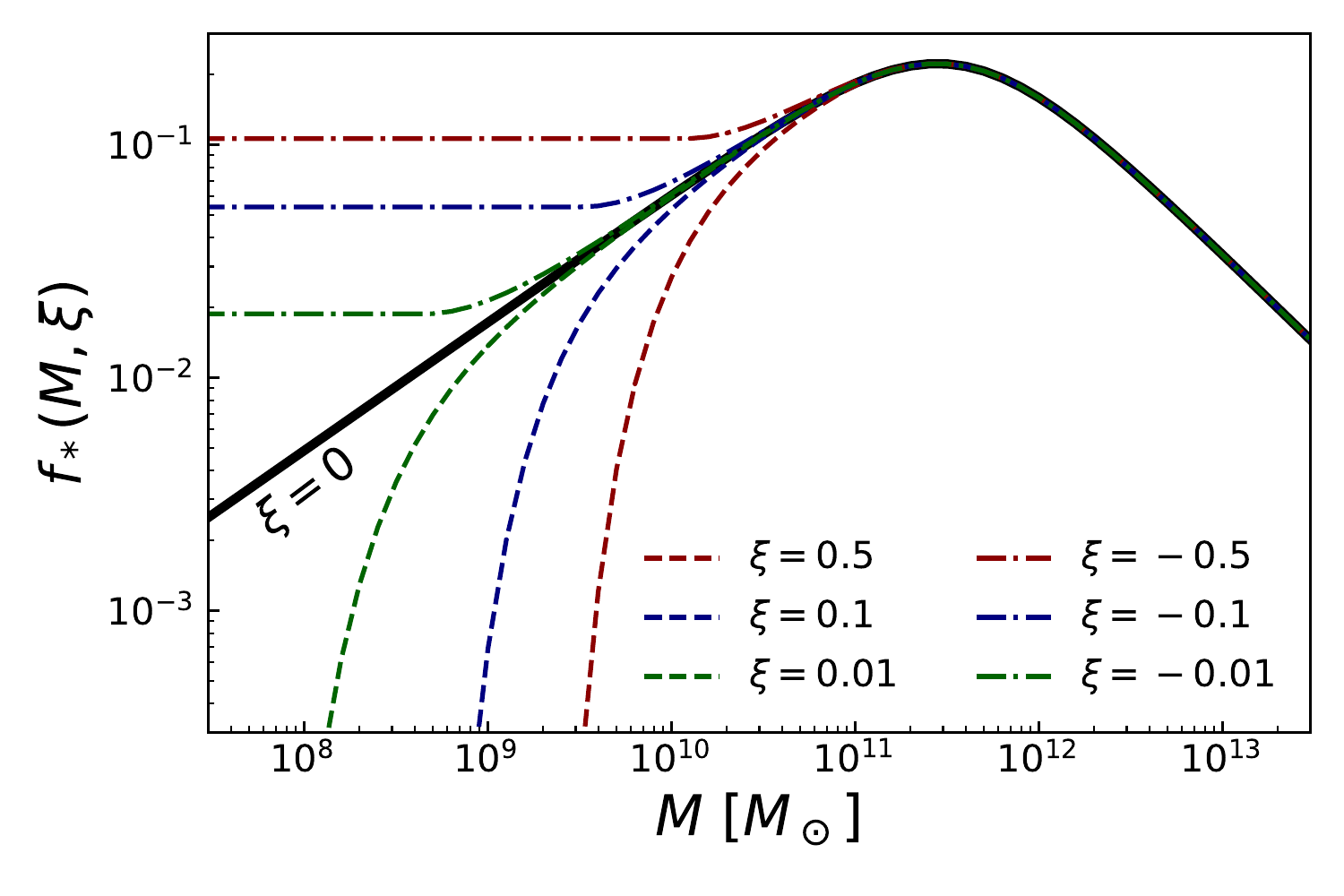}
\caption{The star formation efficiency $f_*$ as a function of halo mass. Curves corresponding to different choices of $\xi$, as defined in Equation~(\ref{eq:sfe}), are shown to illustrate how our model captures the uncertainty in the mass dependence at the low-mass end. }
\label{fig:sfe}
\end{figure}

As discussed in Section~\ref{sec:model-eor}, the SFE $f_*$, which together with the mass accretion rate of dark matter halos determine the SFH, is an essential quantity in our modeling framework. The low-mass end behavior of $f_*$ is particularly important because, in the context of luminosity function, the cosmic star formation rate and therefore the total budget of ionizing photons are determined by both the steepness of the faint-end slope and the cutoff luminosity. As a result, our $f_*$ model aims to maximize the flexibility to explore the degeneracy between these two quantities by extending the low-mass end unconstrained by abundance matching differently, ranging from an asymptote to a constant value to an exponential decay. Specifically, we define a redshift-independent SFE that can be expressed as
\begin{equation}
f_*(M) = \frac{f_{*,0}}{\left( \frac{M}{M_{\rm p}} \right)^{\gamma_{\rm lo}(M)} + \left( \frac{M}{M_{\rm p}} \right)^{\gamma_{\rm hi}}}~,
\label{eq:sfe}
\end{equation}
where $f_{*,0} = 0.22$ is twice the maximum possible SFE peaking at $M_{\rm p} = 3.6 \times 10^{11}\,M_\odot$, and $\gamma_{\rm hi} = 0.77$ specifies the mass dependence of the high-mass end. The low-mass end is allowed to deviate from perfect power law by
\begin{equation}
\gamma_{\rm lo}(M) = -0.55 \times 10^{\xi / (M/M_{\rm c})}~,
\label{eq:defn_xi}
\end{equation}
where $M_{\rm c} = 3 \times 10^9\,M_\odot$ is the characteristic halo mass for a deviation from power law at the low-mass end. $\xi$ is a free parameter defining the level of deviation, with $\xi = 0$ corresponding to the best-fit double power-law model of $f_*$ calibrated against the observed UVLFs at $5<z<9$ after the dust correction \citep[see also][]{Mirocha_2017}. Note that when $\xi < 0$, we impose a ceiling on $f_*$ such that it asymptotes to a constant value rather than blowing up. A few sample $f_*(M)$ curves with different choices of $\xi$ are shown in Figure~\ref{fig:sfe}. 

\section{Window Function} \label{sec:wf}

Following \citet{Dodelson_2003}, we can express the \textit{dimensionless} covariance matrix of a pair of instrument modes, whose wavenumbers are denoted by $\vec{K}_i$ and $\vec{K}_j$ to distinguish from the wavenumber $\vec{k}$ of the sky mode before being filtered by the survey geometry, as
\begin{align}
C_{ij}^S \left( \vec{K}_i, \vec{K}_j \right) & = \langle \delta_{\mathbf{K}_i}^* \delta_{\mathbf{K}_j} \rangle \nonumber \\
& = \int \frac{\mathrm d k \mathrm d \theta \mathrm d \phi}{(2\pi)^3}\, k^2 \sin \theta P(k) \tilde{\psi}_{i} \tilde{\psi}_{j}^{*} \nonumber \\
& = \int \mathrm d k \frac{k^2}{2\pi^2} P(k) W_{ij} \left( k, \vec{K}_i, \vec{K}_j \right)~, \label{eq:p3d_to_cov}
\end{align}
where the weighting function $\tilde{\psi}(k)$ is the Fourier transform of the real-space selection function $\psi(x)$ which describes the actual geometry with survey volume $V_S = \int \mathrm d^3 x = L_x L_y L_z$ and satisfies $\int \mathrm d^3 x \psi(x) = 1$. Note that for simplicity we have assumed the actual fluctuations on sky to be isotropic such that we can replace sky modes $\vec{k}$ with $k$. We consequently define the window function $W_{ij}$ to be the angular average of the inner product of two weighting functions, $\tilde{\psi}^{}_{i} \tilde{\psi}^{*}_{j}$, which satisfies $V_{S} \int \mathrm d k k^2 W_{ij}(k) / 2\pi^2 = 1$ for a 3D survey due to the unitarity of Fourier transform. Effectively, Equation~(\ref{eq:p3d_to_cov}) can be interpreted as a projection from the sky frame to the observing frame that results in mode mixing, where $W_{ij}$ serves as the projection kernel. 

For a line intensity mapping experiment like TIME, the survey volume within which fluctuations of the intensity field are measured can be effectively approximated with a three-dimensional box of dimensions $L_x$, $L_y$ and $L_z$. Specifically, in the case of a 2D line scan, we require that $L_y \ll L_x, L_z$. The corresponding selection function can be specified by a product of top-hat functions, which implies a weighting function in $k$ space of the form
\begin{align}
\tilde{\psi}_{i} \left( \vec{k}, \vec{K}_i \right) & = \int \mathrm d^3 x e^{i(\vec{K}_i - \vec{k}) \cdot \vec{x}} \psi \left( \vec{x} \right) \nonumber \\ 
& = \int \mathrm d^3 x e^{i(\vec{K}_i - \vec{k}) \cdot \vec{x}} \prod_{m=x,y,z} \frac{1}{L_m} \Theta \left(\frac{L_m}{2} \pm m\right) \nonumber \\
& = j_0\left( q_x L_x/2 \right) j_0\left( q_y L_y/2 \right) j_0\left( q_z L_z/2 \right)~,
\label{eq:kK_mapping}
\end{align}
where $\vec{K}_i = (K_{i,x}, 0, K_{i,z})$, $q_m = K_{i,m} - k_m$ and $j_0(x)=\sin(x)/x$ is the spherical Bessel function of the first kind. The covariance matrix $C_{ij}^S$ is thus related to the observed 2D power spectrum (in units of area) by
\begin{equation}
\mathcal{P}\left( \vec{K}_i \right) = L_x L_z C_{ii}^S \left( \vec{K}_i \right)~,  
\label{eq:cov_to_p2d}
\end{equation}
where we only consider the diagonal terms assuming the correlation between different instrument modes are negligible. We note that for the line scan considered, we must normalize Equation~(\ref{eq:cov_to_p2d}) by dividing it with $V_{S} \int \mathrm d k k^2 W_{ii}(k) / 2\pi^2 < 1$ to account for the difference between 2D and 3D power.

\section{Uncertainties of Auto- and Cross-Power Spectra} \label{sec:unc_ps}

Here we present a derivation of the errors on auto and cross-power spectra, which is a simplified version of that given by \citet{VL_2010}. For the cross-power spectrum of two real fields $f_1$ and $f_2$, we define the estimator to be the \textit{real part} of the inner product of their Fourier transforms $\tilde{f}^{}_1$ and $\tilde{f}^{}_2$, namely
\begin{equation}
\hat{P}_{1,2} = \frac{V}{2} \left( \tilde{f}^{}_1 \tilde{f}^*_2 + \tilde{f}^*_1 \tilde{f}^{}_2 \right)
\label{eq:cross_est}
\end{equation}
and its variance can be consequently written as
\begin{equation}
var \left(\hat{P}_{1,2} \right) = \delta \hat{P}_{1,2}^2 = \langle \hat{P}_{1,2}^2 \rangle - \langle \hat{P}_{1,2} \rangle^2~,
\end{equation}
where $\langle ... \rangle$ stands for averaging over the statistical ensemble. Expanding the above expression with Equation~(\ref{eq:cross_est}), we have
\begin{equation}
\delta P_{1,2}^2 = \langle \hat{P}_{1,2}^2 \rangle - \langle \hat{P}_{1,2} \rangle^2 = \left\langle \frac{V^2}{4} \left( \tilde{f}^{}_1 \tilde{f}^*_2 + \tilde{f}^*_1 \tilde{f}^{}_2 \right)^2 \right\rangle - P_{1,2}^2 \label{eq:real_part} = \frac{V^2}{2} \left\langle \tilde{f}^{}_1 \tilde{f}^{}_1 \tilde{f}^*_2 \tilde{f}^*_2 \right\rangle + \frac{V^2}{2} \left\langle \tilde{f}^{}_1 \tilde{f}^*_1 \tilde{f}^{}_2 \tilde{f}^*_2 \right\rangle - P_{1,2}^2~.
\end{equation}
We now use Wick's theorem to rewrite the four-term product as the sum of three cross products
\begin{equation}
\left\langle \tilde{f}_1 \tilde{f}_1 \tilde{f}^*_2 \tilde{f}^*_2 \right\rangle = \left\langle \tilde{f}_1 \tilde{f}_1 \right\rangle \left\langle \tilde{f}^*_2 \tilde{f}^*_2 \right\rangle + 
2 \left\langle \tilde{f}_1 \tilde{f}^*_2 \right\rangle^2 = 2 \left\langle \tilde{f}_1 \tilde{f}^*_2 \right\rangle^2, 
\end{equation}
and
\begin{equation}
\left\langle \tilde{f}_1 \tilde{f}^*_1 \tilde{f}_2 \tilde{f}^*_2 \right\rangle = \left\langle \tilde{f}_1 \tilde{f}_2 \right\rangle \left\langle \tilde{f}^*_1 \tilde{f}^*_2 \right\rangle + 
\left\langle \tilde{f}_1 \tilde{f}^*_1 \right\rangle \left\langle \tilde{f}_2 \tilde{f}^*_2 \right\rangle + 
\left\langle \tilde{f}_1 \tilde{f}^*_2 \right\rangle \left\langle \tilde{f}^*_1 \tilde{f}_2 \right\rangle = \left\langle \tilde{f}_1 \tilde{f}^*_1 \right\rangle \left\langle \tilde{f}_2 \tilde{f}^*_2 \right\rangle + 
\left\langle \tilde{f}_1 \tilde{f}^*_2 \right\rangle \left\langle \tilde{f}^*_1 \tilde{f}_2 \right\rangle. 
\end{equation}
Note that for the Fourier transform $\tilde{f}$ of a real field $f$, the first terms in the two expressions above vanish because of the Hermitianity condition $\tilde{f}^*(k) = \tilde{f}(-k)$ and the fact that different $k$ modes are statistically independent. For the ensemble average, we should have $\left\langle \tilde{f}_1 \tilde{f}^*_2 \right\rangle = \left\langle \tilde{f}^*_1 \tilde{f}_2 \right\rangle$. As a result, the variance becomes
\begin{equation}
\delta P_{1,2}^2 = \frac{V^2}{2} \left\langle \tilde{f}_1 \tilde{f}_1 \tilde{f}^*_2 \tilde{f}^*_2 \right\rangle + \frac{V^2}{2} \left\langle \tilde{f}_1 \tilde{f}^*_1 \tilde{f}_2 \tilde{f}^*_2 \right\rangle - P_{1,2}^2 = \frac{3V^2}{2} \left\langle \tilde{f}_1 \tilde{f}^*_2 \right\rangle^2 + \frac{V^2}{2} \left\langle \tilde{f}_1 \tilde{f}^*_1 \right\rangle \left\langle \tilde{f}_2 \tilde{f}^*_2 \right\rangle - P_{1,2}^2~. 
\end{equation}
Using the definitions of auto- and cross-power spectra (as the ensemble average of the Fourier pair product), we finally obtain
\begin{equation}
\delta P_{1,2}^2 = \frac{3}{2} P_{1,2}^2 + \frac{1}{2} P_1 P_2 - P_{1,2}^2 = \frac{1}{2} \left( P_{1,2}^2 + P_1 P_2 \right)
\end{equation}
For the auto power spectrum, the variance given by Equation~(\ref{eq:real_part}) simply becomes
\begin{equation}
\delta P_1^2 = V^2 \left\langle  \tilde{f}_1 \tilde{f}_1 \tilde{f}^*_1 \tilde{f}^*_1 \right\rangle - P_1^2 = 2 V^2 \left\langle \tilde{f}_1 \tilde{f}^*_1 \right\rangle^2 - P_1^2 = P_1^2
\end{equation}
 
\end{document}